\documentclass[a4paper,11pt]{article}
\pdfoutput=1 

\usepackage{jheppub} 

\usepackage[T1]{fontenc} 

\usepackage{graphicx}
\usepackage{epsfig}
\usepackage{rotating}
\usepackage{amssymb}
\usepackage{subfigure}
\usepackage{dsfont}
\usepackage{psfrag}
\usepackage{amsmath,euscript,array,mathrsfs}
\usepackage{bbold}
\usepackage{epsf}

\usepackage{grffile}

\newcommand{\beq}{\begin{equation}}
\newcommand{\eeq}{\end{equation}}
\newcommand{\beqs}{\begin{eqnarray}}
\newcommand{\eeqs}{\end{eqnarray}}

\newcommand{\gsim}{\mathrel{\raisebox{-
.6ex}{$\stackrel{\textstyle>}{\sim}$}}}
\newcommand{\Tr}{{\rm Tr}}

\def\hbar{\hspace{0pt}\raisebox{1pt}{$-$} \hspace{-7pt} h}

\def\di{\mbox{d}}
\def\r{\rho}

\newcommand{\be}{\begin{equation}}
\newcommand{\ee}{\end{equation}}

\newcommand{\bea}{\begin{eqnarray}}
\newcommand{\eea}{\end{eqnarray}}
\newcommand{\nn}{\nonumber}

\def\lbldef#1#2{\expandafter\gdef\csname #1\endcsname {#2}}

\def\href#1#2{#2}

\newcommand{\ber}{\begin{eqnarray}}
\newcommand{\eer}{\end{eqnarray}}

\newcommand{\beqar}{\begin{eqnarray}}

\newcommand{\eeqar}{\end{eqnarray}}

\def\calE{{\mathcal{E}}}

\def\calW{{\mathcal{W}}}

\newcommand{\dsl}
  {\kern.06em\hbox{\raise.15ex\hbox{$/$}\kern-.56em\hbox{$\partial$}}}

\newcommand{\eeqarr}{\end{eqnarray}}
\newcommand{\ZZ}{{\rm \kern 0.275em Z \kern -0.92em Z}\;}

\def\CC{{\mathchoice
{\rm C\mkern-8mu\vrule height1.45ex depth-.05ex
width.05em\mkern9mu\kern-.05em}
{\rm C\mkern-8mu\vrule height1.45ex depth-.05ex
width.05em\mkern9mu\kern-.05em}
{\rm C\mkern-8mu\vrule height1ex depth-.07ex
width.035em\mkern9mu\kern-.035em}
{\rm C\mkern-8mu\vrule height.65ex depth-.1ex
width.025em\mkern8mu\kern-.025em}}}

\def\RR{{\rm I\kern-1.6pt {\rm R}}}

\def\ZZ{{\rm Z}\kern-3.8pt {\rm Z} \kern2pt}
\def\IB{\relax{\rm I\kern-.18em B}}
\def\ID{\relax{\rm I\kern-.18em D}}
\def\II{\relax{\rm I\kern-.18em I}}
\def\IP{\relax{\rm I\kern-.18em P}}

\newcommand{\bear}{\begin{eqnarray}}
\newcommand{\eear}{\end{eqnarray}}

\def\to{\rightarrow}

\def\to{\rightarrow}

\def\r{\rho}                                     

\def\6{\partial}

\def\bea{\begin{eqnarray}}
\def\eea{\end{eqnarray}}

\def\beqx{\begin{displaymath}}
\def\eeqx{\end{displaymath}}

\newcommand{\bmat}{\left(\begin{array}}
\newcommand{\emat}{\end{array}\right)}

\def\r{\rho}

\newcommand \Nf {\ensuremath{N_\mathrm{f}}}

\def\bo{{\raise-.3ex\hbox{\large$\Box$}}}               
\def\face{{\raise.2ex\hbox{$\displaystyle \bigodot$}\mskip-2.2mu \llap {$\ddot
        \smile$}}}                                   
\def\>{\rangle}                                      
\def\<{\langle}                                      

\def\leftrightarrowfill{$\mathsurround=0pt \mathord\leftarrow \mkern-6mu
        \cleaders\hbox{$\mkern-2mu \mathord- \mkern-2mu$}\hfill
        \mkern-6mu \mathord\rightarrow$}        
\def\dvec#1{\vbox{\ialign{##\crcr
        \leftrightarrowfill\crcr\noalign{\kern-1pt\nointerlineskip}
        $\hfil\displaystyle{#1}\hfil$\crcr}}}           
\def\Tr{{\rm Tr \,}}                                    

\def\-{\hphantom{-}}

\def\dee{\mathrm{d}}

\title{$Sp(4)$ gauge theory on the lattice: towards $SU(4)/Sp(4)$ composite Higgs (and beyond)}

\author[a,b]{Ed Bennett,}
\author[c]{Deog Ki Hong,}
\author[c,d]{Jong-Wan Lee,}
\author[e]{C.-J. David Lin,}
\author[f]{Biagio Lucini,}
\author[a]{Maurizio Piai,}
\author[a,g]{Davide Vadacchino}

\affiliation[a]{Department of Physics, College of Science, Swansea University,
Singleton Park, SA2 8PP, Swansea, Wales, UK}
\affiliation[b]{Swansea Academy of Advanced Computing, Swansea University,
Singleton Park, SA2 8PP, Swansea, Wales, UK}
\affiliation[c]{Department of Physics, Pusan National University, Busan 46241, Korea}
\affiliation[d]{Extreme Physics Institute, Pusan National University, Busan 46241, Korea}
\affiliation[e]{Institute of Physics, National Chiao-Tung University, 1001 Ta-Hsueh Road, Hsinchu 30010, Taiwan}
\affiliation[f]{Department of Mathematics, College of Science, Swansea University,
Singleton Park, SA2 8PP, Swansea, Wales, UK}
\affiliation[g]{INFN, Sezione di Pisa, Largo Pontecorvo 3, 56127 Pisa, Italy}

\emailAdd{e.j.bennett@swansea.ac.uk}
\emailAdd{dkhong@pusan.ac.kr}
\emailAdd{jwlee823@pusan.ac.kr}
\emailAdd{dlin@mail.nctu.edu.tw}
\emailAdd{b.lucini@swansea.ac.uk}
\emailAdd{m.piai@swansea.ac.uk}
\emailAdd{davide.vadacchino@pi.infn.it}

\abstract{The $Sp(4)$ gauge theory
with two Dirac fundamental flavours
provides a candidate for the microscopic origin of composite-Higgs models
based on the $SU(4)/Sp(4)$ coset.
We employ a combination of two different, complementary strategies for the numerical lattice calculations,
based on the Hybrid Monte Carlo  and on the Heat Bath algorithms.
We perform pure Yang-Mills,  quenched computations and 
exploratory studies with dynamical Wilson fermions.

We present the first results in the literature for the spectrum of glueballs of the pure $Sp(4)$ Yang-Mills theory,
an EFT framework for the interpretation of the masses and decay constants of the lightest pion, vector and axial-vector mesons,
and a preliminary calculation of the latter in the quenched approximation.
We show  the first numerical evidence of a bulk phase transition in the lattice theory with dynamical Wilson fermions,
and perform  the technical steps necessary to set up future investigations of the mesonic 
spectrum of the full theory.
}

\preprint{PNUTP-17/A07}

\begin{document}
\maketitle
\flushbottom



\section{Introduction and motivation}
\label{Sec:Introduction}

The study of $Sp(2N)$ gauge theories has the potential to unveil 
many new phenomena of general relevance in  field theory
and its phenomenological applications within high energy physics.
The recent progress in lattice gauge theory makes it an ideally suited 
non-perturbative instrument for this type of investigation.
The literature on the subject is somewhat limited (see for instance~\cite{Holland:2003kg}).
There is a large number of questions that we envision can be answered 
with dedicated lattice studies, and in this introduction we list them and discuss  the long-term 
research programme that this work initiates, before specialising to the 
investigations and results we will report upon in this paper.

\subsection{The $Sp(2N)$ research programme}

In the context of physics beyond the standard model (BSM), the discovery of the Higgs particle~\cite{Aad:2012tfa,Chatrchyan:2012xdj},
combined with the absence of evidence for new physics at the TeV scale from LHC direct searches,
exacerbates the little hierarchy problem. If the mass of the Higgs particle
has a common dynamical origin with hypothetical new physics at multi-TeV scales, the low-energy
effective field theory (EFT) description of the system is in general unnatural (fine-tuned).
The framework of Higgs compositeness we refer to in this paper~\cite{Kaplan:1983fs,Georgi:1984af,Dugan:1984hq,Agashe:2004rs,Katz:2005au,Contino:2006qr,Barbieri:2007bh,
Lodone:2008yy,Ferretti:2013kya,Cacciapaglia:2014uja,Arbey:2015exa,Vecchi:2015fma,Panico:2015jxa,Ferretti:2016upr,
 Agugliaro:2016clv,Alanne:2017rrs,Feruglio:2016zvt,Fichet:2016xvs,Galloway:2016fuo,Alanne:2017ymh,Csaki:2017cep,Chala:2017sjk}
addresses this problem by postulating the existence of a new underlying strongly-coupled
theory, in which an internal global symmetry is broken spontaneously by the dynamically-generated condensates,
resulting in a set of parametrically light pseudo-Nambu-Goldstone bosons (PNGBs). 
One writes their EFT description in terms of  
scalar fields,
and weakly couples it 
to the standard-model gauge bosons and fermions. Four of the PNGBs of the resulting EFT are
interpreted as the Higgs doublet, hence providing an elegant symmetry
argument for the lightness of the associated particles.

Particular attention has been devoted to models based on the 
 $SU(4)/Sp(4)$ coset~\cite{Katz:2005au,Gripaios:2009pe,Barnard:2013zea,Lewis:2011zb,
Hietanen:2014xca,Arthur:2016dir,Arthur:2016ozw,Pica:2016zst,Detmold:2014kba,Lee:2017uvl,Cacciapaglia:2015eqa,Hong:2017smd,Golterman:2017vdj}, 
as EFT arguments suggest that the
resulting phenomenology is both realistic and rich enough to motivate a more systematic study of the underlying dynamics.
This coset emerges naturally in gauge theories with pseudo-real representations, such as $Sp(2N)$ gauge theories
with two massless Dirac fermions in the fundamental representation of the gauge group.
Phenomenological arguments---ultimately related to the fact that if
fundamental fermions carry $SU(3)_c$ (colour) quantum numbers, one can further implement 
partial top compositeness---select $SU(2)\sim Sp(2)$
and $Sp(4)$ as most realistic viable candidates for BSM physics. 
A number of studies has considered the dynamics of $SU(2)$
(see for instance~\cite{Lewis:2011zb,Hietanen:2014xca,Arthur:2016dir,Arthur:2016ozw,Pica:2016zst,Detmold:2014kba,Lee:2017uvl}), while here we focus on $Sp(4)$.

The primary objectives of our research programme include the study of the mass spectrum of mesons and glueballs,
and the precise determination of decay constants and couplings of all these objects by means of
lattice numerical techniques.\footnote{An alternative non-perturbative approach
is followed for example in~\cite{Bizot:2016zyu}.}
Eventually, we want to gain quantitative control over a large set of measurable quantities of relevance for
phenomenological (model-building) considerations, which include also more ambitious determinations of the width of excited states, 
and of the values of the condensates in the underlying dynamics.

A separate set of objectives relates to the physics of baryons and 
composite fermions. In  $Sp(2N)$ with fundamental matter, baryons are bosonic objects,
and hence not suitable  as model-building ingredients for top (partial) compositeness.
Composite fermions could be realised for example by adding 2-index representations 
to the field content of the dynamics (see for example~\cite{Gripaios:2009pe,Barnard:2013zea,Cacciapaglia:2015eqa}).
We envision to start soon a non-trivial programme of study of their dynamical implementation  on the lattice.

A third set of  dynamical questions that our programme wants  to address in the long run
pertains to the thermodynamic properties of the $Sp(2N)$ theories at finite temperature $T$ and chemical
potential $\mu$. It is of general interest to study the symmetry-restoration pattern 
of these models at high temperature (see~\cite{Lee:2017uvl} for a step in this direction in the case of $SU(2)$).
Furthermore, the pseudo-real nature of $Sp(4)$ makes it possible to study the
phase-space of the theory, while avoiding the sign problem.

Finally, there is a different field-theoretical reason for studying $Sp(2N)$ gauge theories.
It is known that  the Yang-Mills theories based on $SU(N)$, $SO(N)$ and $Sp(2N)$ 
all agree with one another on many fundamental physical quantities when the limit of large $N$ is taken.
Lattice results allow for non-trivial comparisons
with field-theory and string-theory studies in approaching the large-$N$ limit.
While there is a substantial body of literature on  $SU(N)$ theories on the lattice~\cite{Lucini:2012gg}, 
for example for the calculation of the glueballs, and some literature on $SO(N)$ models
 (see for instance~\cite{Lucini:2004my,Athenodorou:2015nba,Lau:2017aom} and references therein), 
 there is no systematic, dedicated study of the $Sp(2N)$  gauge theories.
We aim at comparing with results in Yang-Mills theories based on other groups, 
and with conjectures such as those put forwards  
in~\cite{Athenodorou:2016ndx}  and~\cite{Hong:2017suj}. 

\subsection{Laying the foundations of the $Sp(4)$ lattice studies}

With this paper, we start the programme of systematic lattice studies of the dynamics 
of such gauge theories. 
We focus here on the $Sp(4)$ gauge group,
which is of  relevance for the phenomenology of composite-Higgs models.
We perform preliminary studies of the lattice theories of interest,
 a first exploratory computation of the meson spectrum in the quenched approximation
 and a first test of the same calculation with dynamical fermions.
 
 We aim at gaining a quantitative
understanding of the properties of the bound states, possibly describing them within the EFT framework. 
Starting from the leading-order chiral-Lagrangian description of
the PNGBs, we  extend it to include heavier mesons, aiming at providing dynamical information useful for collider searches.
As is well known, the description of the spin-1 composite particles is weakly-coupled only in the large-$N$ limit:
we do not expect the EFT to fare particularly well for $Sp(4)$, yet it is interesting to use it
also in this case, in view of possible future extensions to $Sp(2N)$.
We also begin the analysis of the next-to-leading-order corrections to the chiral Lagrangian 
of the model, as a first preliminary step towards understanding realistic model building in the BSM context.
While still beyond the purposes of this paper, we find it useful also to briefly summarise
the main goals of the exploration of the dynamics of composite fermions emerging 
from introducing on the lattice matter in different representations.

We devote a significant fraction of this paper to the study of the pure Yang-Mills theory.
We perform our $Sp(4)$ lattice calculations in such a way that the
technology we use can be easily generalised to any $Sp(2N)$ theories, in view of implementing 
a systematic programme of exploration of the large-$N$ behaviour. 
We present  the spectrum of glueballs, and study the effective string-theory description, 
for  $Sp(4)$ pure Yang-Mills, with no matter fields.
Our results have a level of accuracy that is comparable to the current state-of-the-art for $SU(N)$ gauge groups
in four dimensions. This both serves as an interesting test of the algorithms we use, 
but also nicely complements existing literature on related subjects.

The paper is organised as follows.
In Sec.~\ref{Sec:Preliminary1} we define the field theory of interest, and introduce its low-energy EFT description
in terms of PNGBs. We also extend the EFT to include the lightest spin-1 states in the theory 
(see also Appendix~\ref{Sec:grouptheory} and~\ref{Sec:TC}).
We define the framework of partial top compositeness for these models, 
and the lattice programme that we envision to carry out
in the future along that line.

In Sec.~\ref{Sec:Preliminary2} we describe in details the lattice actions, as well as
the Heat Bath (HB) and Hybrid Monte Carlo (HMC) algorithms used in the numerical studies.
In Sec.~\ref{Sec:Phasespace} we focus on scale setting and topology. These two technical Sections, together with Appendix~\ref{Sec:grouptheory} and~\ref{Sec:projection}, 
set the stage not only for this paper, but also for future physics studies we will carry out.
In Sec.~\ref{Sec:confinement} we present  the spectrum of glueballs for $Sp(4)$. We also explain in details the process leading to this measurement, that
will be employed in the future also for the spectrum of $Sp(2N)$ with general $N$.
Section~\ref{Sec:mesons} is devoted to the spectrum of mesons of $Sp(4)$ in the quenched approximation, the extraction 
of the masses and decay constants and a first attempt at comparing to the low-energy EFT. 
Preliminary (exploratory) results for 
the full dynamical simulation are presented in Sec.~\ref{Sec:mass}. In particular, we exhibit  the first (to the best of our knowledge) evidence
that a bulk phase transition is present for $Sp(4)$ with fundamental matter.
We conclude the paper with Sec.~\ref{Sec:outlook}, summarising the results and highlighting the future avenues of exploration
that this work opens.

\section{Elements of field theory,  group theory and effective field theory}
\label{Sec:Preliminary1}

The  $Sp(4)$ gauge theory of interest has matter
content consisting of two (massive) Dirac fermions $Q^{i\,a}$, where $a=1\,,\,\cdots\,,\,4$ is the colour index
and $i=1,2$ the flavour index, or equivalently  four 2-component spinors $q^{j\,a}$ with $j=1\,,\,\cdots\,,\,4$.
The Lagrangian density is
\beqs
{\cal L}&=&i\,\overline{Q^i}_{\,a}\,\gamma^{\mu}\,(D_{\mu}Q^i)^a\,-\,m\,\overline{Q^i}_{\,a}Q^{i\,a}\,-\,\frac{1}{2}\Tr V_{\mu\nu} V^{\mu\nu}\,,
\eeqs
where the summations over flavour and colour indices are understood, and where the field-strength tensors are defined 
by $V_{\mu\nu}\equiv \partial_{\mu}V_{\nu}-\partial_{\nu}V_{\mu}+i g \left[V_{\mu}\,,\,V_{\nu}\right]$.
\begin{table}
\begin{center}
\begin{tabular}{|c|c|c|c|}
\hline\hline
{\rm ~~~Fields~~~} &$Sp(4)$  &  $SU(4)$\cr
\hline
$V_{\mu}$ & $10$ & $1$ \cr
$q$ & $4$ & $4$ \cr
\hline
$\Sigma_0$ & $1$ & $6$\cr
$M$ & $1$ & $6$\cr
\hline\hline
\end{tabular}
\end{center}
\caption{The field content of the theory.
$Sp(4)$ is the gauge group, while $SU(4)$ is the global symmetry. 
The elementary fields $V_{\mu}$ are gauge bosons, $q$ are 2-component spinors.
$\Sigma_0$ is the composite scalar defined in Eq.~(\ref{Eq:composite}), the VEV of which
is responsible for the breaking $SU(4)\rightarrow Sp(4)$.
 The mass matrix $M$ is treated  as a scalar spurion, formally transforming as $\sim 6$ of $SU(4)$. }
\label{Fig:fields}
\end{table}

In the $m\rightarrow 0$ limit, the global symmetry  
is $U(1)_A\times SU(4)$. The presence of a finite mass $m\neq 0$ is allowed within the context of composite-Higgs models,
and may play an important (model-dependent) role. 
We write the  symplectic matrix $\Omega$ as
\beqs
\Omega&=&
\left(\begin{array}{cccc}
0 & 0 & 1 & 0\cr
0 & 0 & 0 & 1\cr
-1 & 0 & 0 & 0\cr
0 & -1 & 0 & 0\cr
\end{array}\right)\,,
\label{Eq:symplectic}
\eeqs
and define the composite operator $\Sigma_0$ as
\beqs
\Sigma_0^{\,\,nm}&\equiv&\Omega_{ab} q^{n\,a\,T} \tilde{C} q^{m\,b}\,,
\label{Eq:composite}
\eeqs
so that in 2-component spinor language the mass  matrix is $M\equiv m\, \Omega$.
We collect in Appendix~\ref{Sec:grouptheory}  
some useful elements of group theory.
For the most part we ignore the anomalous $U(1)_A$.
We display the field content in Table~\ref{Fig:fields},
where we list also the transformation properties of the composite field $\Sigma_0$,
and the (symmetry-breaking) spurion $M$.

The vacuum  is characterised by the fact that $0\neq \langle \Sigma_0 \rangle\propto \Omega$,
hence realising the breaking $SU(4)\rightarrow Sp(4)$.
In the absence of coupling to the SM fields, the vacuum structure aligns with the mass term, 
which hence contributes to the masses of the
composite states, by breaking the global $SU(4)$ while preserving its global $Sp(4)$ subgroup.
As a result, the lightest mesons of the theory  arrange themselves into irreducible representations of $Sp(4)$: 
the  PNGBs  $\pi$  and axial-vectors $a_1$ transform on the  $5$ representation of the unbroken $Sp(4)$, 
while the scalars $a_0$ and the vectors $\r$ on the $10$ of $Sp(4)$.
There exist also the corresponding scalar, pseudo-scalar, vector and axial-vector $Sp(4)$ singlets, 
but we will not discuss them in this paper.

\subsection{EFT analysis}
\label{Sec:EFT}

The EFT treatment of the lightest mesons depends on the coset, 
with only numerical values of the coefficients depending on the underlying gauge group.
We summarise here some useful information about two different EFTs.
Some of the material collected in this subsection
can also be found in the literature~\cite{Lewis:2011zb,Cacciapaglia:2014uja,Hietanen:2014xca,Arbey:2015exa,Arthur:2016dir,
Arthur:2016ozw,Pica:2016zst,
Katz:2005au,Lodone:2008yy,Gripaios:2009pe,Barnard:2013zea,Ferretti:2013kya,
Vecchi:2015fma,Ferretti:2016upr,Agugliaro:2016clv,Panico:2015jxa,Lee:2017uvl}. 
We construct the chiral Lagrangian for the $SU(4)/Sp(4)$ coset,
and its generalisation in the sense of Hidden Local Symmetry (HLS)~\cite{
Bando:1984ej,Casalbuoni:1985kq,Bando:1987br,Casalbuoni:1988xm,Harada:2003jx} (see also~\cite{Georgi:1989xy,Appelquist:1999dq,Piai:2004yb,Franzosi:2016aoo}).
The former assumes that only the pions are dynamical fields in the low-energy EFT, while the latter 
includes also  $\r$ and $a_1$ as weakly-coupled fields. We will comment in due time on the 
regime of validity of the two.

\subsubsection{EFT description of pions.}
\label{Sec:pions}

The low-energy EFT description of the pions $\pi$ is constructed by introducing the real composite field $\Sigma$
that obeys the non-linear constraint $\Sigma\Sigma^{\dagger}=\mathbb{I}_4$, and transforms as the antisymmetric representation
$\Sigma\rightarrow U \Sigma U^T$,
under the action of an element $U$ of $SU(4)$.
The  antisymmetric vacuum expectation value (VEV) $\langle \Sigma \rangle \propto \Omega$ breaks $SU(4)$ to the $Sp(4)$ subgroup,
and as a result five generators $T^A$, with $A=1,\cdots,5$, are broken, while $10$ other $T^A$, with $A=6,\cdots,15$, are not.
We normalise them all as $\Tr T^AT^B=\frac{1}{2}\delta^{AB}$.

The field $\Sigma$ contains  the  PNGB fields $\pi=\pi^AT^A$,  conveniently parametrised as
\beqs
\Sigma &\equiv&e^{\frac{i\pi}{f}}\Omega e^{\frac{i\pi^T}{f}}\,=\,e^{\frac{2i\pi}{f}}\Omega\,=\,\Omega \,e^{\frac{2i\pi^T}{f}}\,,
\eeqs
in terms of which, at leading-order,  the EFT has the   Lagrangian density
\beqs
{\cal L}_0&=&\frac{f^2}{4}\Tr\left\{\frac{}{}\partial_{\mu}\Sigma \,(\partial^{\mu}\Sigma)^{\dagger}\right\}\\
&=&\Tr \left\{\partial_{\mu}\pi\partial^{\mu}\pi\right\}\,
+\,\frac{1}{3f^2}\Tr\left\{\frac{}{}\left[\partial_{\mu}\pi\,,\,\pi\right]\left[\partial^{\mu}\pi\,,\,\pi\right]\right\}\,+\,\cdots\,.
\eeqs
The pion fields are canonically normalised,
hence $f=f_{\pi}$ is the pion decay constant.

If it were promoted to a field, the spurion $M$ would transform as $M\rightarrow U^{\ast} M U^{\dagger}$, so that
the combination $\Tr \,M\, \Sigma$ would be manifestly invariant under the $SU(4)$ global symmetry.
The (symmetry-breaking) mass term is hence written as
\beqs
{\cal L}_m&=&-\frac{v^3}{4}\Tr \left\{M\,\Sigma\right\}\,+\,{\rm h.c.}\\
 &=& 2 m v^3 \,-\,\frac{m v^3}{f^2}\Tr \pi^2\,
 +\,\frac{m\,v^3}{3f^4}\Tr (\pi\pi\pi\pi)+\cdots\,,
\eeqs
which  confirms that the $5$ pions are degenerate
in the presence of the explicit breaking given by the Dirac mass for the fermions,
because of the unbroken $SO(5)\sim Sp(4)$ symmetry.
The GMOR relation is automatically satisfied, and takes the form:
\beqs
m_{\pi}^2f_{\pi}^2&=&{m \,v^3}{}\,. 
\label{Eq:GMOR0}
\eeqs

As is the case for the chiral-lagrangian description of low-energy QCD,
we are making use of two expansions: the derivative expansion, that suppresses terms
of higher dimension, and that is reliable provided we consider observables at
energies $E\,\ll\,4\pi f_{\pi}$, and the chiral expansion, 
reliable when the mass of the pion satisfies $m_{\pi}\,\ll\,4\pi f_{\pi}$.

At the sub-leading order, we could for example add  to the chiral Lagrangian the contribution
\beqs
\label{Eq:subleading}
{\cal L}_s&=&v_0 \Tr \left( M \Sigma\right)\Tr\left(\partial_\mu \Sigma\partial^{\mu} \Sigma^{\dagger} \right)\,+\,{\rm h.c.}\\
&=&-32 \frac{m v_0}{f^2}\,{\cal L}_0\,+\,\frac{16 m v_0}{f^4}\Tr\left\{\pi\pi\right\} \Tr\left\{\partial\pi\partial\pi\right\}\,+\,\cdots \,.
\eeqs
The first term of the expansion comes from setting $\Sigma= \Omega$, and amounts to a $m$-dependent rescaling of the interacting ${\cal L}_0$.
No correction to the mass term appears, but just an overall rescaling of both $f$ and $\pi$,
so that the GMOR relation is respected. However, an additional quartic pion coupling is generated, that contributes to the
$\pi\pi$ scattering amplitude. 
In the massless limit $f_{\pi}$ can be equivalently extracted from either 2-point functions or from 
the $\pi\pi$ scattering amplitude.
For $m\neq 0$ there are subtleties:  in the following we will always extract $f_{\pi}$ from the  $q^2\rightarrow 0$ limit of the 2-point functions, 
and we will highlight this fact by denoting the pion decay constant (squared) as $f_{\pi}^2(0)$  in the rest of the paper.

\subsubsection{Hidden Local Symmetry description of $\r$ and $a_1$}
\label{Sec:vectors}

Hidden Local Symmetry provides a way to include spin-1 excitations such as the $\rho$ mesons into the EFT treatment,
hence extending the validity of the chiral Lagrangian
(see for instance~\cite{Bando:1984ej,Casalbuoni:1985kq,Bando:1987br,Casalbuoni:1988xm,Harada:2003jx} 
and also~\cite{Georgi:1989xy,Appelquist:1999dq,Piai:2004yb,Franzosi:2016aoo}). While very appealing on aesthetics grounds, when applied to QCD such idea 
shows severe limitations: the heavy mass and non negligible width of the $\rho$ mesons imply that the weak-coupling treatment 
is not fully reliable. Yet, this description offers a nice way to classify operators and it
 is expected to become more reliable  at large-$N$~\cite{Harada:2003jx}. 
  As we envision future studies with larger $Sp(2N)$ groups, it is useful to show  the 
 construction already in the programmatic part of this paper, and test it on $Sp(4)$. 

\begin{figure}[h]
  \begin{center}
   \rotatebox{-1}{\includegraphics[width=.20\textwidth,angle=-89]{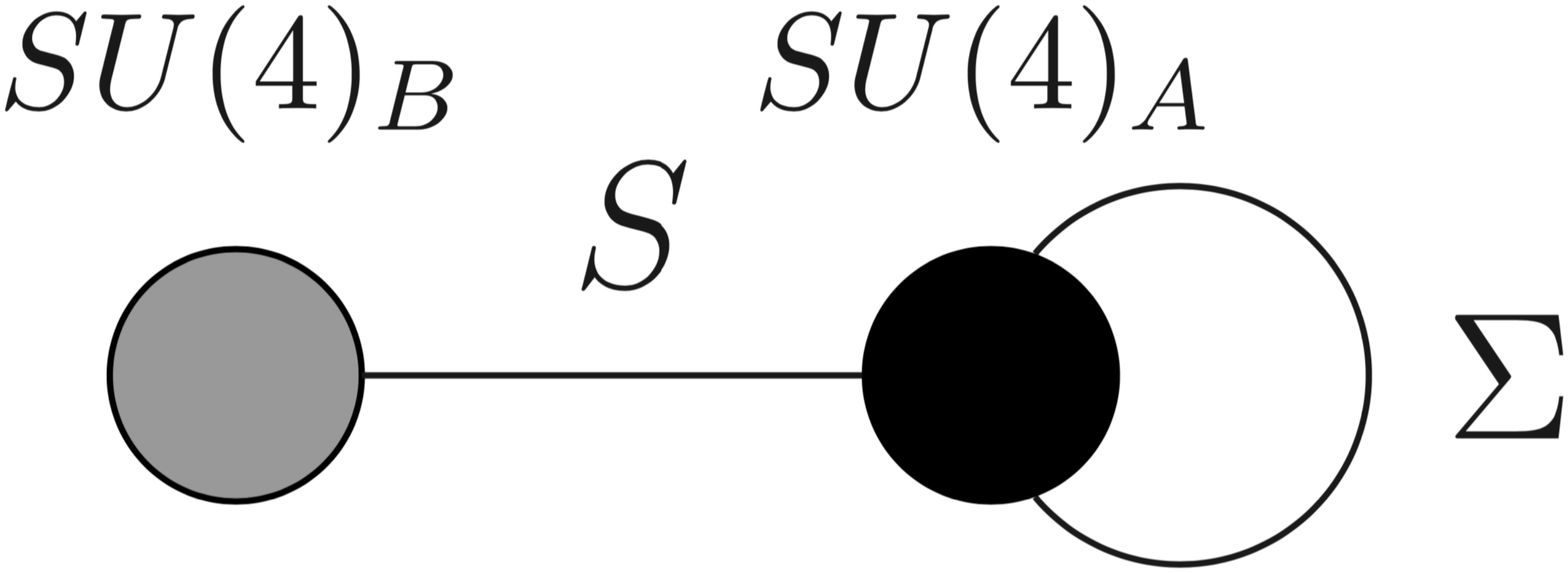}}
    \end{center}
\caption{The moose diagram representing the  low-energy EFT description  of the model in 
Eq.~(\ref{Eq:L}), along the lines of HLS~\cite{
Bando:1984ej,Casalbuoni:1985kq,Bando:1987br,Casalbuoni:1988xm,Harada:2003jx} (see also~\cite{Georgi:1989xy,Appelquist:1999dq,Piai:2004yb,Franzosi:2016aoo}). The two sites represent the  $SU(4)_A\times SU(4)_B$
global symmetry. The scalar $S$ transforms on the bifundamental representation, 
while $\Sigma$ is antisymmetric. The $SU(4)_A$ group is gauged with coupling $g_{\r}$, while  the
 $SU(2)_L\times U(1)_Y$ SM subgroup of $SU(4)_B$ can be weakly gauged, with couplings $g$ and $g^{\prime}$. 
 In most of this paper we set $g=0=g^{\prime}$, and 
hence ignore the interactions of the strongly coupled  dynamics with the standard-model fields.
  }
\label{Fig:EFT}
\end{figure}

The full set of $\r$ and $a_1$ mesons spans the adjoint representation of the global $SU(4)$ symmetry. 
An EFT  description of their long-distance dynamics can be built starting from the diagram
 in Fig.~\ref{Fig:EFT}. The $15$ spin-1 fields are introduced as gauge fields of $SU(4)_A$. Two scalars, the
antisymmetric  $\Sigma$ of $SU(4)_A$, and the bi-fundamental $S$, transform as
\beqs
\Sigma\,\rightarrow\, U_A \Sigma U_A^T\,,~~~~~S&\rightarrow & U_B\, S \,U_A^{\dagger}\,,
\eeqs
under the action of $U_A \in SU(4)_A$ and $U_B\in SU(4)_B$.
 The VEV of $S$ breaks the enlarged symmetry and provides mass for all the vectors.
$S$ is subject to the constraints $ S^{\dagger} S=  \mathbb{I}_4$, that are solved by
parametrising 
$S=e^{\frac{2i\sigma}{F}}$,
with $\sigma=\sigma^AT^A$ and $F$ the decay constant. 
At the same time, we parametrise $\Sigma=e^{\frac{2i\pi}{f}}\Omega$, in such a way that the 
two scalars together implement the breaking $SU(4)_A\times SU(4)_B\rightarrow Sp(4)$,
and describe the $15$ exact Goldstone bosons that are higgsed away into the longitudinal components of the massive
spin-1 states, as well as the remaining $5$ (massive) PNGB, denoted as $\bar{\pi}^A$ in the following.  

 In composite Higgs models, the SM gauge group $SU(2)_L\times U(1)_Y$ is a subgroup of $SU(4)_B$,
 and it is chosen to be a subgroup of the  unbroken global $Sp(4)$. 
The covariant derivative of $S$ is given by
\beqs
D_{\mu}S&=&\partial_{\mu}S+i  \left(g W_{\mu} + g^{\prime}B_{\mu}\right)S-iSg_{\r} A_{\mu}\,,
\eeqs
with $A_{\mu}=A_{\mu}^AT^A$ and $T^A$ the generators of $SU(4)_A$, while $W_{\mu}=W_{\mu}^it_L^i$ and $B_{\mu}$
are the gauge bosons of $SU(2)_L\times U(1)_Y$.
The covariant derivative of $\Sigma$ is
\beqs
D_{\mu}\Sigma&=&
\partial_{\mu} \Sigma + i \left[\frac{}{}(g_{\r}A_{\mu}) \Sigma\,+\,\Sigma (g_{\r}A_{\mu})^T\right]\,\\
&=&\left\{\partial_{\mu} e^{\frac{2i\pi}{f}} + i \left[\frac{}{}(g_{\r}A_{\mu})e^{\frac{2i\pi}{f}}\,-\,e^{\frac{2i\pi}{f}} \Omega  (g_{\r}A_{\mu})^T \Omega \right]\right\} \Omega\,,
\eeqs
where we have made use of the fact that $\Omega^2=-\mathbb{I}_4$.
From this point onwards, we set $g=0=g^{\prime}$, and focus on the dynamics of the strongly-coupled new sector in isolation from the SM fields.

We write the Lagrangian density describing the 
$15$ gauge bosons $A_{\mu}^A$,
as well as the $20$ pseudo-scalar fields $\pi^A$ and $\sigma^A$, as 
\beqs
{\cal L}&=& 
-\frac{1}{2}\Tr\left.A_{\mu\nu}A^{\mu\nu}
-\frac{\kappa}{2}\Tr\left\{ A_{\mu\nu} \Sigma (A^{\mu\nu})^T \Sigma^{\ast}\right\}\right.\nonumber\\
&&+\frac{f^2}{4}\Tr\left\{\frac{}{}D_{\mu}\Sigma\,(D^{\mu}\Sigma)^{\dagger}\right\}\,+\,\frac{F^2}{4}\Tr\left\{\frac{}{}D_{\mu}S\,(D^{\mu}S)^{\dagger}\right\}\nonumber\\
&&+b \frac{f^2}{4}\Tr\left\{D_{\mu}(S\Sigma )\left(D^{\mu}(S\Sigma)\right)^{\dagger}\right\}\,+\,
c\frac{f^2}{4} \Tr\left\{D_{\mu}(S\Sigma S^T)\left(D^{\mu}(S\Sigma S^T)\right)^{\dagger}\right\}\,\nonumber\\
&&-\frac{v^3}{8}\Tr\left\{\frac{}{}M\, S\, \Sigma\,S^{T} \right\}\,+\,{\rm h.c.} \label{Eq:L}\\
&&- \frac{v_1}{4} \Tr\left\{\frac{}{} M\, (D_{\mu} S)\, \Sigma \, (D^{\mu} S)^T \right\}\,
 -\frac{v_2}{4} \Tr\left\{\frac{}{} M\, S\,(D_{\mu}  \Sigma) \, (D^{\mu} S)^T \right\}\,+\,{\rm h.c.}\,\nonumber\\
 &&
 -\frac{y_3}{8}\Tr\left\{A_{\mu\nu}\Sigma\left[(A^{\mu\nu})^TS^T M S-S^T M S A^{\mu\nu}\right]\right\}\,+\,{\rm h.c.}\nonumber\\
  &&
 -\frac{y_4}{8}\Tr\left\{A_{\mu\nu}\Sigma\left[(A^{\mu\nu})^TS^T M S+S^T M S A^{\mu\nu}\right]\right\}\,+\,{\rm h.c.}\,\nonumber
 \\
&&
-\frac{v_5^2}{128}\left(\Tr M S \Sigma S^T \,+\, {\rm h.c.}\frac{}{}\right)^2\,.\nonumber
\eeqs
The first line of Eq.~(\ref{Eq:L}) depends on the field-strength tensor $A_{\mu\nu}$ of the gauge 
group, and includes the symmetry-breaking term controlled by $\kappa$, that would be omitted from 
the linear-sigma model version of the same EFT.
The covariant derivatives of combinations of $S$ and $\Sigma$ are defined in the obvious way, 
generalising the covariant derivatives of $S$ and $\Sigma$.
The mass deformations are introduced via a new spurion $M$
and via combination of fields such as $S\Sigma S^T$, that transforms as $S \Sigma S^{T}\rightarrow U_B S \Sigma S^{T} U_B^T$.
The spurion  differs from the one in the chiral Lagrangian
as it formally transforms as $M\rightarrow U_B^{\ast} M U_B^{\dagger}$. 
In this way the whole Lagrangian is manifestly $SU(4)_A$ gauge invariant.\footnote{If part of the $SU(4)_B$ were gauged,
as in technicolor models, then one might be forced to work in the $m=0$ limit. But further discussion of this point can be found in Appendix~\ref{Sec:TC}.}

In the expansion, we include two sets of operators.
We call leading order (LO) the ones appearing in the first four lines, controlled by the parameters
$F$, $f$, $b$, $c$, $\kappa$, $g_{\r}$ and $v$. This is an exhaustive set of operators, at this order.
 We call next-to-leading order (NLO) those in the last four lines,  controlled by the parameters
$v_1$, $v_2$, $y_3$, $y_4$ and $v_5$. As we will discuss shortly,
the list of NLO operators is incomplete. In total there are $12$ parameters.
This Lagrangian has to be used with caution. The appearance of  $\r$ and $a_1$ fields in the EFT
 is fully justified only if the coupling $g_{\r}$ is small, which must be discussed  a posteriori,
yet is expected to hold  in the large-$N$ limit, and as long as $m$ is small.

The last four lines of Eq.~(\ref{Eq:L}) contain terms that are sub-leading in the power-counting.
 Because we are going to perform lattice simulations at finite mass $m$, a priori we do not know how important
 such terms are, and hence we include them. While non-vanishing values of $m$ are allowed within the composite-Higgs framework,
 the EFT is useful only when $m$ is small enough that truncating at this order is justified.

We do not include the full set of sub-leading four-derivative terms, because they are not important for our current purposes.
These terms would become important when a complete analysis of 3-point and 4-point functions is performed, for example.
We also omit topological terms.
Furthermore, we do not include in $\cal L$ terms with the structure of Eq.~(\ref{Eq:subleading}),
 such as
\beqs
\label{Eq:subleading2}
\Tr\left\{\frac{}{}M\, S\, \Sigma\,S^{T} \right\}
\Tr\left\{D_{\mu}(S\Sigma S^T)\left(D^{\mu}(S\Sigma S^T)\right)^{\dagger}\right\}\,.
\eeqs
We will comment later in the paper on the implications of all these omissions.

We conclude this subsection with a technical comment.
Some of the terms in the Lagrangian density in Eq.~(\ref{Eq:L}) involve only nearest-neighbour interactions, 
in the sense of the diagram in Figure~\ref{Fig:EFT}, while other couplings 
introduce non-nearest-neighbour interactions. Such additional interactions might for example emerge from
 the process of integrating out heavier degrees of freedom.
One of the big limitations of the HLS language is that the number of independent, admissible such non-nearest-neighbour interactions 
grows rapidly with the number of fields in the theory, and hence by introducing more resonances the EFT Lagrangian density loses predictive power
because of the proliferation of new free parameters. 
In the special Lagrangian we wrote, such interactions are controlled by the parameters $\kappa$, $b$, $c$,
as well as $v$, $v_1$, $v_2$, $y_3$, $y_4$ and $v_5$. If only nearest-neighbour couplings were to be allowed, the set of parameters would be restricted to 
just $f$, $F$ and $g_{\r}$, at this order in the expansion.

\subsubsection{2-point functions}
\label{Sec:2}

To compute masses and decay constants of the mesons, we use the language of the $SU(2)^t_L\times SU(2)^t_R$
symmetry that would be
of direct relevance if we were to treat this as a Technicolor model. 
In particular, this symmetry is not a subgroup of the unbroken $Sp(4)$ global symmetry,
and the condensate breaks it. 
We treat this as a technical tool, that is convenient in order  to extract physical quantities 
from the correlation functions. Yet our results hold also for finite $m$, 
and apply as well to the composite-Higgs scenario,
as we never include in the calculations the effects of the couplings to the external (weakly-coupled) SM fields.
The Left-Left current-current correlator is (see Appendix~\ref{Sec:TC})
\beqs
\Sigma(q^2)
&=&f_0^2+\frac{M_{\r}^2f_{\r}^2}{q^2-M_{\r}^2}+\frac{M_{a_1}^2 f_{a_1}^2}{q^2-M_{a_1}^2}\,,
\label{Eq:Sigma}
\eeqs
from which one can read that the masses and decay constants are given by
\beqs
M_{\r}^2&=&\frac{1}{4(1+\kappa+m\,y_3)} {g_{\r}}^2 \left(b f^2+F^2+2 m v_1\right)\,,\\
M_{a_1}^2&=&\frac{1}{4(1-\kappa-m \,y_4)} {g_{\r}}^2 \left(b f^2+F^2+2 m v_1\right)+\\
&&\nonumber+\frac{g_{\r}^2}{1-\kappa-m \,y_4} \left(f^2+m
   (v_2-v_1)\right)\,,\\
   f_{\r}^2&=&
   \frac{1}{2} \left(b f^2+F^2+2 m v_1\right)\,,\\
   f_{a_1}^2&=&
   \frac{\left(b f^2-F^2+2 m (v_1-v_2)\right)^2}{2 \left((b+4) f^2+F^2-2 m
   v_1+4 m v_2\right)}\,,\\
   f_0^2&=&\frac{}{}F^2+(b+2c)f^2\,.
\eeqs
and that the pion decay constant is
\beqs
f_{\pi}^2(0)&=&\lim_{q^2\rightarrow 0}\Sigma(q^2)\,=\,f_0^2-f_{\r}^2-f_{a_1}^2\,.
\eeqs
As anticipated, the notation explicitly specifies that $f_{\pi}^2(0)$ is extracted from 2-point functions evaluated at $q^2=0$.
The fact that $f_0^2=f_{\pi}^2(0)+f_{\rho}^2+f_{a_1}^2$ is independent of $m$ 
is the accidental consequence of the truncation we made, in particular of the omission of
the operator in~(\ref{Eq:subleading2}). Whether or not this is justified, 
depends on the range of $m$ considered and on the size of the EFT coefficients,
as emerging from  lattice data.

One can compute the right-hand-side of the first and second Weinberg sum rules, within the EFT, to obtain
\beqs
f_{a_1}^2-f_{\r}^2+f_{\pi}^2(0)&=&2(c f^2-m v_1)\,,\\
f_{\r}^2M_{\r}^2-f_{a_1}^2M_{a_1}^2&=&
\frac{g_{\r}^2}{8}  \left(\frac{(-b {f^2}+{F^2}+2 {m}
   ({v_2}-{v_1}))^2}{\kappa+{m} {y_4}-1}+\frac{(b {f^2}+{F^2}+2
   {m} {v_1})^2}{\kappa+{m} {y_3}+1}\right),
   \eeqs
hence showing explicitly that the non-nearest-neighbour couplings $b$, $c$, $\kappa$, $y_4$, $y_3$, $v_1$ and $v_2$ yield 
to direct violations of the Weinberg sum rules, within the EFT.
As anticipated, this is not surprising: non-nearest-neighbour interactions are expected to emerge from integrating out 
heavy degrees of freedom, and result in the violation of the Weinberg sum rules because their rigorous derivation involves
summing over all possible resonances. The additional couplings, in effect, parameterise the contribution to
the sum rules of heavier resonances that have been omitted.

\subsubsection{Pion mass and $g_{\r\pi\pi}$ coupling}
\label{Sec:pi}

To compute the physical mass and couplings of the pions, 
 it is convenient to fix the unitary gauge,
by setting $\sigma^A=0$ along the unbroken $A=6\,,\,\cdots\,,\,15$ generators, and 
\beqs
\sigma^A&=&S\bar{\pi}^A\,=\,\
\frac{((2+b)f+m v_2/f)F}{\cal N}\bar{\pi}^A\,,
\label{Eq:defineS}\\
\pi^A&=& C\bar{\pi}^A\,=\,
\frac{F^2- bf^2-2m(v_1-v_2)}{\cal N} \bar{\pi}^A\,,
\label{Eq:defineC}
\eeqs
for $A=1\,,\,\cdots\,,\,5$ along the broken generators,
with the normalisation factor ${\cal N}$ chosen so that the physical $\bar{\pi}^A$ are canonically normalised:
\beqs
{\cal N}^2&=&
\frac{\left((b+4) f^2+F^2-2 m {v_1}+4 m {v_2}\right)}{f^2} 
\left\{2 f^2 m {v_2} (b+2
   c)-m^2 {v_2}^2+\frac{}{}\right.\\
&&\nonumber\left.\frac{}{}+f^2 \left(b c f^2-2
   m {v_1} (b+c+1)+b f^2+b F^2+4 c f^2+c F^2+F^2\right)\right\} \,.\nonumber
\eeqs

The $5$ degenerate pions have mass
\beqs
m^2_{\pi}&=&\left(\frac{}{}m v^3+m^2 v_5^2\right) \frac{((4+b)f^2+F^2-2m v_1+4m v_2)^2}{2f^2{\cal N}^2}\,,
\eeqs
which modifies  the GMOR relation  to read 
\beqs
m_{\pi}^2f_{\pi}^2=m (v^3+m v_5^2)\,.
\label{Eq:GMOR}
\eeqs

The $g_{\r\pi\pi}$ coupling is conventionally defined by the Lagrangian density
\beqs
{\cal L}_{\r\pi\pi}&=&-2i g_{\r\pi\pi}\Tr\left(\frac{}{}\rho^{\mu}[\partial_{\mu}\bar{\pi}\,,\,\bar{\pi}]\right)\,,
\eeqs
so that the width  (at tree level) is
$\Gamma_{\r}=\frac{g^2_{\r\pi\pi}}{48 \pi}M_{\r}\,\left(1-4\frac{m_{\pi}^2}{M_{\r}^2}\right)^{3/2}$.
We find that
\beqs
g_{\r\pi\pi}&=&
\frac{{g_{\r}}}{2 f^2 {\cal N}^2\sqrt{1+\kappa+m\,y_3}} \left\{\frac{}{}m^2 v_2^2 \left((5 b-8) f^2-3 F^2+10 m
   v_1\right)\right.\nonumber\\
   &&\left.+(b+2) f^2 \left(\left(2 f^2+F^2\right) \left(b f^2+F^2\right)\right.\right.\nonumber\\
   &&\left.\left.-2 m
   v_1 \left(2 (b+1) f^2+F^2\right)\right)+2 m v_2 \left(2 b (b+3) f^4+m
   v_1 \left(F^2-3 b f^2\right)
   \right.\right.\nonumber\\
   &&\left.\left.
   +2 (b+1) f^2 F^2-2 m^2 v_1^2\right)-8 m^3
   v_2^3\frac{}{}\right\}\,.
   \label{Eq:grpp}
\eeqs

We conclude with a comment about unitarity. While the calculations performed here make use of the unitary gauge,
we must check that the kinetic terms of all the Goldstone bosons be positive before setting to zero the linear combinations 
providing the longitudinal components of the vectors. We call the relevant normalisations $k_{10}$, $k_5$ and $k_5^{\prime}$,
coming from the kinetic term of  $\sigma^A$ with $A>5$, as well as  the trace and the determinant of the kinetic matrix
mixing $\sigma^A$ and $\pi^A$ with $A<6$.
Such combinations are explicitly given by:
\beqs
k_{10}&=&\frac{b f^2+F^2+2 m {v_1}}{F^2}\,,\nonumber\\
k_{5}&=&2+b+c+(b+4c)\frac{f^2}{F^2}-2 m v_1\,,\\
k_{5}^{\prime} F^2&=&b \left((c+1) f^2+F^2-2 m v_1+2 m v_2\right)\nonumber +\\
&&+c \left(4 f^2+F^2-2 m
   v_1+4 m v_2\right)-\frac{m^2 v_2^2}{f^2}+F^2-2 m v_1\,.\nonumber
\eeqs
We require that $k_{10},k_5,k_5^{\prime}>0$.
Furthermore, for the kinetic terms of the vectors to be positive definite one must impose $\kappa+m \,y_4<1$
and $\kappa+m \,y_3>-1$.

\subsubsection{On the regime of validity of the EFT}

In the EFT we wrote to include the $\r$ and $a_1$ particles, we are making use of several expansions.
Besides the derivative expansion and the expansion in the mass of the fermions $m$, 
appearing also in the chiral Lagrangian,
there is a third expansion, that involves the coupling  $g_{\r}$ and deserves discussing in some detail.

From lattice calculations of 2-point functions, one extracts the decay constants 
of $\pi$, $\r$ and $a_1$, in addition to the masses.
In the $m=0$ limit,  the expressions for the five quantities 
$M_{\r}$, $M_{a_1}$, $f_{\r}$, $f_{a_1}$ and $f_{\pi}$ 
depend on the six free parameters  $f$, $F$, $b$, $c$, $\kappa$ and $g_{\r}$, 
that hence cannot all be determined. Let us choose to leave $\kappa$ undetermined, for example, and
 solve the algebraic relations for  the other five parameters in terms of the physical quantities.
The  $g_{\r\pi\pi}$ coupling can then be written as
\beqs
\lim_{m\rightarrow 0} g_{\r\pi\pi}^2&=&\lim_{m\rightarrow 0}\frac{M_{\r}^2(f_{\r}^2M_{a_1}^2(-1+\kappa)+f_{a_1}^2M_{\r}^2(1+\kappa))^2}{2f_{\pi}^4f_{\r}^2M_{a_1}^4(1-\kappa)^2}\,.
\eeqs

If one were restricted to the massless theory, only by gaining access to 3-point functions  could one measure $\kappa$.
Yet, detailed information  about the $m$-dependence of 2-point functions
can be used to  predict $g_{\r\pi\pi}$, and the width of the $\r$ meson.
In principle, the width of the $\rho$ meson $\Gamma_{\r}$ could be compared with
the physical width extracted from lattice calculations~\cite{Luscher:1991cf,Feng:2010es}. 
In this way we would be able to adjudicate explicitly whether the weak-coupling 
assumption that underpins this EFT treatment is justified. 
However, the direct extraction of $\Gamma_{\r}$ from lattice data is highly non-trivial,
and will require a future dedicated study.

There is no reason a priori to expect that $g_{\r}$, or $g_{\r\pi\pi}$, be small,
except in the large-$N$ limit.
The fact that from 2-point functions we can infer some of the properties of the EFT that 
enter the 3-point functions holds only provided  the coupling is small, with $g_{\rho\pi\pi}^2/(48\pi) \ll 1$.
Furthermore, if $M_{\r}$ and $M_{a_1}$ happen to be  large in respect to $f_{\pi}$,
bringing them close to the natural cut-off set by the derivative expansion, 
it would again signal  a break-down of the perturbative expansion within the EFT.

Nevertheless, even in the regime of large $g_{\r}$, we can still learn something from
the expansion in small mass $m$. In particular, we should be able to use the EFT
to reproduce the $m$-dependence of masses and decay constants, at least  in the small-$m$ regime.
In the future, we envision repeating the study performed in the following sections 
for $Sp(2N)$ theory with dynamical fermionic matter, and with larger
values of $N$, and hence  track the $N$-dependence of the individual coefficients.

\subsection{Spin${-}1/2$ composite fermions and the top partner}
\label{baryons}

In composite-Higgs models, the generation of the SM fermion masses is
often supplemented by the mechanism of partial compositeness (PC).  
The SM fermions, in particular the top quark, mix with spin-$1/2$ bound states emerging from the novel 
strong-interaction sector (the composite sector), and phenomenologically this allows both
to enhance the fermion mass (as in precursor top-color models) and  to trigger
electroweak symmetry breaking via vacuum (mis)alignement.
As an example, we borrow some of the construction in~\cite{Katz:2005au} and~\cite{Barnard:2013zea}.
So many other, equally compelling, examples exist in the literature, 
that we refer the reader to the review~\cite{Panico:2015jxa} and to the references therein.\footnote{{   See also the approach
based on an extended EFT in~\cite{DeGrand:2016pgq}.}}

Let us assume that the microscopic theory admits the existence of $Sp(4)$-colour singlet  operators $\hat{\Psi}_i$ and $\hat{\Psi}^c_i$,
that have spin-$1/2$, carry $SU(3)_c$ colour and, combined, span vectorial representations of the SM gauge group.
The index $i=1,2$ refers to the $SU(2)_L$ singlets and doublets, respectively, and the notation refers to the fact that we write the operators as 2-component fermions.
Let us now consider the low-energy description of the lightest particles excited from the vacuum by such operators,
and write it in terms of new 2-component spinorial fields $\Psi_i$ and $\Psi^c_i$ with the same quantum numbers as $\hat{\Psi}_i$ and $\hat{\Psi}^c_i$.
Coarse-graining over model-dependent details,  $\Psi_i$ and $\Psi^c_i$  have the correct quantum numbers to couple to the SM 
quarks,
in particular to the SM top quark, represented  by the 2-component  Weyl fermions $t$ and $t^c$,
provided $\Psi_i$ transforms on the fundamental of $SU(3)_c$ and $\Psi^c_i$ on its conjugate. 

Below the electroweak symmetry-breaking scale $v_W$,
the mass terms take the  form
\beqs
\label{eq:L_int}
 {\mathcal{L}}^{{\mathrm{mix}}} &=&
 -\frac{1}{2}\left\{
    \lambda_1 M_{\ast} \left(\frac{M_{\ast}}{\Lambda}\right)^{d_{\Psi} - 5/2} \Psi_1^T \tilde{C} t^c
    +    \lambda_2 M_{\ast}\left( \frac{M_{\ast}}{\Lambda}\right)^{d_{\Psi^c} - 5/2} t^T \tilde{C}\Psi_2^c+\right.\nonumber\\
    &&\left.
    + \lambda M_{\ast} \left[\Psi_1^T \tilde{C} \Psi_1^c\frac{}{}+\,\Psi_2^T \tilde{C} \Psi_2^c\right] 
    + y v_W \left[\Psi_1^T \tilde{C} \Psi_2^c\frac{}{}+\,\Psi_2^T \tilde{C} \Psi_1^c\right]^{\frac{}{}}  \right\}\,+\,{\mathrm{h.c.}}\,,
\eeqs
where $\lambda_1$, $\lambda_2$, $\lambda$ and $y$ are dimensionless couplings, $M_{\ast}$ represents the typical scale of the masses of  composite
fermions in the $Sp(4)$ gauge theory and $\Lambda$ represents the underlying scale at which (third-generation) flavour physics arises (see also~\cite{Katz:2005au}).
$d_{\Psi}=d_{\Psi^c}$  is the dimension of the operators $\hat{\Psi}$ and $\hat{\Psi}^c$ in the underlying theory.

Diagonalisation of the resulting mass matrix, under the assumption that $y v_W$ be small in respect to the other scales, 
yields two heavy Dirac masses approximately given by
\beqs
m^2_1&\simeq&\left(\lambda^2  \,+\,  \lambda_1^2  \left(\frac{M_{\ast}}{\Lambda}\right)^{2d_{\Psi} - 5} \right) M_{\ast}^2\,,\\
m^2_2&\simeq&\left(\lambda^2 \,+\,  \lambda_2^2  \left(\frac{M_{\ast}}{\Lambda}\right)^{2d_{\Psi^c} - 5} \right) M_{\ast}^2\,,
\eeqs
and finally the mass (squared) of the top is given approximately by
\beqs
m_t^2&\simeq&\frac{  \lambda_1^2  \lambda_2^2  
y^2 \left(\frac{M_{\ast}}{\Lambda}\right)^{2d_{\Psi}+2d_{\Psi^c}-10} v_W^2 M_{\ast}^4}{m_1^2m_2^2}\,.
\eeqs

In order to assess the viability of these models, one needs to provide a microscopic origin for all of the
parameters appearing in Eq.~(\ref{eq:L_int}).
To do so, one must specify the (model-dependent) microscopic details controlling the nature of the composite fermions.
Spin{-}$1/2$ composite $Sp(4)$-neutral particles arise in the presence of
fermions in higher-dimensional irreducible
representations.  As an example, ref.~\cite{Barnard:2013zea} proposes
to extend the field content of the microscopic theory 
in Table~\ref{Fig:fields} to include 2-component 
elementary fermions $\chi$ (and $\chi^c$) in the antisymmetric representation of the gauge $Sp(4)$,
transforming as singlets of the global $SU(4)$, and on the fundamental (and anti-fundamental) representation of the $SU(3)_c$ 
gauge symmetry of QCD.

The $\chi$ and $\chi^c$  fermions carry QCD colour charge, 
which allows 
to construct coloured composite states in the antisymmetric,
six-dimensional representation of the global $SU(4)$ group, by coupling 
them to a pair of fundamental fermions $q$.
For example, the operators $\hat{\Psi}$ and $\hat{\Psi}^c$ aforementioned can be obtained as
\begin{equation}
\label{eq:top_partner_list}
\hat{\Psi}^{ab\alpha} \equiv \left ( q^{a} \chi^{\alpha} q^{b} \right)\,,\\
\hat{\Psi}^{c\,ab}_{\alpha} = \left (q^{a} {\chi}^c_{\alpha} q^{b} \right)\,,
\end{equation}
where {   summations over} $Sp(4)$ gauge indices  are understood,
while we show explicitly the (antisymmetrised) global $SU(4)$ indices $a$ and $b$, and the $SU(3)_c$ colour index $\alpha$.

One of our long-term goals is to study the PC mechanism with lattice
simulations, which  requires generalising the lattice study we will report upon  in the following sections
to the case in which the field content contains at least two species of fermions transforming in different representations of the fundamental gauge
group.  The example we outlined here, though incomplete,
immediately highlights how, from the phenomenological perspective, the determination of the masses
of the top partners (the scale $M_{\ast}$ and couplings such as $\lambda$, as a function of the 
elementary-fermion mass $m$)
in the PC mechanism are of direct interest, as they represent 
a way to test  the theory. At the same time,
they are accessible on the lattice, even without 
introducing (model-dependent) couplings to the SM fields.

The other additional, essential, input from
non-perturbative dynamics of the microscopic theory is the anomalous
dimension of the top-partner operators, such as $\hat{\Psi}$ and $\hat{\Psi}^c$ in the example.   
 For the PC mechanism to be valid,
in principle one needs the operator dimensions to be small, for example $d_{\Psi} \le 5/2$,
which implies that the operator $\hat{\Psi}_1^T \tilde{C}  t^c$ is relevant in the IR,
and that the anomalous dimensions of the  candidate operators 
have to be non-perturbatively large. 
In practice, since $\Lambda/M_{\ast}$ is not infinity, this requirement may be relaxed, at the price of admitting some degree of fine-tuning.

Finally, the (model-dependent) extension of the field content, required by the PC mechanism, also implies the enlargement of the global 
symmetry, and additional light PNGB's, some of which are neutral, some of which carry $SU(3)_c$ colour, and many of which may be lighter than the typical scale
of the other composite particles. Lattice calculations of the masses of such particles would offer the opportunity to connect with the phenomenology derived from direct
particle searches at the LHC.

\section{Numerical lattice treatment}
\label{Sec:Preliminary2}

In this Section, we present the discretised Euclidean action and Monte Carlo techniques 
used in the numerical studies. 
We adapt the state-of-the-art lattice techniques established for QCD 
to the two-flavour $Sp(4)$ theory. 
Pioneering lattice studies of $Sp(4)$ gauge theory without matter can be found in~\cite{Holland:2003kg}. 
Numerical calculations are carried out by modifying the HiRep code~\cite{DelDebbio:2008zf}.

\subsection{Lattice action}

For the numerical study of $Sp(2N)$ gauge theory on the lattice, we consider the standard plaquette action 
\beqs
S_g[U]=\beta\sum_x\sum_{\mu<\nu} \left(1-
\frac{1}{2N}\textrm{Re}~\textrm{Tr}\ \mathcal{P}_{\mu\nu}(x)\right),
\eeqs
where $\beta=4N/g^2$ is the lattice bare gauge coupling, and $N=2$ in the $Sp(4)$ case of this paper. 
The plaquette $\mathcal{P}_{\mu\nu}$ is given by
\beqs
\mathcal{P}_{\mu\nu}(x)=U_\mu(x) U_\nu(x+\hat{\mu}) U_\mu^\dagger(x+\hat{\nu}) U_\nu^\dagger(x)\,,
\eeqs
where the link variables $U_\mu(x)$ are $Sp(4)$ group elements in the fundamental representation,
while $\hat{\mu}$ and $\hat{\nu}$ are unit vectors along the $\mu$ and $\nu$ directions.

In the dynamical simulations with two Dirac fermions in the fundamental representation, 
we use the (unimproved) Wilson action
\beqs
S_f[U,\bar{\psi},\psi]=a^4\sum_x \bar{\psi}(x)D_m \psi(x)\,,
\eeqs
where the massive Wilson-Dirac operator is given by
\beqs
D_m\psi(x)&\equiv&(D+m_0)\psi(x) \nonumber\\
&=&(4/a+m_0)\psi(x)
-\frac{1}{2a}\sum_\mu \left\{\frac{}{}
(1-\gamma_\mu)U_\mu(x)\psi(x+\hat{\mu})+\right.\\
&& \left.\nonumber
+(1+\gamma_\mu)U_\mu(x-\hat{\mu})\psi(x-\hat{\mu})\frac{}{}
\right\},\nonumber
\eeqs
where $a$ is the lattice spacing and $m_0$ is the bare fermion mass. 

\subsection{Heat Bath}
\label{Sec:HB}

\begin{figure}[h]
    \begin{center}
    \includegraphics{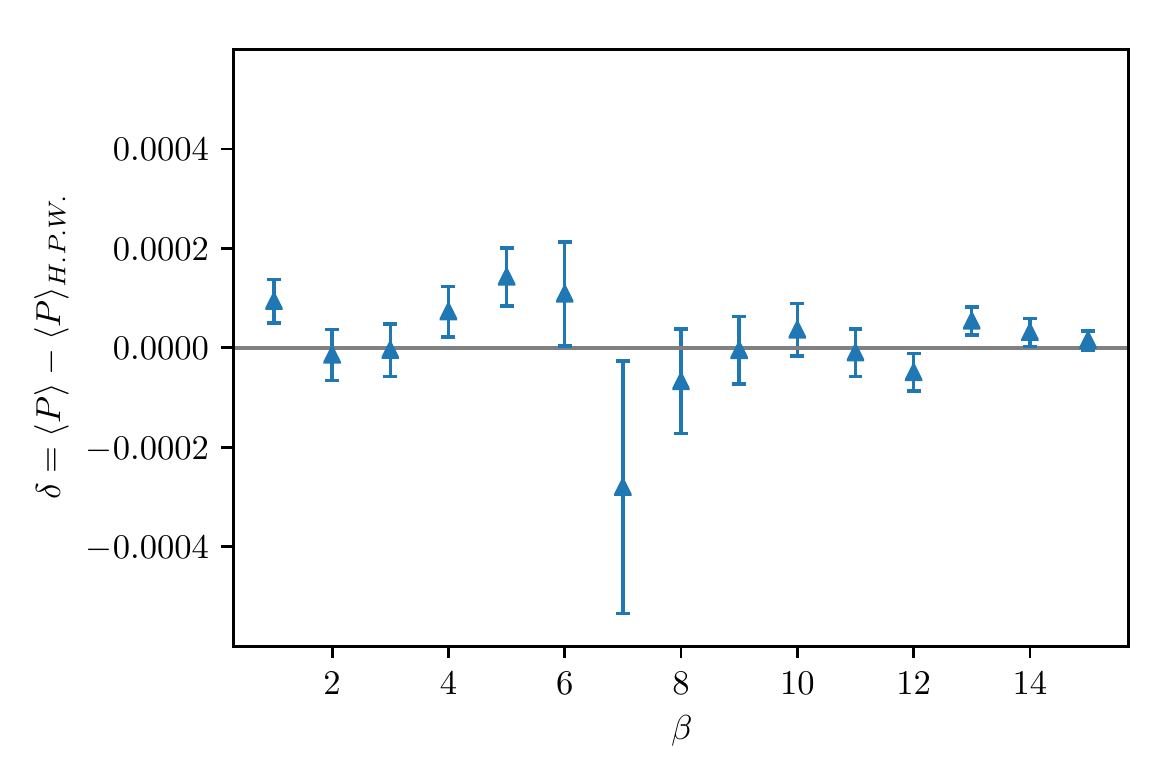}
    \end{center}
    \caption{
    Difference between the averaged plaquette obtained at various values of $\beta$ in this work and in Ref.~\cite{Holland:2003kg}.  
    The symbols $\langle P \rangle$ and $\langle P \rangle_{H.P.W.}$ denote the measurements from this work and those reported in Ref.~\cite{Holland:2003kg}, respectively.
Our lattice  volume  is $V=8^4$, and both calculations use the HB algorithm with over-relaxation, as explained in Sec.~\ref{Sec:HB}. Each point 
    has been obtained from $5000$ measurements, and the errors are corrected for autocorrelations.
    }
    \label{Fig:plaqvsbeta}
\end{figure}

As a powerful way to perform calculations in the pure  $Sp(4)$ gauge theory, 
we implemented a heat bath (HB) algorithm with micro-canonical over-relaxation updates, to improve the decorrelation 
of successive configurations. 
As in the case of $SU(N)$~\cite{Cabibbo:1982zn}, the algorithm acts in turn on  $SU(2)$ subgroups,
the choice of which can be shown to strongly relate to the ergodicity of the update pattern.

A sufficient condition to ensure ergodicity is to update the minimal set of $SU(2)$ subgroups to cover
the whole $Sp(2N)$ group. This condition can be suitably translated to the algebra of the group
  and generalised to any $Sp(2N)$. In the  $Sp(4)$ case, of relevance to this paper, 
  we choose to update a redundant set of subgroups,
in order to  improve the decorrelation of configurations. 
We provide below a possible partition of the generators used to cover all of the $Sp(4)$ gauge group, 
written with the notation of~\cite{Lee:2017uvl}.

\begin{itemize}
\item $SU(2)_L$ subgroup, with generators $T_L^i$ in Eq.~(B.6) of~\cite{Lee:2017uvl}.
\item $SU(2)_R$ subgroup, with generators $T^i_R$ in Eq.~(B.7) of~\cite{Lee:2017uvl}.
\item $SU(2)_{\tau}$ subgroup, with generators expressed in terms of B.4 in~\cite{Lee:2017uvl}:
\begin{flalign}
\tau^1 = T^{11}  \  ;\qquad \tau^2 = T^7 \ ;  \qquad \tau^3 = T^{15} \ . 
\end{flalign}
\item $SU(2)_{\tilde{\tau}}$ subgroup, with generators expressed in terms of B.4 in~\cite{Lee:2017uvl}:
\begin{flalign}
\tilde{\tau}^1 = T^{13} \  ;\qquad \tilde{\tau}^2 = T^8 \ ;  \qquad \tilde{\tau}^3 = T^{14} \ .
\end{flalign}
\end{itemize}
The set of $10$ generators $T^i_L$, $T^i_R$, $\tau^{1\,,\,2}$ and $\tilde{\tau}^{1\,,\,2}$ spans the whole $Sp(4)$.
The minimal set of $5$ elements that generate the whole group by closure consists for example of any two elements $T^i_L$,
any two elements $T^{j}_R$ and one additional element among $\tau^{1\,,\,2}$ and $\tilde{\tau}^{1\,,\,2}$.

As a check of correctness of the algorithm we employed, we compared the average  
of the elementary plaquette to the results obtained in~\cite{Holland:2003kg}, 
as shown in Fig.~\ref{Fig:plaqvsbeta}, confirming that they  
are compatible  within the statistical errors.

\subsection{Hybrid Monte Carlo}
\label{Sec:HMC}

\begin{figure}
\begin{center}
\includegraphics{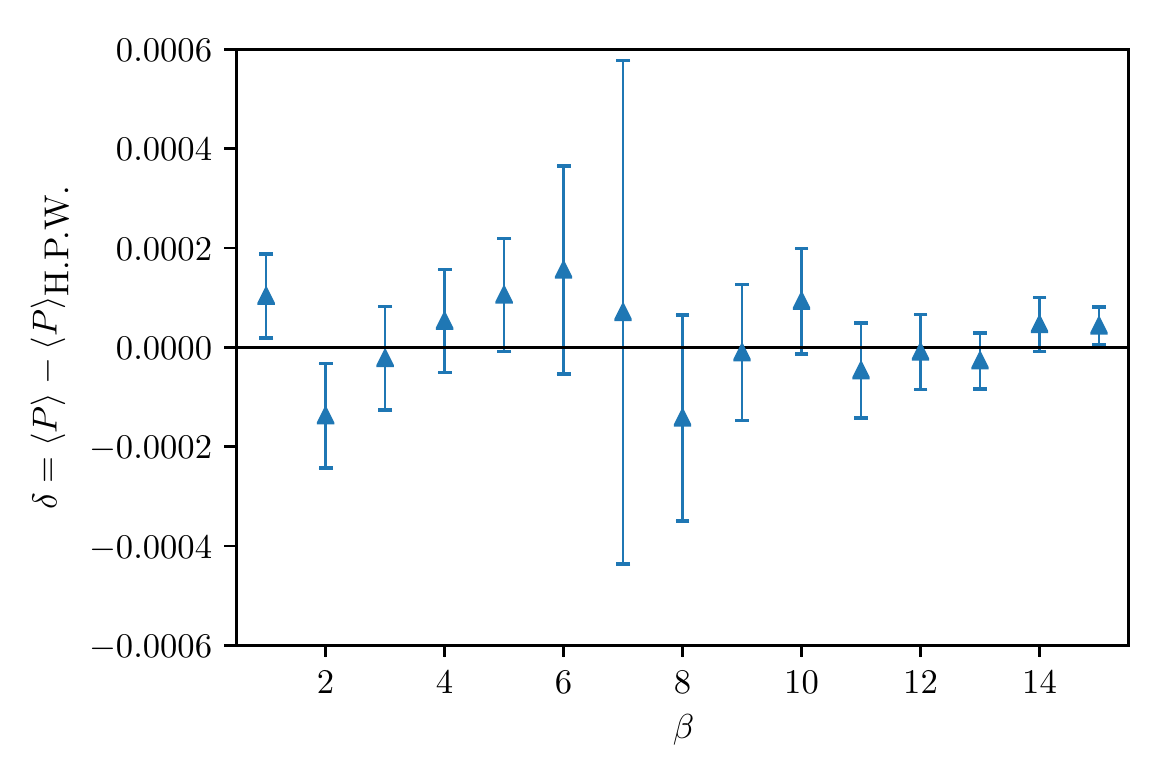}
\end{center}
\caption{Differences between the average plaquette values $\langle P \rangle$ obtained by using the HMC algorithm 
described in Sec.~\ref{Sec:HMC} with heavy quarks ($a m_0=10.0$) 
and the HB for pure $Sp(4)$ theories from the literature~\cite{Holland:2003kg}. 
}
\label{Fig:plaq_diff}
\end{figure}

In the study of $N_f=2$ dynamical Dirac fermions, we make use of the hybrid Monte Carlo (HMC) algorithm.  
As $Sp(4)$  is a subgroup of $SU(4)$, most of the numerical techniques used for $SU(N)$ with an
even number of fermions can straightforwardly be extended to our study. 
However, there are two distinguishing features.

First of all, in contrast to the HB algorithm, 
the explicit form of the group generators of ${ Sp}(4)$ 
is necessary for the molecular dynamics (MD) update. 
For instance, the MD forces for the gauge fields are given by 
\beqs
F^A_G(x,\mu)=-\frac{\beta}{2 N T_F} {\textrm {Re}}~{\textrm {Tr}} [i T^A U_\mu(x) V^\dagger_\mu(x)],
\eeqs
where $V_\mu(x)$ is the sum of forward and backward staples around the link $U_\mu(x)$. 
The generators  $T^A$ with $A=6\,,\,\cdots\,,\,15$ are given in  Appendix B of~\cite{Lee:2017uvl}. 
The group invariant $T_F$ is defined as $\Tr(T^A T^B)=T_F \delta^{AB}$, which in our case yields $T_F=1/2$,
so that for $Sp(4)$ the normalization is $2 N T_F=2$.

Secondly, due to machine precision, 
it is not guaranteed that the link variables stay in the $Sp(4)$ group manifold.
In analogy with the re-unitarization process implemented in $SU(N)$ studies, 
we perform a re-symplectisation at the end of each MD step. 
We describe in Appendix~\ref{Sec:projection}  the procedure, 
based on $Sp(4)$ projection that makes use of the quaternion basis.

{  
As a further  test of this implementation of
the HMC algorithm, we calculated the expectation value of the difference of the auxiliary Hamiltonian at the beginning and the 
end of a MD trajectory $\langle \Delta H \rangle$ for various values of the integration step size $\epsilon$, in the case with $\beta=6.9$ and $am_0=-0.85$ 
on a $24\times 12^3$ lattice. We found that $\langle \Delta H \rangle$ is proportional to $\epsilon^4$, as expected for a second order Omelyan integrator~\cite{Takaishi:1999bi}, 
and Creutz's equality $\langle \textrm{exp}(-\Delta H) \rangle=1$~\cite{Creutz:1988wv} is satisfied. 
We also checked that the average plaquette values are consistent 
with each other for all values of the step size.}

The HiRep code~\cite{DelDebbio:2008zf} is designed to implement $SU(N)$ gauge theories with 
a generic number of colours and flavours, with fermions in any two-index representation. 
One of its crucial features is that the gauge group and the representation 
can be fixed at compile time 
by using preprocessor macros. 
This provides us with great flexibility in implementing the aforementioned features of $Sp(4)$.

\begin{figure}
\begin{center}
\includegraphics{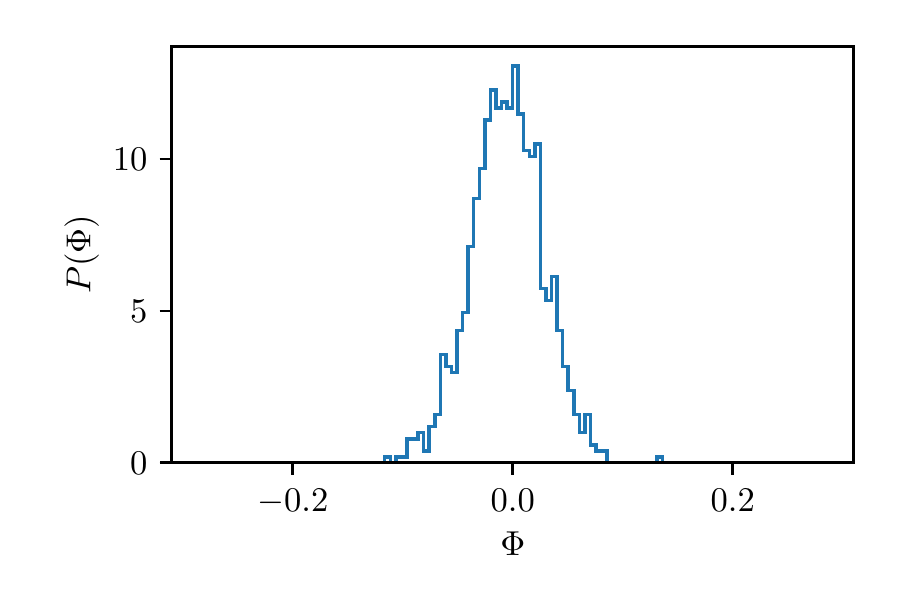}
\includegraphics{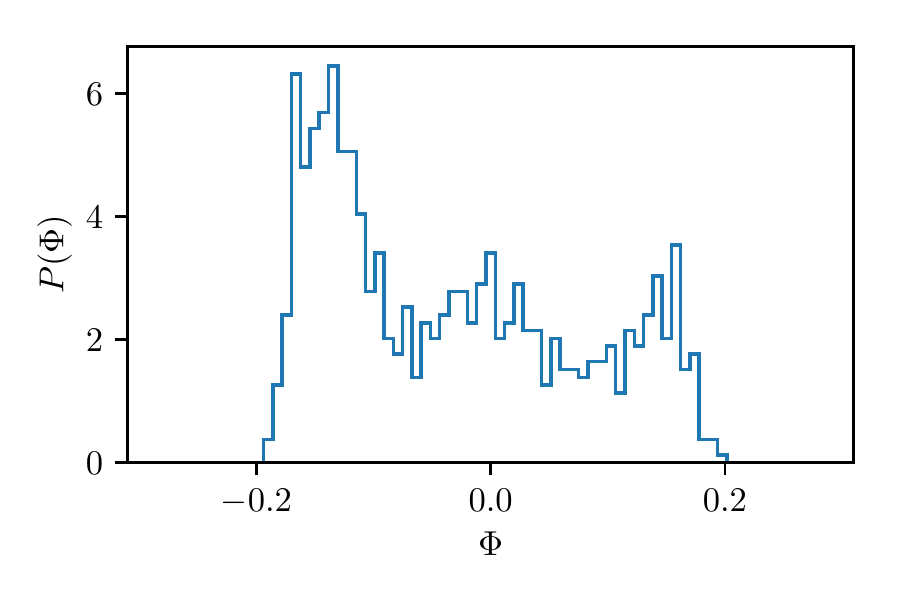}
\includegraphics{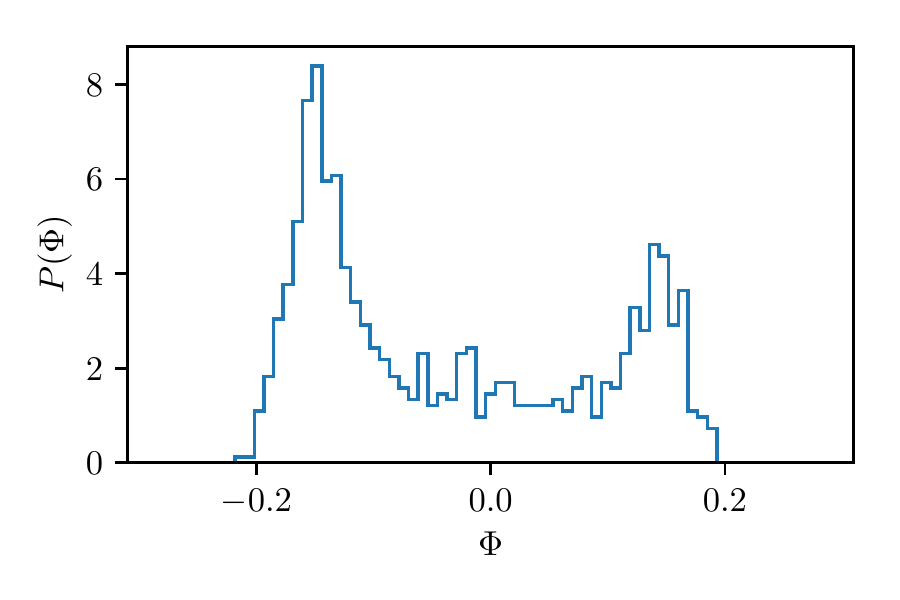}
\end{center}
\caption{
Probability distribution $P(\Phi)$ of the expectation value of the Polyakov loop $\Phi$ averaged over the space-like points,
 defined in Eq.~(\ref{Eq:Phi}), with the normalisation of $\int d\Phi P(\Phi)=1$,
 measured on a lattice with size $4\times12^3$, by making use of the HMC algorithm. 
 The lattice couplings are $\beta=7.3$, $7.339$ and $7.345$ (top to bottom panel),
roughly corresponding to temperatures $T$ below, at, and above the critical temperature $T_c$, respectively. 
}
\label{Fig:deconfinement}
\end{figure}

As a nontrivial test of the HMC code, we first calculate the expectation value of the plaquette 
of the theory with two degenerate, very heavy fundamental fermions ($a\, m_{0} = 10.0$) 
and compare the results with the pure $Sp(4)$ results from~\cite{Holland:2003kg}. 
In Fig.~\ref{Fig:plaq_diff}, we plot the differences of the average plaquette values 
between the two calculations for various values of $\beta$.
The two series are compatible with each other, within the statistical errors. 

It is known that the pure $Sp(4)$ theory in $3+1$ dimensions exhibits a first-order 
deconfinement phase transition~\cite{Holland:2003kg}. 
Although a finite-size scaling analysis is needed to confirm the existence of the first-order phase transition, 
for the purpose of a consistency check of the code it is worth showing 
numerical evidence of
the coexistence of the confined and deconfined phases. 
To this end, we calculate the expectation value of the Polyakov loop averaged over the space-like points, defined by
\beqs
\Phi&\equiv&\frac{1}{N_s^3}\sum_{\vec{x}}{\textrm {Tr}}\left(\prod_{t=0}^{N_t-1}U_0(t,\vec{x})\right).
\label{Eq:Phi}
\eeqs
The temperature $T$ in lattice units is identified with the inverse of 
the extent of the temporal lattice, $1/N_t$. 
Near the critical temperature $T_c$, the probability distribution of $\Phi$ indeed shows the coexistence 
of two phases, as in the second panel of Fig.~\ref{Fig:deconfinement}. 
In agreement with expectations, the numerical results also show 
that the expectation value of the Polyakov loop averaged over space
 is dominated by configurations at $\Phi=0$ in the confining phase 
(first panel of Fig.~\ref{Fig:deconfinement}), 
while it is dominated by two non-zero values of $\Phi$ in the deconfinement phase
(third panel of Fig.~\ref{Fig:deconfinement}).

\section{Lattice calibration}
\label{Sec:Phasespace}

This Section is devoted to discuss two lattice technicalities that are important 
in order to extract the correct continuum physics: we address the 
problem of setting the scale, using the gradient flow, and study the topology of the ensembles 
generated by our numerical process, to verify that there is no evidence of major problems in 
the lattice calculations.

\subsection{Scale setting and gradient flow}
\label{sec:scale_setting}

Lattice computations are performed by specifying dimensionless bare
parameters in the simulation, and all dimensionful results are extracted in units
of the lattice spacing.  These results have to be extrapolated to the
continuum limit to make impact on phenomenology.  It is also desirable
to express them in natural units.  Such demands make the scale setting an important
task in lattice calculations.  To carry out this task, the most straightforward approach is to compute a physical
quantity on the lattice, and then compare with its experimental
measurement.  In the absence of experimental results for the $Sp(4)$ gauge theory, one can still accomplish
reliable continuum extrapolations by employing a scale defined on theoretical grounds,
such that one can determine the ratio $a_{1}/a_{2}$ of the
lattice spacings in two simulations 
performed at different choices of the bare parameters.

The gradient flow in quantum field theories, as revived in recent
years by Martin L\"{u}scher in the context of the trivialising
map~\cite{Luscher:2009eq}, is a  popular
method for scale setting~\cite{Luscher:2010iy, Luscher:2011bx}.  
To study the gradient flow in a field theory, one first adds an extra
dimension, called {\it flow time} and denoted by $t$. 
An important point articulated by L\"{u}scher is that a field theory
defined initially with a cut-off can be  renormalised at
non-vanishing flow time.
In addition, choosing carefully the bulk equation governing the
gradient flow, the theory does not generate new operators
along the flow time (counter-terms), keeping the renormalisation of the
five-dimensional theory simple.\footnote{See
  Ref.~\cite{Fujikawa:2016qis} for a choice of the flow
  equation that induces the need for extra care of renormalisation in the 
   $\phi^{4}$ scalar field theory.}

{  
The Yang-Mills gradient flow of the gauge field $B_\mu(t,x)$ is
implemented {via} the equation
\begin{eqnarray}
  \label{eq:flow}
  && \frac{\dee B_\mu(t,x)}{\dee t} = D_{\nu}G_{\nu\mu}(t,x)\,,                 
   {\mathrm{with}} \mbox{ } B_\mu(t,x)|_{t=0} = A_\mu(x), 
\end{eqnarray}
where $G_{\mu\nu}$ is the field strength tensor} associated with
$B_{\mu}(t,x)$, $D_\mu = \partial_\mu + [B_\mu,\cdot]$ the
corresponding covariant derivative, and $A_\mu(x)$ the initial gauge
field in the four-dimensional theory.  Noticing that Eq.~(\ref{eq:flow})
describes a diffusion process, the flow time $t$ therefore has
length-dimension two.  It has been shown that, to all
orders in perturbation theory, any gauge invariant composite
observable constructed from $B_\mu(t,x)$ is
renormalised at $t>0$~\cite{Luscher:2011bx}.  In particular,
L\"{u}scher demonstrated that the action density can be related to the
renormalised coupling, $\alpha(\mu)$, at the leading order in
perturbation theory through
\begin{equation}
\label{eq:alpha_to_action}
  \alpha (\mu) = k_{\alpha} t^2\langle E(t)\rangle \equiv k_{\alpha}
  {\calE}(t) \,,
\end{equation}
with $\mu = \frac{1}{\sqrt{8t}}$, and
\begin{equation}
  E(t) = -\frac{1}{2}{\rm Tr}(G_{\mu\nu}G_{\mu\nu})\,.
\label{eq:energy_density}
\end{equation}
The dimensionless constant $k_{\alpha}$ is analytically computable~\cite{Luscher:2010iy}.  
Equation~(\ref{eq:alpha_to_action}) can actually serve as the
definition of a renormalisation scheme:   the
gradient-flow (GF) scheme.   Furthermore, since $t^2\langle
E(t)\rangle \equiv {\calE}$ is proportional to the GF-scheme coupling, this quantity
can be used to set the scale.  In other words, if one imposes the
condition
\beq
\label{eq:E_scale_setting}
  {\calE}(t)|_{t = t_{0}} = {\calE}_{0}\, ,
\eeq
where ${\calE}_{0}$ is a constant that one can choose, then $\sqrt{t_{0}}$ should be a
common length scale, assuming  lattice artefacts are under
control.  In practice, one measures $\sqrt{t_{0}}$ in lattice units.
That is, one computes $\sqrt{t_{0}}/a \equiv \sqrt{\hat{t}_{0}}$.
This allows the determination of the ratio of lattice spacings from
simulations performed at different values of the bare parameters.

It is worth mentioning that the diffusion radius in Eq.~(\ref{eq:flow}) is $\sqrt{8 t}$, and
it is convenient to define the ratio
\beq
\label{eq:c_tau_def}
 c_\tau = \sqrt{8t}/L \,,
\eeq
where $L$ is the lattice size.

Given that the right-hand side of Eq.~\eqref{eq:flow} is the gradient
of the Yang-Mills action, the most straightforward way to latticise it  is\footnote{The precise meaning of the Lie-algebra valued
  derivative $\partial_{x,\mu}$ is given in Ref.~\cite{Luscher:2010iy}.
}
\beq
 \frac{\partial V_{\mu} (t,x)}{\partial t} = - g_{0}^{2} \left \{
   \partial_{x,\mu} S^{({\mathrm{flow}})}_{{\mathrm{latt}}} \left [ V_{\mu} \right
   ]\right \} V_{\mu} (t,x) , \mbox{ }\mbox{ } V_{\mu} (0,x) = U_{\mu}
 (x)\, ,
\label{eq:WF_equation}
\eeq
where $V_{\mu}(t,x)$ is the gauge link at flow time 
$t$, and $S^{({\mathrm{flow}})}_{{\mathrm{latt}}}$ is a lattice gauge action.  Notice that
$S^{({\mathrm{flow}})}_{{\mathrm{latt}}}$ does not have to be the same as the gauge action
used in the Monte Carlo simulations.   We employ the
Wilson flow where $S^{({\mathrm{flow}})}_{{\mathrm{latt}}}$ is the
Wilson plaquette action.

The gradient flow serves as a smearing procedure for the gauge
fields.  This means the larger the flow time, the smoother the
resultant gauge configurations will be.  In other words, the larger
the flow time is, the smaller the ultraviolet fluctuations of
flown observables.  On the other hand, it
also means the gauge fields become more extended
objects as the flow time grows.  This results in longer
autocorrelation time, and makes the statistics worse.   Furthermore,
having $c_{\tau} > 0.5$ can lead to significant finite-volume
effects.  These are issues one would have to consider carefully when
choosing a value for the constant ${\calE}_{0}$ in Eq.~(\ref{eq:E_scale_setting}).

\begin{figure}
	\null\hfill\includegraphics{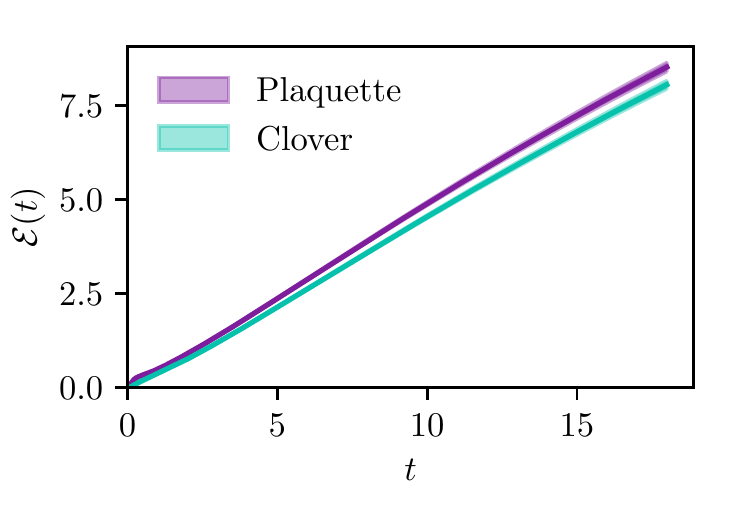}\hfill
	\includegraphics{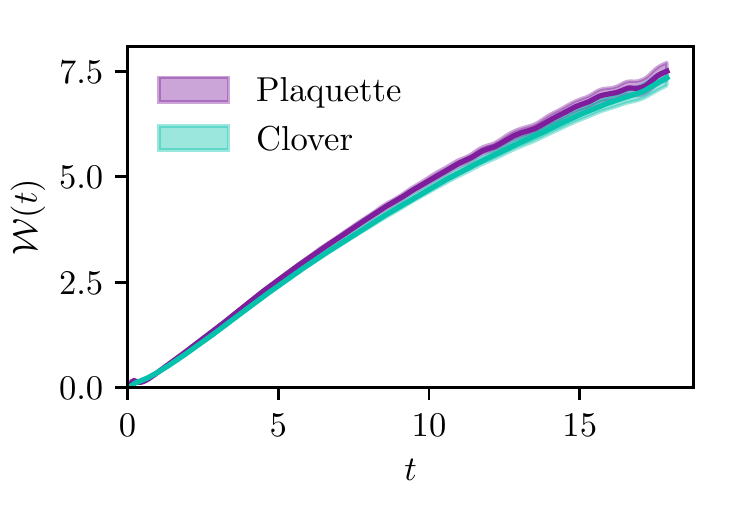}\hfill\null
	\caption{The Wilson flow functions ${\calE}(t)$ in Eq.~(\ref{eq:alpha_to_action}) (left panel)
	and ${\calW}(t)$ in Eq.~(\ref{eq:Wt_def}) (right panel)
	for $\Nf=2$, $\beta=6.9$, $a m_0=-0.90$ and $L=12$, 
	as a function of the flow time $t$, 
	computed by using 
	the  methods described in Sec.~\ref{sec:scale_setting}.}
	\label{fig:flow}
\end{figure}

The action density ${\calE}(t)$ at non-vanishing flow time is
obtained from the diffusion process in Eq.~(\ref{eq:WF_equation}), 
starting from the bare gauge fields.  To further reduce the cut-off
effects in the scale-setting procedure, an alternative quantity was
proposed in Ref.~\cite{Borsanyi:2012zs}.  Define
\begin{equation}
\label{eq:Wt_def}
 {\calW}(t) \equiv t \frac{\dee {\calE}(t)}{\dee t} .
\end{equation}
Then the scale can be set by
\begin{equation}
\label{eq:w0_scale_def}
 {\calW}(t)|_{t=w_{0}^{2}} = {\calW}_{0} ,
\end{equation}
where ${\calW}_{0}$ is again a dimensionless constant that one
can choose.

\begin{figure}
	\includegraphics{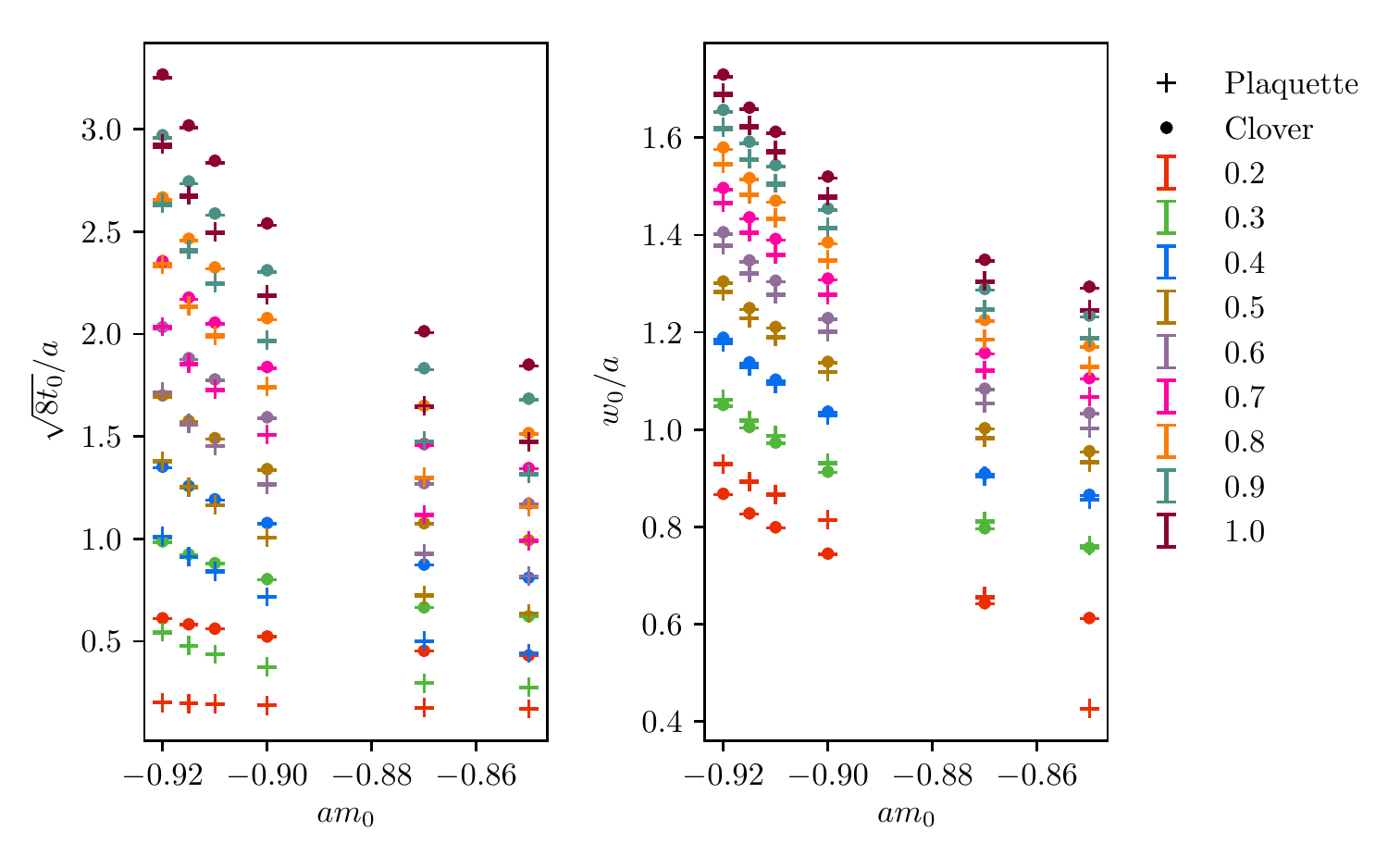}
	\caption{The gradient flow scales $t_0$ defined in Eq.~(\ref{eq:E_scale_setting}) (left panel)
	and $w_0$ defined in Eq.~(\ref{eq:w0_scale_def}) (right panel), normalised to the lattice spacing $a$,
	 as a function of the fermion mass $a m_0$ (on the horizontal axis),
	for various choices of the scales $\calE_0$ and $\calW_0$, and for the two choices of plaquette or clover (as indicated in the legend). {  
	Lattices with size $32\times16^3$ were used for $-0.92 \le a m_0 \le -0.89$ and $24\times12^3$ for $-0.87 \le a m_0 \le -0.85$.}}
	\label{fig:twm}
\end{figure}

On the lattice, the calculation of $E(t)$ depends on a definition of $G_{\mu\nu}$,
for which a variety of choices are available. The most obvious is to associate 
it with the plaquette ${\cal P}_{\mu\nu}$; an alternative is to define a four-plaquette 
clover, which has a greater degree of symmetry~\cite{Luscher:2010iy}. In the continuum, all definitions
should become equivalent, and at finite lattice spacing 
the relative difference between the two decreases
at large $t$. The shape of $E(t)$ at very small $t$ is dominated by ultraviolet
effects, and so differs strongly between the two methods; this introduces further
constraints into the choice of $\calE_0$.  Figure \ref{fig:flow} shows $\calE(t)$ and $\calW(t)$, calculated both via the 
plaquette and the clover. As anticipated from~\cite{Borsanyi:2012zs}, the
discretisation effects are smaller in $\calW(t)$ than $\calE(t)$; this is visible in the 
splitting between plaquette and clover curves being smaller in the
$\calW(t)$ case.\footnote{The relative size of discretisation effects in two different observables 
can also depend on the actions used in the Monte Carlo simulations and the implementation of the 
gradient flow~\cite{Fodor:2014cpa, Ramos:2015baa}, as well as the flow time~\cite{Lin:2015zpa}.}

In the continuum theory,  $B_\mu(x)$ are elements of the $Sp(4)$ gauge group; however,
it is possible that the finite precision of the computer could introduce some numerical artefact
that would cause the integrated $B_\mu (x)$ to leave the group. Since the integration is an 
initial value problem, any such artefact introduced would compound throughout the flow, 
giving potentially significant distortions at large flow time. For this reason we have introduced 
the re-symplectisation procedure described in Appendix~\ref{Sec:projection} after each 
integration step. We find  no appreciable difference between the flow with and without 
re-symplectisation.

We now proceed
to set the values of $\calE_0$ and $\calW_0$, such that $t_0$ and $w_0$ avoid
both the regions of finite lattice spacing and finite volume artefacts. In order to 
obtain a single value for the lattice spacing corresponding to a particular value of
$\beta$, and allow comparisons to be drawn with pure gauge theory, we must 
also be in the vicinity of  the chiral limit. Findings in~\cite{Borsanyi:2012zs} indicate
that when the fermions are light enough any mass dependence in $w_0$ should be small.

Figure~\ref{fig:twm} shows the fermion-mass dependence in
$\sqrt{8t_{0}}$ and $w_{0}$ at $\beta=6.9$, choosing
${\mathcal{E}}_{0} = {\mathcal{W}}_{0} \in [0.2, 1.0]$.  The
discretisation effects are significantly smaller in $w_{0}$.  
We see a relatively strong dependence on the fermion mass in both $t_0$ and $w_0$; this is contrary
to expectations from studies of QCD with light quarks~\cite{Borsanyi:2012zs}.
Presently we are studying this fermion-mass dependence.  
Results of this
study will be detailed in future publications.
It should be noted that if the behaviour highlighted here persists also in proximity of the chiral limit, extra
care will be needed in the process of taking the continuum limit.

\subsection{Topological charge history}

\begin{figure}
\begin{center}
		\includegraphics[width=0.80\linewidth]{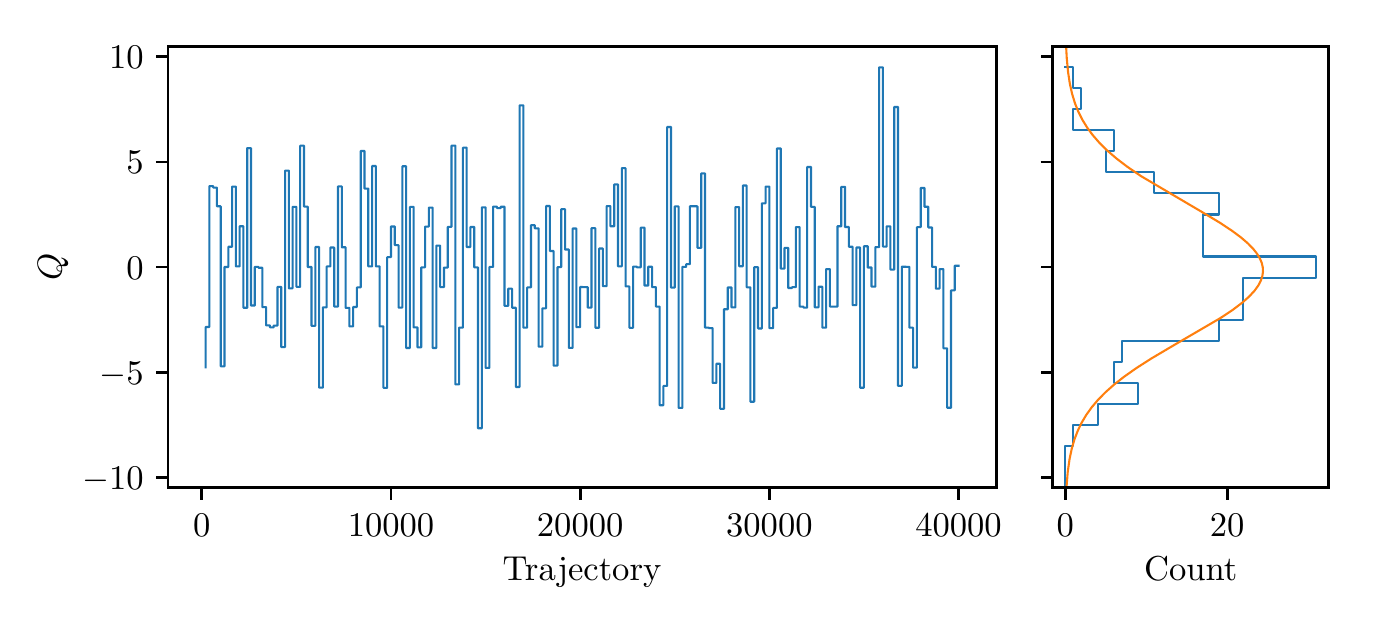}	
		\includegraphics[width=0.80\linewidth]{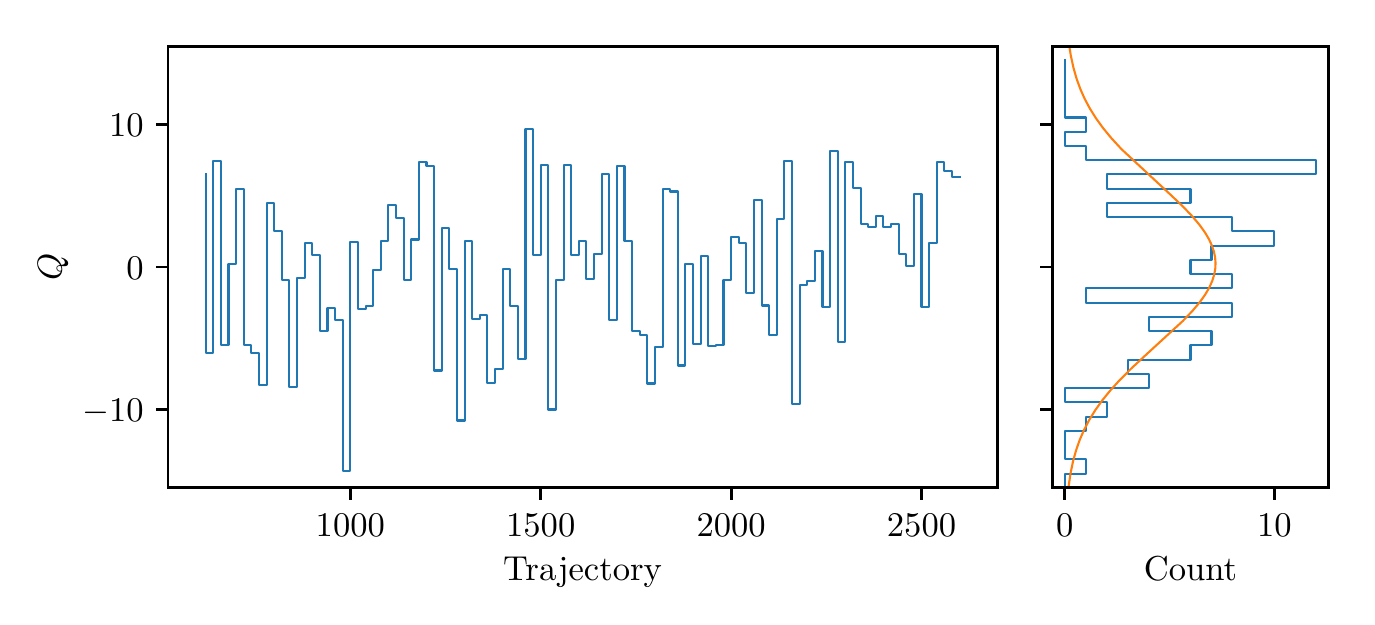}		
		\includegraphics[width=0.80\linewidth]{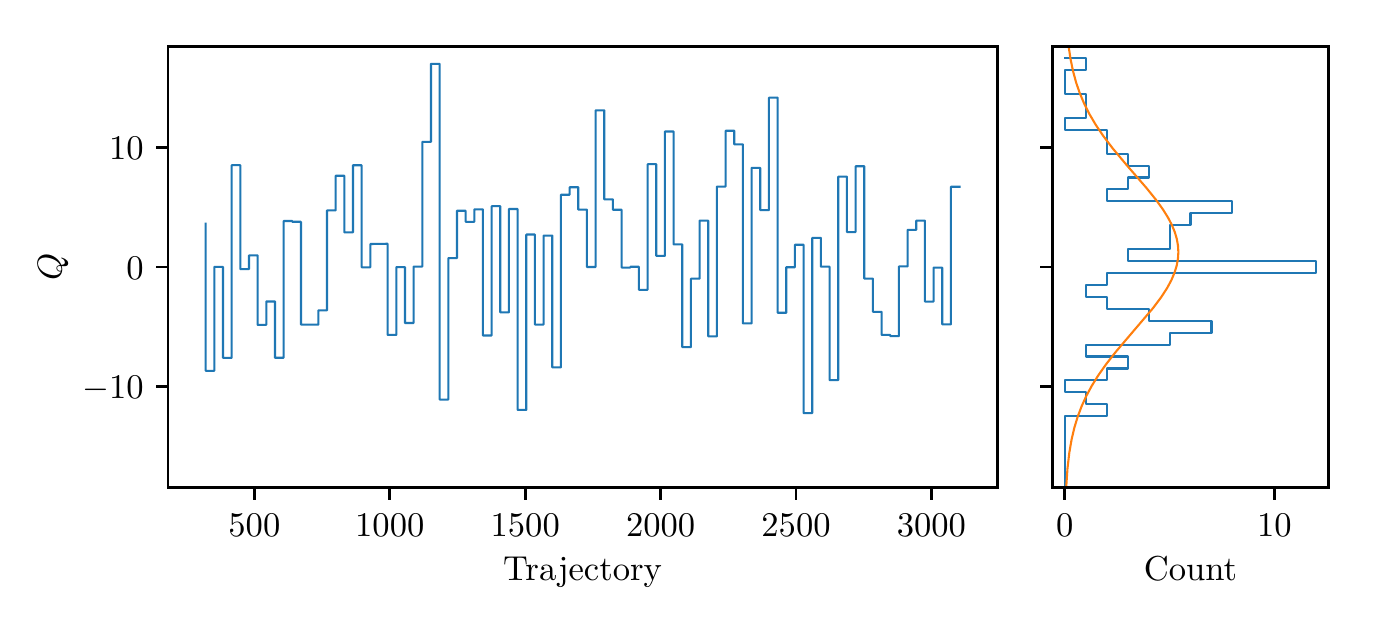}
	\caption{Histories and statistical distributions of the topological charge in Eq.~(\ref{Eq:Q}) for three selected ensembles.
	Top to bottom, the figures show the result for the following choices.
	The  pure gauge case, with $\beta=7.62$ and $L=16$ yields the average charge  $Q_0 = -0.17 \pm 0.24$ and the 
	standard deviation is $\sigma = 3.24 \pm 0.24$, with $\chi^2/{\rm d.o.f}=0.88$.
	For $\Nf = 2$, with $\beta=6.9$, $L=12$ and $a m_0=-0.85$ we find $Q_0 = 0.23 \pm 0.88$ and $\sigma = 5.75 \pm 0.89$, with $\chi^2/{\rm d.o.f}=2.51$.
	For $\Nf = 2$ with $\beta=6.9$, $L=16$ and $a m_0=-0.92$ we obtain
	$Q_0 = 1.26 \pm 0.97$ and $\sigma = 6.61 \pm 0.98$, with $\chi^2/{\rm d.o.f}=1.28$. 
	}
	\label{fig:qhist}
\end{center}
\end{figure}
As the lattice extent is finite in all directions, a given configuration will fall into one of a number of topological sectors, 
labelled by an integer (or, at finite $a$, near-integer) topological charge $Q$, which is expected to have a 
Gaussian distribution about zero. Since it is probabilistically unfavourable to change a discrete global 
observable using a small local update, $Q$ can show very long autocorrelations; as the continuum limit 
(i.e.\ the limit of integer $Q$) is approached, $Q$ can ``freeze'', ceasing to change at all. 

It is necessary to check that $Q$ is not frozen, and instead moves hergodically, for two reasons. Firstly, the 
exponential autocorrelation time of the Monte Carlo simulation as a whole scales as one of the longest autocorrelation time in the system 
(see e.g.~\cite{Luscher:2011kk}). 
Secondly, the values of physical observables depend on which topological sector a configuration is 
in \cite{Galletly:2006hq}; sampling a single $Q$ or an unrepresentative distribution of $Q$s will introduce 
an uncontrolled systematic error. It is therefore necessary to verify that $Q$ not only moves sufficiently rapidly, but also displays the expected Gaussian histogram.

The topological charge $Q$ is computed on the lattice as
\beqs
	Q = \sum_x q(x)\,, {\rm with}\,\,  q(x) = \frac{1}{32\pi^2} \epsilon_{\mu\nu\rho\sigma} \operatorname{Tr} \left\{U_{\mu\nu}(x)U_{\rho\sigma}(x)\right\}\,,
	\label{Eq:Q}
\eeqs
and where $x$ runs over all lattice sites. For gauge configurations generated by 
Monte Carlo studies, this observable is dominated by ultraviolet fluctuations; therefore it is necessary to 
perform some sort of smoothing to extract the true value. The gradient flow (described in the 
previous subsection) is used for the purposes of this work.

We have examined the topological charge history for all our existing ensembles,
 including both pure gauge and those with matter. 
In most cases, $Q$ is found to move with no noticeable autocorrelation, and shows the expected Gaussian distribution centred on $Q = 0$. 
Samples of these histories are shown in Fig.~\ref{fig:qhist}. Some marginal deviation is visible for example in the second of the three series.

\section{The spectrum of the Yang-Mills theory}
\label{Sec:confinement}

In this Section, we focus our attention on the $Sp(4)$ Yang-Mills theory.
We start by reminding the reader about several technical as well as conceptual 
points related to the physics of glueballs and to the description of confinement in terms of 
effective string theory. We then summarise the specific methodology
we adopt in the process of extracting physical information from the lattice data.
We conclude this section by presenting our numerical results on the glueballs, and commenting on their general implications.

\subsection{Of glueballs and strings}
\label{Sec:glueandstring}

At zero temperature, $Sp(4)$ Yang-Mills theory is expected to confine. 
The particle states are colour-singlet gluon bound states, referred to in the literature as {\em glueballs},
and labelled by their (integer) spin $J$ and (positive or negative) parity $P$ quantum numbers as $J^P$.\footnote{Since the gauge group is pseudo-real, 
charge conjugation is trivial.}
To distinguish between states with 
the same $J^{P}$ assignment but different mass, in the superscript of  the $n$-th excitation we add $n$ asterisks ($\ast$). 
For instance, $2^{+\ast\ast}$ denotes the second excitation with $J = 2$ and $P = +$, 
while $2^{+}$ is the ground state in the same channel.

The calculation of the mass spectrum of glueballs requires a fully non-perturbative treatment of
 the strongly interacting dynamics.
 We follow the established procedure that extracts glueball masses from the Monte Carlo 
 evaluation of two-point functions of gauge-invariant operators $O(x)$. 
The operators $O(x)$ transform according to irreducible representations 
of the rotational group and either commute or anti-commute with the parity operator,
hence having well-defined $J^P$. Given  $O(x)$ defined at any spacetime point $x = (t, \vec{x})$,
 we  separate the space-like and time-like components $\vec{x}$ and $t$,\footnote{Not to be confused with 
 the flow time in Sec.~\ref{sec:scale_setting}.} and define the zero-momentum operator $O(t)$ as 
\begin{equation}
O(t) = \sum_{\vec{x}} O(t, \vec{x}) \ , 
\end{equation}
where the sum runs over all spatial points $\vec{x}$ at fixed $t$. 
The lowest-lying glueball mass in the $J^P$ channel is then given by 
\begin{equation}
m_{J^P} = - \lim_{t \to \infty} \frac{\log \langle O^{\dag}(0) O(t) \rangle} {t}\,.
\end{equation}

Assuming only contributions from poles (an hypothesis that certainly holds at large $N$), 
 we can insert a complete set of single-glueball states $|g_n(J,P)\rangle$ carrying the same quantum numbers of $O(t)$ in the
correlator $\langle O^{\dag}(0) O(t) \rangle$, and arrive to
\begin{equation}
\label{eq:ym:eqg0}
\langle O^{\dag}(0) O(t) \rangle = \sum_n |c_{J^P,n}|^2 e^{- m_{J^P,n} t} \ ,
\end{equation}
with $c_{J^P,n} = \langle  g_n(J,P) | O(0) | 0 \rangle$ being the overlap of the state $|g_n(J,P) \rangle$ 
with the state $O(0) | 0 \rangle$, created by acting with $O(0)$ on the vacuum $| 0 \rangle$.  
The correlator $\langle O^{\dag}(0) O(t) \rangle $ contains information not only on the ground state but also 
on all excitations with non-null overlap with $O(0) | 0 \rangle$ in the given $J^P$ channel. 

Glueballs are not the only interesting observables in Yang-Mills theory. 
In the presence of infinitely massive, static quarks, the spectrum contains also confining flux tubes. 
While  flux tubes are exposed by the static probes, their physics is fully determined by the Yang-Mills dynamics and 
plays a crucial role in the study of confinement.
Consider a static quark $Q_S$ and the corresponding antiquark $\bar{Q}_S$, a distance $\Delta x$ apart. In a confining theory, the
 static quark-antiquark pair is bound by a linearly rising potential
\begin{equation}
	V(\Delta x) = \sigma \Delta x\,,
\end{equation}
where the quantity $\sigma$ (having dimension of a mass squared) is the (confining) {\em string tension}, and provides a measurement of the dynamically generated 
 confinement scale. In Yang-Mills theory there is only one dynamically generated dimensionful quantity, 
 hence the square root of the string tension also sets the scale of the glueball masses, besides providing a fundamental test of confinement.

The semiclassical cartoon associated with linear confinement explains the latter as arising from the formation of a thin
(flux) tube in which the conserved colour flux is being channeled. Over distances much bigger than the transverse size of the confining flux tube,
 the latter can be represented by a string 
 of tension $\sigma$ binding quark and antiquark together. 

At zero temperature, a signature of  confinement is the area law:
\begin{equation}
	\label{eq:arealaw}
	\langle W (\Delta x, \Delta t) \rangle \simeq e^{- \sigma A} \,,
\end{equation}
where the Wilson loop $W(\Delta x, \Delta t)$ is  defined as
\begin{equation}
\label{eq:ym:2}
W(\Delta x, \Delta t) = \mathrm{Tr} \left( \mathrm{P}e^{i g \oint_{\cal R} A_{\mu} \mathrm{d} x^{\mu}} \right) \,.
\end{equation}
The contour integral of the gauge field $A_{\mu}$ extends over a rectangular path ${\cal R}$ of sides $\Delta x$ along one spatial direction and $\Delta t$ in the temporal direction.
 In Eq.~(\ref{eq:ym:2}), $g$ is the coupling, $\mathrm{Tr}$ indicates the trace and the exponential is path-ordered along ${\cal R}$. 
 The potential is then obtained as 
\begin{equation}
\label{eq:ym:1}
V(\Delta x) = - \frac{1}{\Delta t}\ln \langle\frac{}{} W (\Delta x,\Delta t) \frac{}{} \rangle \,.
\end{equation}

At finite temperature, the temporal direction of size $\tau$ is compactified on a circle, 
and the resulting thermal field theory has temperature $T=1/\tau$. 
The order parameter for confinement can be identified with the expectation value of 
the {\em Polyakov loops}:
\begin{equation}
\Phi(\vec{x}) \equiv \mathrm{Tr} \left( \mathrm{P}e^{i g \oint_{\cal C} A_0(t,\vec{x}) \mathrm{d} t} \right) \,, 
\label{Eq:Polyakov}
\end{equation}
with ${\cal C}$ being the circle (of circumference $\tau$) at fixed spatial point $\vec{x}$.\footnote{
To avoid confusion with the average over spatial directions  $\Phi$ in Eq.~(\ref{Eq:Phi}),
when referring to Polyakov loops we explicitly indicate the $\vec{x}$-dependence,
being it understood that the average over the other space-like coordinates is taken.
For example, we will later indicate as
$\Phi(z)$ the average of $\Phi(\vec{x})$ over two space-like directions $x$ and $y$.}
The expectation value of this quantity vanishes in the confined phase. 
This observable has the advantage that it makes transparent the fact that 
the transition is associated with the breaking of the centre symmetry of the gauge group.
 In this respect, the $Sp(2N)$ theories play a useful complementary r\^ole with respect to $SU(N)$, the centre of the former being $\mathbb{Z}_2$ for every $N$,
 as opposed to the  $\mathbb{Z}_N$ centre  of the latter. 
 In this set-up, the propagation of a pair of static conjugated quarks is represented 
 by two oppositely-oriented Polyakov loops
  and their correlator $\langle \Phi^{\dag}(\vec{0}) \Phi(\vec{x}) \rangle$ probes 
    strings  attached to  two static lines at $\vec{0}$ and $\vec{x}$.
In the language of string theory, the confining string stretching between static sources is an {\em open string}. 

Yet, in Euclidean space 
we can reinterpret the zero-th direction as a compact spatial dimension and (for instance) the third direction as Euclidean time. 
From this point of view, the string is not attached to any static source but closes upon itself. 
For this reason, it can be also interpreted as a {\em closed string}. 
Choosing $\vec{x} = (0,0,z)$ and inserting a complete set of eigenstates $| l_n \rangle$ of the transfer matrix (the time-translation operator)
 in the third direction yields
\begin{equation}
\label{eq:ym:6}
\langle \Phi^{\dag}(\vec{0})  \Phi(\vec{x}) \rangle = \sum_n c^l_n e^{- E_n z} \ , \qquad c^l_n = | \langle 0 | \Phi^\dag(\vec{0}) |  l_n \rangle |^2  ,  
\end{equation}
with $c_n^l$ the overlap between the state $\Phi(\vec{0}) |0\rangle$ and the $n$-th eigenstate of the Hamiltonian along $z$ and $E_n$  the
corresponding energy eigenvalue. 
In this case the Polyakov loop correlator probes
(closed) string states wrapping along the compact direction, created at $\vec{0}$ and annihilated at $\vec{x}$. 
 
 The fact that the  same correlator can be interpreted in terms of either propagating closed or  open strings expresses the open-closed string duality,
a key observation that has profound physical implications. 
 Among them, the most direct and practically relevant for our study is the fact that the string tension can be extracted in the closed string channel from 
 correlators of Polyakov loops. 
 This is related to the fact that
  the topology of the world-sheet swept by the string is cylindrical.\footnote{In the case of zero temperature, where the relevant observable
   is the Wilson loop, the world-sheet has a disk topology.}
 
 If we instead consider the operator obtained by averaging $\Phi(\vec{x})$ along two dimensions
\begin{equation}
\Phi(z) = \frac{1}{N_s^2} \sum_{(x,y)} \phi(x,y,z) \,, 
\label{Eq:PolyakovAverage}
\end{equation}
where the sum runs over the two spatial coordinates in the directions orthogonal to $z$, for the correlator we obtain
\begin{equation}
\langle \Phi^{\dag}(0) \Phi(z)\rangle = \sum_n c_n^l e^{- m^l_n z} \,, 
\label{Eq:PolyakovCorr}
\end{equation}
and open-closed string duality implies that 
\begin{equation} 
\label{eq:ym:7}
\sigma = \lim_{\tau \to \infty} \frac{m^l_0}{\tau} \,. 
\end{equation}
The state corresponding to $m^l_0$ (where the subscript $l$ stands for {\em loop}) can be 
interpreted as the ground state mass of a {\em torelon}, a stringy (flux tube) state 
that wraps around the compact direction. 
In general, torelon states can be labelled by their length $\tau$, the absolute value of their angular and longitudinal momenta $J$ and $q$, 
their  (transverse) 
parity $P_t$ in a plane transverse to their symmetry axis, and their longitudinal parity $P_t$  along the wrapping direction. 
As for glueballs, the gauge group being pseudo-real, charge conjugation is always positive, and furthermore we are interested only in torelons 
with both transverse momenta equal to zero and both  positive parities.

The quantum fluctuations around the classical world-sheet solution corresponding to the area law in Eq.~(\ref{eq:arealaw}) generate a spectrum 
of modes for the flux tube that can be computed using an effective string theory description. 
The relevant degrees of freedom are identified as the $D-2$  Goldstone bosons 
 living in the $2$-dimensional world-sheet of the flux tube
that breaks  
the $D$-dimensional Poincar\'e group $ISO(D)$ according to:
\begin{equation}
	\label{eq:sym_pattern}
	ISO(D) \longrightarrow ISO(2)\times SO(D-2)\,.
\end{equation}
If the theory has a mass gap, as is the case for Yang-Mills theory, 
and no other degrees of freedom are present on the world-sheet, 
the most general effective action $S_\text{\tiny eff}[X]$ describing the dynamics  can  be expressed 
as an expansion in derivatives of $X^{\mu}=\{\xi^a, X^i\}$ with respect to the world-sheet parameters $(\xi^0, \xi^1)$,
\begin{multline}\label{Eq:effstring}
	S_\text{\tiny eff}[X] = \int_0^\tau d\xi^0 \int_0^l d\xi^1 \left[ \frac{}{}\sigma + ~C_0~(\partial_a X^i )^2~ +~ C_2~ (\partial_a X^i \partial_a X^i )^2~  +~  \right. \\ 
	\left. + ~C_3~( \partial_a X^i \partial_b X^j)^2~+~ 
	C_4~(\partial_c X^k)^2(\partial_a \partial_b X^i \partial^a \partial^b X_i ) +  \dots\right]
\end{multline}
where  $a,b,c=0,1$, while $i,j=2,\dots,D-1$ and summation over repeated indices is understood. 
This action can be naturally recast as an expansion in powers of $1/(\sigma \tau l)$, as a low momentum expansion around an infinitely long string.
This expansion is meaningful as long as $l\tau\gg1/\sigma$. In turn, 
a flux tube is string-like provided the  long-string expansion is valid ($l\gg \tau$),
and hence provided   $l\gg1/\sqrt{\sigma}$.

In lattice calculations, 
spacetime is a box of finite extent.  
When taking limits such as the r.h.s.~of Eq.~(\ref{eq:ym:7}), the extraction of $m^l_0$ is contaminated by
short-distance contributions that can be non-negligible. 
Let us specialise to the closed string channel, for which the world sheet 
is a cylinder  with time direction collinear to its axis.  
The mass can be systematically approximated as
\begin{equation}
	m_0^l = \sigma\tau \left( 1 + \sum_{k=1}^{\infty} \frac{d^l_k}{(\sqrt{\sigma} \tau)^k}\right)~.
\end{equation}
The dimensionless quantities $d^l_k$---related to the $C_k$ in 
Eq.~(\ref{Eq:effstring})---enclose all the short-distance effects 
that have been integrated out in this effective description. They are not completely independent: 
the request that $S_\text{\tiny eff}$ inherits the same space-time symmetry as the underlying Yang-Mills field theory 
imposes constraints (open/closed string duality being an example),  and only few of the $d^l_k$---equivalently, the 
$C_k$---are left as free parameters. This 
striking 
effective string-theory feature determines the  universal nature of the quantum 
corrections to the area law, to which we devote the
remainder of this subsection.

In one of the earliest works on the subject~\cite{Luscher:1980fr}, 
it was shown that the first non-null coefficient appears at ${\cal O}\left(\frac{1}{\sigma \tau^2 }\right)$, 
and has a universal nature:\footnote{An alternative description reaching 
similar conclusions and based on the long-distance restoration of conformal
 invariance has been proposed by Polchinski and Strominger in~\cite{Polchinski:1991ax}.}
\begin{equation}\label{eq:effs_LO}
	m^l_{0}(\tau)_{LO} = \sigma \tau  - c \frac{\pi (D-2) }{6 \tau} \,.
\end{equation}
The correction
 is known as the {\em L\"uscher term}~\cite{Luscher:1980ac}, and can be interpreted as the Casimir energy of the string massless modes. 
 The coefficient $c$ captures the central charge of the effective string theory, hence encoding 
 information on its nature. 
 For example, the purely bosonic string theory in non-critical dimension has $c=1$,
  and  there is evidence that it can be used to describe 
  $SU(N)$ Yang-Mills gauge theory
    (see e.g.~\cite{Lucini:2001nv,Necco:2001xg,Athenodorou:2010cs}).

The next-to-leading-order correction coming from the effective field theory has been studied in~\cite{Luscher:2004ib,Aharony:2009gg} 
(see also~\cite{Drummond:2004yp,HariDass:2006sd,Drummond:2006su,Dass:2006ud} for a related derivation in the Polchinski-Strominger approach). 
It arises at ${\cal O}(1/(\sigma \tau^2)^2)$. Duality arguments (in particular the annulus-annulus duality~\cite{Aharony:2009gg}) 
can be used to prove the universality of this correction. At this order, the effective mass appearing in Polyakov loop correlators is given by 
\begin{equation}\label{eq:effs_NLO}
	m_0^l(\tau)_{NLO}  = \sigma  \tau - \frac{\pi (D - 2)}{6 \tau} - \frac{1}{2}\left( \frac{\pi (D - 2)}{6}\right)^2 \frac{1}{\sigma \tau^3} \ . 
\end{equation}
Interestingly, the same result can be obtained by expanding in powers of $\frac{1}{\sigma \tau^2}$  the light-cone spectrum obtained from the Nambu-Goto action in non-critical dimensions ($D \neq 26$): 
\begin{equation}\label{eq:effs_NG}
	m_0^l(\tau)_{NG} = \sigma \tau \sqrt{1 - \frac{(D-2) \pi}{3\sigma \tau^2}}~.
\end{equation}

Going for a moment beyond the specific purposes of our paper, we recall that establishing the nature of the string that forms between a static quark-antiquark pair 
in a confining gauge theory is a very interesting programme in itself, as it can shed some crucial light on the nature and on the mechanism of colour confinement. 
The recent revival in its interest 
has resulted in new fundamental advances, among which the key observation that the constraints mentioned above 
are in fact particular cases of a more general viewpoint  
allowing to extend universality considerations to higher orders. 
Because the effective action is still  Poincar\'e invariant (despite spontaneous symmetry breaking), 
the difference between the number of derivatives and the number of fields (called  {\em weight}) is an invariant. The expansion of $S_\text{\tiny eff}$ can be organised according to the weight, and (non-linearly realised) Poincar\'e invariance imposed upon each of the terms. 
The  result is the emergence of recurrence relations  among  $d_k^l$ terms of the same weight. 
The unique weight-$0$ invariant action is precisely the Nambu-Goto action, with the leading correction appearing at weight-$4$. 
This explains, for example, why up to order $1/\tau^5$ and in $D=4$, the predictions of the light-cone spectrum for the ground state energy are fully universal. 
For a more detailed analysis, we refer the reader to~\cite{Aharony:2013ipa,Dubovsky:2015zey} and references therein.

Coming back to our lattice calculation, we build on the results available for $SU(N)$ and assume that the nature of the confining string
 is reproduced also in $Sp(4)$. We  use (rather than derive) the expressions for the bosonic string in order to extract 
 numerical values for the string tension $\sigma$. Subject to the validity of such working assumption 
 this enables us to remove large finite-size corrections from the extraction of the string tension.

\subsection{A variational approach to the mass spectrum}

In this subsection, we outline the lattice methods used to extract both $\sigma$ and the spectrum of glueball masses~\cite{Berg:1982kp,Morningstar:1997ff,Morningstar:1999rf}. 
It is important to note from the outset that because of centre symmetry, 
glueball and flux-tube states do not mix in the confined phase. 
Glueballs and flux tubes are sourced by products of link matrices around contractible 
and non-contractible loops $\mathcal{C}$, respectively, with their geometrical symmetry properties 
determining $J^{P}$ for the former and size $L$, momentum $q$, 
angular momentum $J$ and parities $P_l$ and $P_t$ for the latter. 
We limit our study of flux tubes to the state with $J=q=0$ and $P_t=P_l=+$.
We refer to~\cite{Athenodorou:2010cs} for a detailed analysis of creation of excited flux tubes in various channels.

The isotropic lattice breaks the continuum rotational group to the octahedral group 
(the 24-elements group of the symmetries of the cube). 
At finite lattice spacing, glueball states are  
classified by the conventional names $R=A_1,A_2,T_1,T_2,E$ of the irreducible representations of the octahedral group,
so that we label the glueballs as $R^{P}$, instead of using the continuum $J^{P}$. 

The irreducible representations of the octahedral group can be decomposed into irreducible representations of the continuum rotational group. 
Since the octahedral group is finite, different continuum spins are associated with a given octahedral irreducible representation. 
For instance, 
the continuum $J=0$ spectrum is found in the $A_1$ representation, which also contains (among others) $J=4$ states. 
While on physical grounds one can assume that the lightest $A_1$ state corresponds to a $J=0$ 
glueball (at least when $P = +$), 
distinguishing different continuum channels in the excited spectrum measured on the lattice is not an easy task. 
Guidance is provided by the degeneracies that are expected in different octahedral representations, 
where different polarisations of the same state can appear. This is for instance the case for the continuum $J=2$ states, 
 two polarisations of which are to be found in the octahedral $E$ representation, and the other three in  $T_1$. 
 Hence, close to the continuum limit, states that are degenerate in the $E$ and $T_1$ channel can be interpreted 
 as would-be continuum states with $J=2$.

Given a lattice path $\cal C$ with given shape and size, located at reference 
coordinates $(t,\vec{x})$ on the lattice,
that transforms in an irreducible representation $R$ of the octahedral group and is an eigenstate of parity, 
the lowest-lying mass in the $R^{P}$ channel can be extracted from the asymptotic behaviour of the correlator 
\begin{equation}\label{eq:glue1}
	\Gamma_{\mathcal{C}}(t) = \frac{\langle O_{\mathcal{C}}^{\dag}(0)O_{\mathcal{C}}(t) \rangle}{\langle O_{\mathcal{C}}^{\dag}(0) O_{\mathcal{C}}(0) \rangle} 
	= \frac{\sum_i \left| \langle i | O_{\mathcal{C}}(0) |  0  \rangle\right|^2 e^{-m_i t}}{\sum_i \left| \langle i | O_{\mathcal{C}}| 0 \rangle\right|^2} \,,
\end{equation}
where $m_i$ is the mass of state $|i\rangle$ and $O_{\cal C}(t)$ 
is the zero momentum operator:\footnote{If ${\cal C}$ is a 
circle in the time direction, then $O_{\cal C} =\Phi$ as defined in Eq.~(\ref{Eq:Phi}).}
\begin{equation}
O_{\mathcal{C}}(t) = \frac{1}{N_s^3}\sum_{\vec{x}} \Tr \left( \prod_{\mathcal{C}} U_l \right) \ .
\label{Eq:op}
\end{equation}

The decay of each exponential appearing in the spectral decomposition is controlled by the squared normalised amplitudes
\begin{equation}
\left| c_j \right|^2 = \frac{\left| \langle j | O_{\mathcal{C}}(0) |  0  \rangle\right|^2}{\sum_i \left| \langle i | O_{\mathcal{C}}| 0 \rangle\right|^2} \,.
\end{equation}
In practice, since the statistical noise is expected to be constant with $t$, the signal-to-noise ratio decays exponentially, 
eventually defying attempts to isolate the ground state. 
Hence, for a generic choice of $\cal C$,
the mass that is extracted at large but finite $t$ suffers from contamination from excited states. 
We notice that, as a consequence of unitarity, this results in an overestimate of the mass.

To improve accuracy in  the extraction of $m_i$, in principle one should choose the operators $O$ to maximise the overlap of $O_{\cal C}(t)|l\rangle$ 
with the desired state $|l\rangle$. While such an operator is not known a priori, 
we can operationally construct a good approximation by performing a variational calculation involving loops $\cal C$ of various shapes and sizes. 
The size $\lambda$ of the loop $C$ must be chosen appropriately: in order for it to capture the infrared physics, 
it should have a size of the order of the confinement scale. This means that in practice the size of $C$ in lattice units must grow as the lattice spacing goes to zero. 
Over the years, various methods to circumvent these potential problems have been suggested. 
In this work, we shall use a variational calculation involving an operator basis constructed with a combination of smearing and blocking operations. 
We briefly review the method used, and we refer to~\cite{Lucini:2004my} for more details, before
presenting our results in Sec.~\ref{sec:glueresults}.

Given a set of $N$ operators $O_i$, defined as in Eq.~(\ref{Eq:op}) for paths ${\cal C}_i$ of different shape and sizes labelled by $i$,
we compute the $N\times N$, normalised correlation matrix
\begin{equation}
	C_{ij}(t) = \frac{ \langle 0|O^\dag_i(0) O_j(t)|0 \rangle}{\langle0| O^\dag_i(0) O_j(0)|0 \rangle}\,.
\end{equation}
Assuming maximal rank, $C_{ij}$ can be diagonalised, and we call $\tilde{C}_{ii}$ the $N$ resulting functions of $t$. 
 The special linear combination
	$\sum_i \alpha_i O_i(t)$,
corresponding to the maximal eigenvalue, 
 has the maximal overlap with the ground state in the given symmetry channel. 
Assuming its mass is the only one present in the given channel, we obtain it by fitting the data with the function
\begin{equation}\label{eq:fitglue}
	\tilde{C}_{ii}(t) = \left| c_i \right|^2 \cosh{\left(m_i t -\frac{N_t}{2}\right)}\,,
\end{equation}
where $|c_i|^2$ and $m_i$ are the fit parameters, and where the appearance of the $\cosh$
 is due to the inclusion of backward propagating particles through the periodic boundary.
In general the data is contaminated by contributions from states with higher mass.  
Hence we must restrict the fit to the range for which we see the appearance of 
a plateau in  the quantity
\begin{equation}\label{eq:glueplat}
m_{eff}(t) = {\rm arcosh} \left( \frac{\tilde{C}_{ii}(t + 1) + \tilde{C}_{ii}(t-1)}{2 \tilde{C}_{ii}(t)}  \right)\,.
\end{equation}

In order to include operators which extend beyond the ultraviolet scale, following~\cite{Lucini:2004my}, 
we subjected the lattice links to a combination of smearing and blocking transformations. 
These are iterative procedures similar to block transformations in statistical mechanics, except that we restrict them to space-like links. 
The operators $O_i$ defined as in Eq.~(\ref{Eq:op}) for paths ${\cal C}_i$ are evaluated using the output links from each iterative smearing and blocking step.
After $S$ iterations, one has a collection of $S \times N$ such operators, where $N$ is the number of basis paths in the given channel.  
We chose to build the operators $O_i$ by  starting from a large set of basic lattice paths. 
In this set, we include all the closed paths with length $\lambda$ up to ten in units of the lattice spacing, 
appropriately symmetrised to transform according to an irreducible representation of the octahedral group and to have definite parity.

We implement the process of smearing along the directions orthogonal to the direction of propagation, by starting from the link $U^{s=0}_i(x) \equiv U_i(x)$
and iteratively defining $\tilde{U}^{s>0}_i(x)$ as follows
\beqs
	\tilde{U}^{s+1}_i(x) &=& U^s_i(x) + p_a \sum_{j\neq i} U^s_j(x) U^s_i(x+\hat{\jmath}) U^{s\dag}_j(x+\hat{\imath})+\\
	&&+ p_a \sum_{j\neq i} U^{s\dag}_j(x-\hat{\jmath}) U^{s}_i(x-\hat{\jmath}) U^{s}_j(x-\hat{\jmath}+\hat{\imath})\,,\nonumber
\eeqs
where $\hat{\jmath}$ and $\hat{\imath}$ refer to the unit-length displacements along the lattice  directions $j$ and $i$, respectively,
while the positive parameter $p_a$ controls how much smearing is taking place at each step.  
The smeared matrices $\tilde{U}^{s>0}_{i}(x)$ 
are not  elements of the gauge group. We project those matrices to the target group by finding the $Sp(4)$ matrix 
$U^{s>0}_{i}(x)$ that maximizes ${\rm Re}\, \Tr\, \tilde{U}^{s\dag}_i(x) U^{s}_i(x)$. 
This is done in two steps: a crude projection is operated by using one of the re-symplectization algorithms presented in 
Appendix~\ref{Sec:projection}, and afterwards a number  of cooling steps~\cite{Hoek:1986nd} ($15$ in our case) is performed on the link.

Similarly, blocking is implemented  by replacing simple links $U^{b=0}_i(x) \equiv U_i(x)$ with super-links $\tilde{U}^{b>0}_i(x)$ that join lattice 
sites $2^b$ spacings apart, where $b$ is the number of blocking iterations, as described by
\beqs
	\tilde{U}^{b+1}_i(x) &=& U^b_i(x)U^b_i(x+2^b\hat{\imath}) +\\
	&&+ p_a \sum_{j\neq i} U^b_j(x) U^b_i(x+2^b\hat{\jmath}) U^b_i(x+2^b\hat{\jmath}+2^b\hat{\imath}) U^{b\,\dag}_j(x+2^b\hat{\imath})+\nonumber\\
	&&+ p_a \sum_{j\neq i} U^{b\,\dag}_j(x-2^b\hat{\jmath}) U^{s}_i(x-2^b\hat{\jmath})U^{s}_i(x-2^b\hat{\jmath}+2^b\hat{\imath}) 
	U^{b}_j(x-2^b\hat{\jmath}+2^b\hat{\imath})\,.\nonumber
	\eeqs
	Again, each such step yields a matrix $\tilde{U}^{b+1}_i(x)$  that does not belong to the $Sp(4)$ group, and hence must be projected
	onto ${U}^{b+1}_i(x)$ within the group in same way as for the smearing.
In practical terms, when performing numerical lattice studies blocking allows to reach the physical size of the glueball 
in fewer steps, while at the physical scale smearing provides a better resolution.
 Due to the fact that the identification of the physical scale is a dynamic problem, an iterative combination of $n$ 
 smearing steps (generally, $n = 1,2$) with a blocking step generally proves to be an efficient strategy~\cite{Lucini:2010nv}.

\subsection{Lattice results}\label{sec:glueresults}

In this Subsection, we report the results of our numerical analyses of the glueball spectrum 
and the string tension of the pure $Sp(4)$ Yang-Mills theory. The calculations have been performed 
on fully isotropic lattices of various sizes and lattice spacings. To investigate the finite size effects, 
we first consider the coarsest lattice with $\beta=7.7$. {   Based on the estimate of the critical 
coupling of the bulk phase transition~\cite{Holland:2003kg}, the choice of
this value should provide a prudent yet reasonable compromise between the practical
necessity of performing a detailed calculation at a lattice coupling
at which the physically large volumes can be reached on a
moderately coarse grid and the physically paramount request that the lattice gauge theory
be in the confining phase connected to the continuum theory as $a \to
0$. Indeed, we have found evidence in our calculations that at $\beta = 7.7$ the lattice theory
is in the physically relevant confined phase.

We started with this $\beta=7.7$ value, and increased the
lattice size, starting from  $L=10a$, until we found the best economical choice at which
the (exponentially suppressed) finite-size effects became much smaller than
the statistical errors. Assuming scaling towards the continuum limit,
this analysis provides a lower bound for the physical 
volume of the system such that finite-size effects are negligible with respect to the
statistical errors, and hence ensures that the calculations 
cannot be distinguished from infinite volume ones.

We repeated the same set of  measurements on progressively
finer lattices (larger $\beta$), always making sure that  the physical volume 
were large enough for the calculation to be considered at infinite volume for all
practical purposes, and  extrapolated the glueball
spectrum to the continuum limit. 

The whole procedure is illustrated
more quantitatively in the following two sub-subsections. Our parameter
choices for the continuum extrapolation are  reported in the first two
columns of Table~\ref{tab:sigmas}. }For each lattice setup 10000
configurations were generated,  and a binned and bootstrapped analysis
of errors was carried out to take care of temporal autocorrelations.
Operators blocked to the level $N_b\leq L$ and with $15$ cooling steps
were used, resulting  in a variational basis of $\sim 200$ operators.

\begin{table}
	\centering
	\begin{tabular}{|c|c|c|c|c|}
\hline \hline
$L/a$ & $\beta$ & $\sqrt{\sigma} a$\\
\hline
$32$ & $8.3$ & $0.1156(3)$ \\
$26$ & $8.2$ & $0.1293(6)$ \\
$20$ & $8.0$ & $0.1563(6)$ \\
$18$ & $7.85$ & $0.1885(7)$\\
$16$ & $7.7$ & $0.227(1)$\\
\hline \hline	
\end{tabular}
\caption{The final estimates for $\sqrt{\sigma}$ at different lattice setups ($L$ and $\beta$),
as discussed in Sec.~\ref{sec:glueresults}.}
\label{tab:sigmas}
\end{table}

\subsubsection{The string tension}

To extract the string tension from measurements of masses of closed flux tubes, we turn to effective string theory. A finite overlap with
flux-tube states can be obtained with lattice operators defined on
non-contractible loops, as described earlier. We produce two
measurements of the string tension, that we denote as $\sigma_t$ and
$\sigma_s$. {   The former is obtained as follows. We first consider loops that wrap the
time-like direction as in Eq.~(\ref{Eq:Polyakov}) and  average them along two
space-like dimensions as in Eq.~(\ref{Eq:PolyakovAverage}). We then  
compute the correlators as in Eq.~(\ref{Eq:PolyakovCorr}), with an
additional statistics-enhancing average over interchanges of $(x,y,z)$, to extract
the lowest mass $m_0^l$ (which in this case we refer to as $m_t$). Finally,
we determine the string tension as in Eq.~(\ref{eq:ym:7}) in
three different ways: by using Eq.~(\ref{eq:effs_LO})  (LO), 
Eq.~(\ref{eq:effs_NLO}) (NLO) or Eq.~(\ref{eq:effs_NG}) (NG). 

A similar procedure is performed for obtaining $\sigma_s$, which is extracted from
the mass $m_s$ associated with correlators in time of Polyakov loops
winding one of the spatial directions---except that in this case there is no average over
interchange of equivalent directions. Because the lattice used is isotropic, we expect $\sigma_t$ to
be compatible with $\sigma_s$, since the latter could be obtained from the
former by interchanging the roles of the time direction and one of the space
directions used for defining the correlation functions associated with $m_t$. }
Notice that because of the averaging over interchanges of spatial directions, the statistical error on
$\sigma_t$ is reduced in respect  to $\sigma_s$.  

\begin{table}
	\centering
	\begin{tabular}{|c|c|c|c|c|c|}
\hline \hline
$L/a$ & $a m_t$  & $a m_s$ & $\sqrt{\sigma_t} a$(NG) & $\sqrt{\sigma_s} a$(NG) & $\sqrt{\sigma} a$\\
\hline
$ 24 $ &  $1.218(13)$ & $1.157(30)$ & $ 0.2294(12) $ & $ 0.2237(28) $ & $ 0.2285(11) $ \\
$ 20 $ &  $0.981(11)$ & $0.995(18)$ & $ 0.2275(12) $ & $ 0.2290(20) $ & $ 0.228(10) $ \\
$ 16 $ &  $0.7570(93)$ & $0.762(19)$ & $ 0.2271(13) $ & $ 0.2278(26) $ & $ 0.2272(11) $ \\
$ 14 $ &  $0.6436(73)$ & $0.647(15)$ & $ 0.2272(11) $ & $ 0.2277(23) $ & $ 0.2273(10) $ \\
$ 12 $ &  $0.5196(53)$ & $0.534(11)$ &$ 0.22623(96) $ & $ 0.2288(20) $ & $ 0.22673(86) $ \\
$ 10 $ &  $0.3744(32)$ & $0.3804(69)$  & $ 0.22215(69) $ & $ 0.2234(15) $ & $ 0.22238(63) $ \\
\hline	\hline
\end{tabular}
\caption{Masses obtained from Polyakov loop correlators winding in
  the time direction ($m_t$) and in a spatial direction ($m_s$),
  together with the corresponding string tensions $\sigma_t$ 
and $\sigma_s$  extracted from the Nambu-Goto (NG) prediction for the
ground state energy of the flux tube of length $L$ at $\beta=7.7$.  In
the last column, we report the result of the weighted average in
Eq.~(\ref{eq:sigma_avr}). } 
\label{tab:sigmas_7.7}
\end{table}

\begin{figure}[htb!]
	\includegraphics{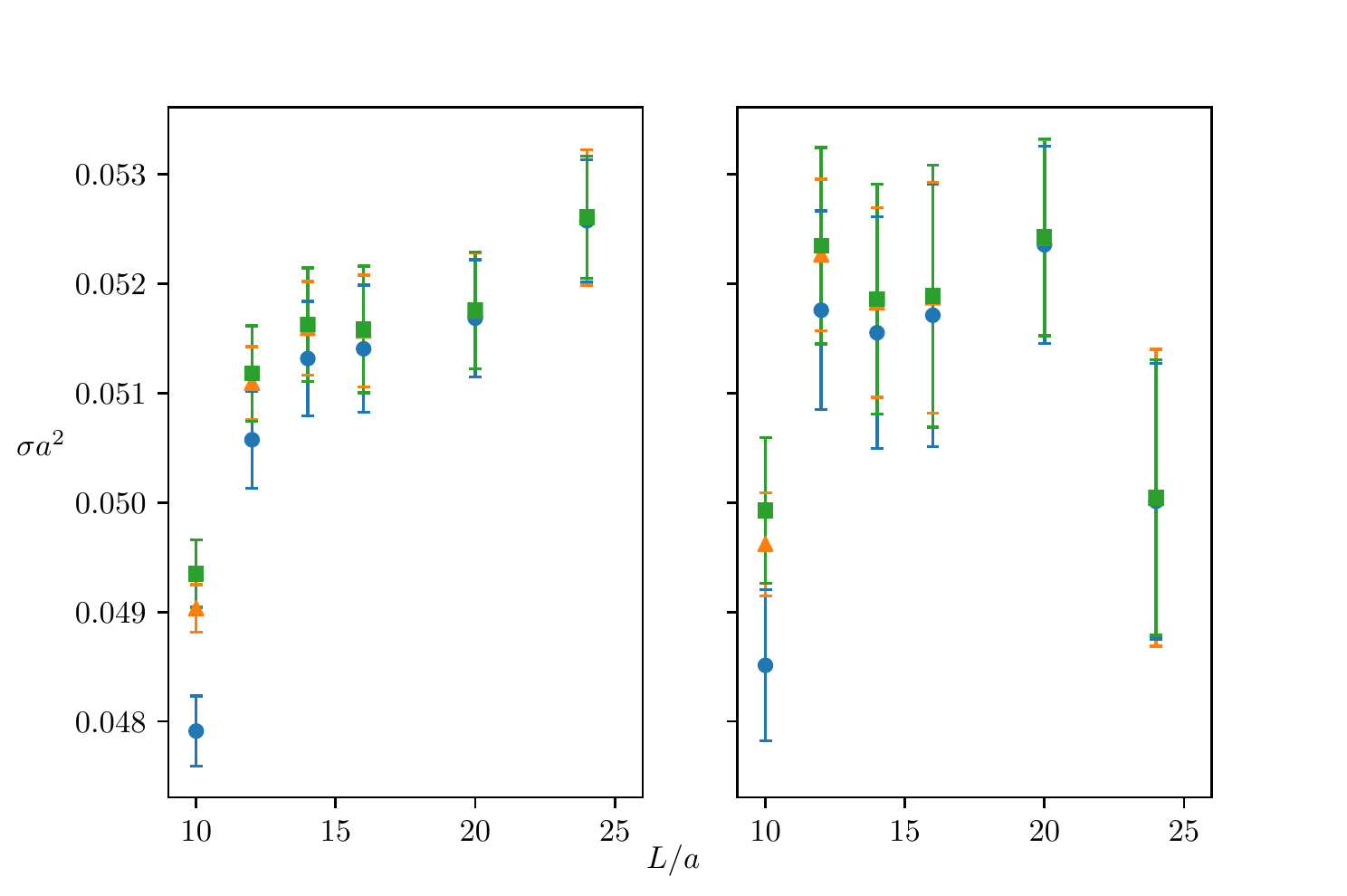}
\caption{
Values of the time-like $\sigma_t a^2$ (left panel) and space-like
$\sigma_s a^2$ (right panel) string tensions extracted at fixed
lattice spacing ($\beta=7.7$) and varying lattice size $L/a$,  using
the LO($\bullet$), NLO($\blacktriangle$) and NG($\blacksquare$)
expressions for the ground state energies from
Eqs.~(\ref{eq:effs_LO}), (\ref{eq:effs_NLO}) and~(\ref{eq:effs_NG}),
respectively.} 
\label{fig:sigma_vs_L_7.7}
\end{figure}

In order to carefully assess finite size effects, we show the results
of the analysis for $\beta=7.7$ in Table~\ref{tab:sigmas_7.7}. {    We
perform a best fit analysis of the data for $m_t$ and $m_s$ by using
Eqs.~(\ref{eq:effs_LO})-(\ref{eq:effs_NG}). We  start from the largest
flux length  $L=24 a$ and gradually add in the fit lower-length values,
until the value of the $\chi^2/{\rm d.o.f}$ becomes larger than a fixed
threshold that we conventionally set at 1.2. We find  
the following best fit results for $\sigma_t$, obtained with the largest possible range for which $\chi^2/{\rm d.o.f}<1.2$:
\begin{equation}
\begin{cases}
\sigma_t(LO) a^2 = 0.05174(29) , \qquad  \ \ 14 \le L/a \le 24 \ ,
\ \chi^2/{\rm d.o.f} \simeq 1.12 \ , \\
\sigma_t(NLO) a^2 = 0.05166(25) , \qquad 12 \le L/a \le 24 \ ,
\ \chi^2/{\rm d.o.f} \simeq 1.19 \ , \\
\sigma_t(NG) a^2 = 0.05169(24), \qquad \ \ 12 \le L/ a \le 24 \ ,
\ \chi^2/{\rm d.o.f} \simeq 1.07 \,.
\end{cases}
\end{equation}
The analogous process yields for $\sigma_s$ the following:
\begin{equation}
\begin{cases}
\sigma_s(LO) a^2 =  0.05164(36), \qquad  \ \ 12 \le L/a \le 24 \ ,
\ \chi^2/{\rm d.o.f} \simeq 0.59 \ , \\
\sigma_s(NLO) a^2 = 0.05187(38), \qquad 12 \le L/a \le 24 \ ,
\ \chi^2/{\rm d.o.f} \simeq 1.19\ , \\
\sigma_s(NG) a^2 = 0.05190(39), \qquad \ \ 12 \le L/a \le 24 \ ,
\ \chi^2/{\rm d.o.f} \simeq 0.70 \ .
\end{cases}
\end{equation}
For both $m_t$ and $m_s$, our requirement for the acceptability of the fit is verified down to
$L = 12 a$ for all the three proposed functional forms, except for the
leading-order ansatz in the case of $\sigma_t$, which requires $L =14 a$. 
We regard this last case as a warning that at $L = 12 a$ the effective string
description at leading order might break, although the results for the
NLO and NG descriptions give us confidence that higher orders cure
this problem. 

The fits provide very good indications that the description we adopted is robust for $L /a \ge 14$. 
Indeed, all the reported fitted values are compatible, regardless of the ansatz used and of the
direction of correlation of the Polyakov loops from which we extract
the relevant mass. Conversely, when we try to extend the fit down to $L = 10 a$, we
typically find a significantly larger $\chi^2/{\rm d.o.f}$,  of order 3-10, indicating
that the effective string description cannot be trusted between $L
= 10 a$ and $L = 12 a$. The only exception is the NG description
of $\sigma_s$, for which we get $\chi^2 /{\rm d.o.f} \simeq 1.75$. While
it would be tempting to interpret this result as evidence that the
NG ansatz provides a better description of the data, in the absence of
confirmation of this hypothesis in the $\sigma_t$ case (for which an extension
down to $L = 10 a$ leads to $\chi^2/{\rm d.o.f} \simeq 8.22$) and given also the
scope of our calculation, we prefer to take a cautious attitude
towards our results and assume that a safe lower bound for all effective
string models analysed to work (and to produce compatible results) is
$L = 14 a$. 

Given all of these considerations, and taking into account all our estimates of $a^2 \sigma$, a
safe infinite-volume value for the latter quantity that encompasses the spread of the
fits is $\sigma a^2= 0.5179(50)$, which translates into $\sqrt{\sigma}
a= 0.2276(11)$. Using this result, in physical units one 
finds that $L \sqrt{\sigma} = 2.731(13)$ for $L = 12 a$ and  $L
\sqrt{\sigma} = 3.186(15)$ for $L = 14 a$. The fact that effective
string description works remarkably well for Polyakov loop correlator masses 
down to at least $L = 14a$ is consistent with the picture of confinement through
the formation of thin flux tubes.

It is of  practical relevance for numerical studies to assess how well the
infinite-volume value of the string tension is represented by the result
extracted inverting Eqns.~(\ref{eq:effs_LO}-\ref{eq:effs_NG}) at a
single finite size $L$, and how this would be affected by varying $L$. 
To provide information about this, we report the results of our procedure in
Table~\ref{tab:sigmas_7.7} and in Fig.~\ref{fig:sigma_vs_L_7.7}. As we
see from the figure, the value of $\sigma a^2$ is constant for a wide range of $L$. This
holds for the LO, NLO and the NG extractions, with the corresponding
values being always compatible within errors. Based on these 
observations, we use the NG approximation to extract our best estimate. We detect finite-size effects for the smallest
lattice volumes $L = 10 a$. Though we do not present the detailed
results,  we also detect  a discrepancy at $L=24a$ between the
space-like and time-like string tensions. This discrepancy may be a
consequence of the systematic error coming from the difficulty in
extracting the asymptotical behaviour of the correlator for very large masses. }

Our final estimate for the value of $\sqrt{\sigma}$ as a function of $L$ is obtained from the first two columns of Table~\ref{tab:sigmas_7.7} by computing the weighted average
\beqs
	\sqrt{\sigma} &=& \frac{ \frac{\sqrt{\sigma_t}}{(\Delta \sqrt{\sigma_t})^2}+ \frac{\sqrt{\sigma_s}}{(\Delta \sqrt{\sigma_s})^2}}{\left(\frac{1}{\Delta \sqrt{\sigma} }\right)^2} \,,
	\label{eq:sigma_avr}
	\eeqs
	where the error is given by
	\beqs
	\nonumber
	\Delta \sqrt{\sigma} &= \frac{1 }{\sqrt{\frac{1}{(\Delta \sqrt{\sigma_t})^2}+ \frac{1}{(\Delta \sqrt{\sigma_s})^2}}}\,.
\eeqs
The resulting values are reported in the last column. From the behaviour of $\sqrt{\sigma}$ we conclude 
that finite size effects are certainly smaller than the statistical
error for $L \ge 14a$, and we take as a final estimate of
$\sqrt{\sigma}$ at this coupling  the value at $L=16a$. {   We also note
that compatible results are obtained for $L = 12 a$.}

{   Assuming scaling towards the continuum, from our finite-size study
we obtain firm evidence that all lattices for which $L \sqrt{\sigma} \ge 3$ provide an estimate of the
string tension that is compatible within the statistical errors with
the infinite-volume value. Hence, we conclude that finite-volume effects are negligible 
once $L \sqrt{\sigma} \gsim 3$. In particular,  we have verified that the condition $L \sqrt{\sigma} \gsim 3$ is safely
fulfilled  when carrying out calculations on lattice ensembles  with larger
$\beta$,  starting from  the finite-size analysis at $\beta =
7.7$. } Table~\ref{tab:sigmas} reports the lattice parameters of
the calculations we have used to extrapolate  to the continuum limit
and the corresponding results for $\sqrt{\sigma}$.

\subsubsection{The mass spectrum of glueballs}
\label{Sec:glue}

{   As for the string tension, we began our analysis of glueball masses
with a study of finite-size effects for lattice coupling $\beta=7.7$.  
We aimed at estimating finite-volume
effects as a function of the lattice size $L$, and bound $L$ such that
 the systematic error due to the finite size be negligible with
respect to statistical error on the measurement of the masses. 

Our results for the mass spectrum at $\beta=7.7$ for various volumes are
reported in the rows of Table~\ref{tab:gluevol}.  
While this particular choice of $\beta$ is suitable for finite-size
studies, as it allows us to reach large lattices in physical units with a
relatively small computational effort, the coarseness of the lattice
spacing pushes most of the masses above the lattice cut-off, making their extraction
numerically challenging. For this reason, we observe
a systematic effect related to the isolation of the ground state on
all irreducible representations other than the lowest-lying one.
While in our tables we quote only the statistical
error, for higher excitations the systematic error coming from the ground state
isolation in a given channel is expected to have a comparable size.

Another systematic effect that affects our calculation
of the glueball masses is contamination of the spectrum by
multi-glueball scattering states and torelon states (the latter being
lattice artefacts associated with state propagation of pairs of
oppositely directed Polyakov loops). Separating the physical spectrum
from those unwanted states requires a more demanding
calculation\footnote{See e.g.~\cite{Lucini:2010nv} and references
  therein for a discussion of a calculation performed along those lines.} that
goes beyond the scope of this first exploration of the glueball
masses in $Sp(4)$. 
What results is a hard-to-control error related to mixing with
spurious states, which is strongly dependent on the volume and manifests itself
in occasional sudden jumps and irregularities of the extracted
masses. Indeed this is visible for some of the most massive states we report
in Table~\ref{tab:gluevol}. 
In view of all these considerations, and  to focus the discussion 
to the main purposes of this paper, we
limit the analysis of finite-size effects at $\beta = 7.7$ to the
ground state (found in the $A1^+$ channel) and to the would-be
continuum $2^+$ glueball (expected to appear in the $E^+$ and in the
$T1^+$ channels). }

\begin{table}
	\centering
	\small
\begin{tabular}[!htb]{ |c||c|c|c|c|c|c| }
\hline \hline
$R^P$ & $L/a=10$ & $L/a=12$ & $L/a=14$ & $L/a=16$ & $L/a=20$ & $L/a=24$ \\
\hline
$ A1^+ $ &   $0.569(13)$ &  $0.728(15)$ &  $0.738(16)$ &  $0.742(16)$ &  $0.764(15)$ &  $0.739(11)$ \\ 
$ A1^- $ &   $1.039(35)$ &  $1.275(41)$ &  $1.406(47)$ &  $1.210(40)$ &  $1.300(34)$ &  $1.323(34)$ \\ 
$ A2^+ $ &   $1.70(11)$ &  $1.706(95)$ &  $1.778(95)$ &  $1.650(76)$ &  $1.771(81)$ &  $1.748(74)$ \\ 
$ A2^- $ &   $2.48(34)$ &  $1.83(17)$ &  $1.74(14)$ &  $2.21(25)$ &  $2.23(24)$ &  $2.252(22)$ \\ 
$ E^+ $ &   $0.623(13)$ &  $1.111(32)$ &  $1.150(32)$ &  $1.159(27)$ &  $1.217(26)$ &  $1.036(59)$ \\ 
$ E^- $ &   $1.402(62)$ &  $1.347(58)$ &  $1.401(48)$ &  $1.509(66)$ &  $1.597(59)$ &  $1.463(45)$ \\ 
$ T1^+ $ &   $1.170(43)$ &  $1.220(36)$ &  $1.202(39)$ &  $1.209(31)$ &  $1.173(26)$ &  $1.82(11)$ \\ 
$ T1^- $ &   $1.465(75)$ &  $1.513(69)$ &  $1.515(66)$ &  $1.522(57)$ &  $1.499(51)$ &  $1.87(12)$ \\ 
$ T2^+ $ &   $1.53(11)$ &  $1.70(12)$ &  $1.99(14)$ &  $1.578(94)$ &  $1.839(96)$ &  $1.179(95)$ \\ 
$ T2^- $ &   $1.60(12)$ &  $2.04(18)$ &  $2.38(27)$ &  $2.07(18)$ &  $1.94(15)$ &  $1.505(46)$ \\ 

\hline \hline
\end{tabular}
\caption{Glueball masses  obtained at coupling $\beta=7.7$, with operators blocked at level 
$N_b\leq N_L$ and with $15$ cooling steps. The quantum number $R^P$
refer to the octahedral group as explained in
Sec.~\ref{Sec:glueandstring}. The calculations are repeated for
various values of the lattice size $L/a$.} 
\label{tab:gluevol}
\end{table}

\begin{figure}[htb!]\begin{center}
	\includegraphics{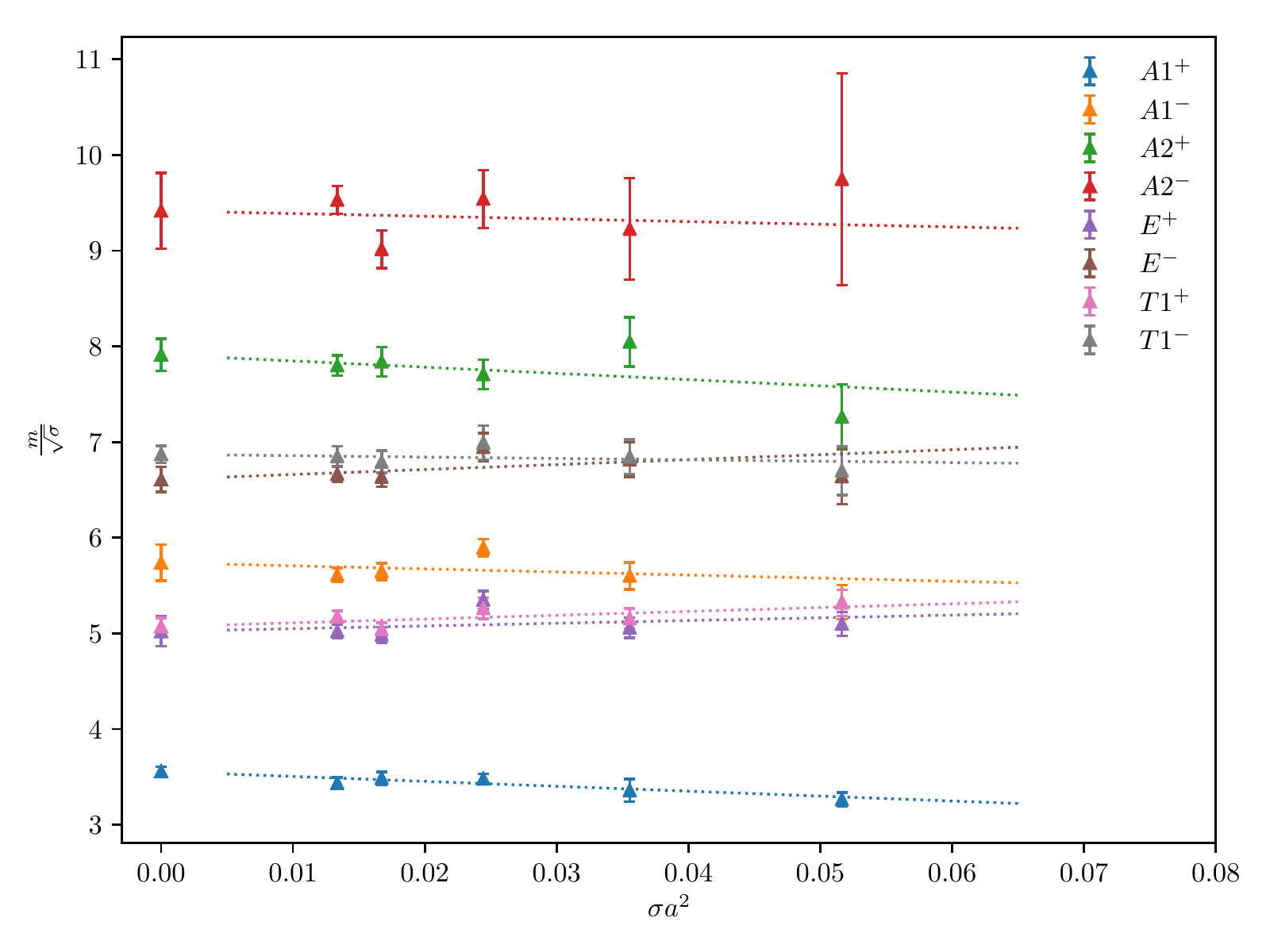}
\caption{Continuum limit extrapolations of the glueball masses $m/\sqrt{\sigma}$, in units of the string tension,  in $Sp(4)$ Yang-Mills,
as described in Sec.~\ref{Sec:glue}. States are labelled by the quantum numbers of the octahedral group, but notice the emergence of degeneracies
in the continuum limit,  in the case of $T_1^+$ and $E^+$, as well as $T_1^-$ and $E^-$.}\label{fig:contlim1}
\end{center}
\end{figure}

\begin{figure}[htb!]\begin{center}
	\includegraphics{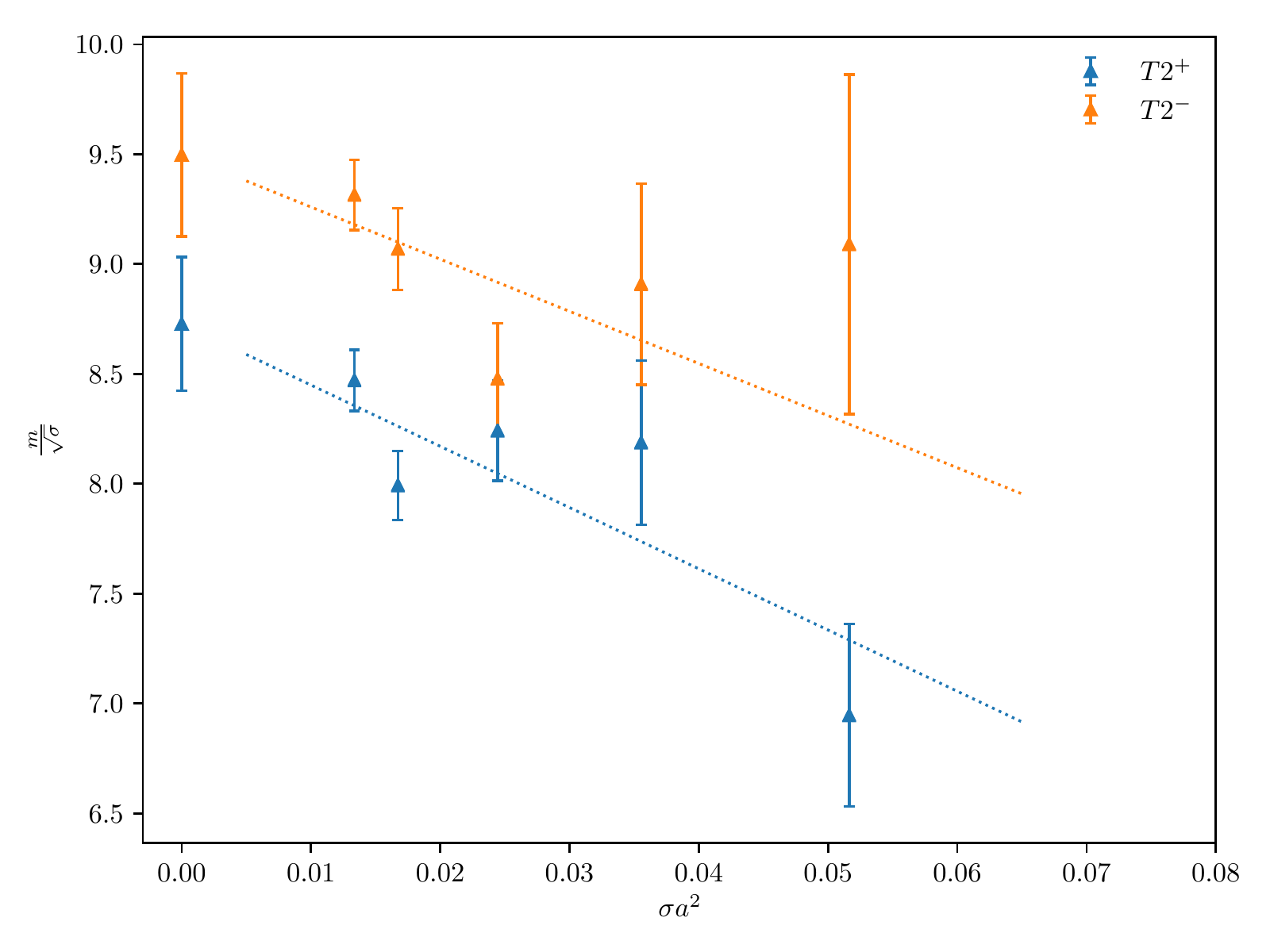}
\caption{Continuum limit extrapolations of the glueball masses $m/\sqrt{\sigma}$, in units of the string tension,  in $Sp(4)$ Yang-Mills,
as described in Sec.~\ref{Sec:glue}. States are labelled by the quantum numbers of the octahedral group.}\label{fig:contlim2}
\end{center}
\end{figure}

\begin{figure}[htb!]\begin{center}
	\includegraphics{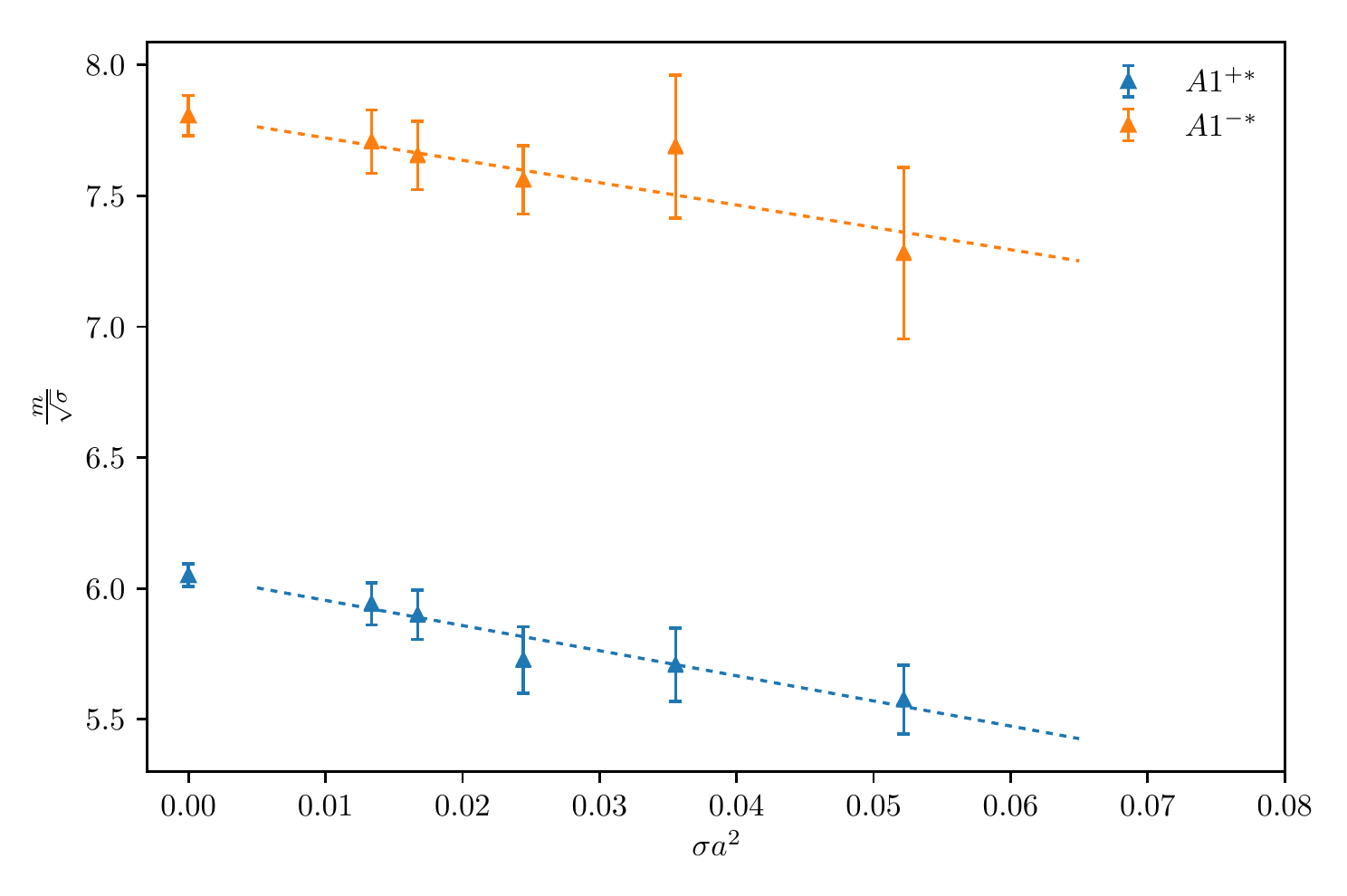}
\caption{Continuum limit extrapolations of the glueball masses for
  selected excited states $m/\sqrt{\sigma}$, in units of the string
  tension,  in $Sp(4)$ Yang-Mills, as described in
  Sec.~\ref{Sec:glue}. States are labelled by the quantum numbers of
  the octahedral group.}\label{fig:contlim3}  
\end{center}
\end{figure}

\begin{table}
\centering
\begin{tabular}{|c|c|}
\hline \hline
$R^P$ & $m(R^P)/\sqrt{\sigma}$\\
\hline
$ A1^+ $ & $ 3.557(52) $ \\
$ A1^{+*} $ & $ 6.05(4) $ \\
$ A1^- $ & $ 5.74(19) $ \\
$ A1^{-*} $ & $ 7.81(8) $ \\
$ A2^+ $ & $ 7.91(17) $ \\
$ A2^- $ & $ 9.42(40) $ \\
$ E^+ $ & $ 5.02(16) $ \\
$ E^- $ & $ 6.61(13) $ \\
$ T1^+ $ & $ 5.070(91) $ \\
$ T1^- $ & $ 6.872(89) $ \\
$ T2^+ $ & $ 8.73(30) $ \\
$ T2^- $ & $ 9.50(37) $ \\

\hline\hline
\end{tabular}
\caption{Continuum  extrapolated  glueball masses in all the symmetry channels.
States are labelled by the quantum numbers of the octahedral group.}
\label{tab:contlim}
\end{table}

{   Starting with the $A1^+$ state, we fit the behaviour of the extracted
mass to the size dependency parameterised as follows:
\begin{equation}
m(L) = m \left( 1 + \frac{b_1 e^{- b_2 m L}}{m L} \right) \,, 
\end{equation}
and we find  its infinite volume limit to be
\begin{equation}
a m = 0.746(6) \ , \qquad \chi^2/{\rm d.o.f} = 0.68 \ . 
\end{equation}
This value for $a m$ is compatible within errors with all the measured masses
in the $A1^+$ channel for $L /a\ge 12$. Hence, as in the case of the
string tension discussed earlier, the systematic error due to neglecting the finite size of
the lattice is found to be comfortably less than the statistical error as long as $ L
\sqrt{\sigma} \ge 3$. Thus, having taken care that this bound be
satisfied at all $\beta$ values simulated for the continuum extrapolation,
in the extraction of the continuum limit we shall use the value
measured at one lattice size. 

It is also interesting to look at the masses in the $E^+$ and in the
$T1^+$ channels, as the lowest-lying states in these two channels are
expected to be degenerate in the continuum limit, since they
correspond to different polarisations of the continuum $2^+$
state. From Table~\ref{tab:gluevol} we infer that this degeneracy is
satisfied for lattice sizes other than the smallest $L/a = 10$ (for
which large finite-size effects are expected to arise differently in
the two channels) and the largest $L/a = 24$, where some uncontrolled systematics
(possibly due to mixing with spurious states) creates visible
anomalies in this measurement. We take the agreement in the
intermediate region as a good indication that $\beta = 7.7$ is
sufficiently close to the continuum limit to justify  its inclusion in the
continuum extrapolation. 

Going towards the continuum, we have measured glueball masses in
the various lattice channels for the couplings and volumes
reported in the first two columns of Table~\ref{tab:sigmas}.   }
Continuum extrapolations for the ratio $m_G/\sqrt{\sigma}$ are obtained from the expectation
that the corrections due to discretisation are linear in $\sigma a^2$.
The results  are reported in Table~\ref{tab:contlim} 
and represented in Figs.~\ref{fig:contlim1} and \ref{fig:contlim2}. 
Some excited states are shown in Fig.~\ref{fig:contlim3}. 
As expected, in the continuum limit the states $T_1^\pm$ and $E^\pm$ are degenerate in pairs. 
The channels $T_2^\pm$ show strong fluctuations and 
discretisation effects that are caused by the difficulty in extracting their masses in lattice units. 
The values obtained for the $0^+$ and $2^+$ states are at the same order of magnitude 
as those obtained in $SU(N)$ theories~\cite{Lucini:2001ej}.

\subsubsection{Epilogue}

In this Section, we have reported on what is (to the best of our knowledge) the first controlled calculation in the 
continuum limit of the glueball spectrum and of the string tension for the $Sp(4)$ pure Yang-Mills theory.
The main purpose of our study is to gain some understanding of the glue dynamics in this theory, 
and progressively aim at providing an interpretation of the results emerging from the theory with dynamical quarks. 
Nevertheless, the outcomes of our investigation in the pure gauge case are interesting in their own right: for instance, 
they provide a first step towards a systematic calculation of the pure Yang-Mills spectrum in the large $N$ limit of $Sp(2N)$
 gauge theories and give us an opportunity to contrast the non-perturbative dynamics of $Sp(2N)$ with that of $SU(N)$ gauge theories.

Albeit expected, the first remarkable outcome of our calculation is that the pure gauge quantities behave not dissimilarly from those of $SU(N)$. 
The mass associated to the decay of Polyakov-loop correlators follows the properties  expected 
from a confining flux tube, hence supporting the confining nature of the  pure gauge dynamics in $Sp(4)$. 
The lowest-lying state in the glueball spectrum is the $0^+$ glueball. 
The ratio $m_{0^+}/\sqrt{\sigma} \simeq 3.55$ is not far from the large-$N$ value from $SU(N)$ groups,  $m_{0^{++}}/\sqrt{\sigma} \simeq 3.3$~\cite{Lucini:2012gg}, 
and is indeed compatible within errors with the $SU(3)$ value of 
$m_{0^{++}}/\sqrt{\sigma}=3.55(7)$~\cite{Lucini:2004my}, with the value for $SU(4)$ of $m_{0^{++}}/\sqrt{\sigma} = 3.36(6)$ 
being appreciably different, albeit close.

This behaviour is consistent with the above ratio being a mildly decreasing function 
of the number of generators---while $SU(4)$ has 15 generators, $SU(3)$ has 8 and  $Sp(4)$ has 10---
and deserves further enquiry by performing numerical studies of $Sp(2N)$ gauge theories at larger $N$. 
An interesting observation has been put forward in~\cite{Hong:2017suj}, by suggesting that
\begin{equation}
\label{eq:casimirscaling}
\frac{m_{0^{++}}^2}{\sigma} = \eta \frac{C_2(A)}{C_2(F)} \ , 
\end{equation}
where $C_2(A)$ and $C_2(F)$ are the quadratic Casimir of the adjoint and of the fundamental representation, respectively, and $\eta$ is a universal constant, in the sense that it depends on the dimensionality of the spacetime, but not on the gauge group. Noting that for $Sp(2N)$
\begin{equation}
\frac{C_2(A)}{C_2(F)} = \frac{4\left(N+1\right)}{2N+1} \ , 
\end{equation}
we find that our determination of the proportionality constant 
\begin{equation}
\eta = 5.27(15) \ , 
\end{equation}
is compatible with the value $\eta = 5.41(12)$ extracted from $SU(N)$ in 3+1 dimensions~\cite{Hong:2017suj}.

The rest of the glueball spectrum also follows a pattern that is broadly similar to that of $SU(N)$. 
Another interesting quantity in the glueball sector is the ratio $m_{2^{++}}/m_{0^{++}}$. 
Using universality arguments, it has been argued in~\cite{Athenodorou:2016ndx} that for
confining theories where the dynamics does not yield large anomalous dimensions, as in pure Yang-Mills,
one should find $m_{2^{++}}/m_{0^{++}} = \sqrt{2}$. Our numerical results give $m_{2^{++}}/m_{0^{++}} = 1.425(32)$, 
a value that is fully compatible with the conjecture of~\cite{Athenodorou:2016ndx}.

Besides being relevant for models of electroweak symmetry breaking based on a 
Pseudo-Nambu-Goldstone interpretation of the Higgs field, the investigation of which is the central leitmotif of this paper, 
studies of $Sp(2N)$ pure gauge theories provide new relevant information on universal aspects of Yang-Mills dynamics. 
We shall develop this latter line of research in future numerical investigations.

\section{Of quenched mesons: masses and decay constants}
\label{Sec:mesons}

In this Section, we present our results for  the masses and decay constants of the
lightest flavoured mesons in the quenched approximation. Our main purpose  is to illustrate the process 
that we envision we will carry out once simulations with dynamical quarks are available.
Although we are aware of the fact that the quenched results may not capture 
in full the features of the theory,
we still expect it to provide some useful information about its qualitative features.
Experience on QCD with light quarks suggests that several quantities are well captured by the quenched approximation,
although we already cautioned the reader  about the fact that such considerations may not extend to these dynamical theories.

The EFT  in Sec.~\ref{Sec:EFT}, within the limitations discussed therein,
describes the continuum limit of the 
dynamical simulations, not the quenched ones. In principle, one could 
make more sense of the comparison  by adopting the approach of quenched chiral perturbation 
theory~\cite{Bernard:1992mk,Sharpe:1992ft} or of partially-quenched chiral perturbation
 theory~\cite{Sharpe:2000bc,Sharpe:2001fh,Bernard:1993sv}, but for present purposes our strategy will suffice,
 though we invite the reader to use caution,  in particular for quantities such as the $g_{\r\pi\pi}$ coupling, 
 that are certainly affected by the quenching procedure.

\subsection{Observables}

As discussed in Sec.~\ref{Sec:EFT}, the observables of most direct phenomenological relevance,
and at the same time the most directly accessible to lattice calculations, are
the masses, $m_M$, and the decay constants, $f_M$, of pseudo-scalar, vector and axial-vector mesons. 
This numerical study focuses on flavoured particles. The interpolating operators 
and quantum numbers are summarised in Table~\ref{Tab:mesons}. 

\begin{table}
\begin{center}
\begin{tabular}{|c|c|c|c|}
\hline\hline
{\rm ~~~Label~~~} & {\rm ~~~Operator ${\cal O}_M$~~~} & {\rm ~~~Meson~~~} & $J^P$\cr
\hline
$PS$ & $\overline{Q^i}\gamma_5 Q^j$ & $\pi$ & $0^-$ \cr
$V$ & $\overline{Q^i}\gamma_\mu Q^j$ & $\rho$ & $1^-$ \cr
$AV$ & $\overline{Q^i}\gamma_5\gamma_\mu Q^j$ & $a_1$ & $1^+$\cr
\hline\hline
\end{tabular}
\end{center}
\caption{
Interpolating operators used for the measurement of the properties of flavoured mesons ($i\neq j$) in 
the pseudo-scalar (PS), vector (V) and axial-vector (AV) cases, associated with the $\pi$, $\rho$ and $a_1$ mesons, respectively.
We also report the Lorentz-group quantum numbers $J^P$. 
The summations over colour and spinor indices are understood.
}
\label{Tab:mesons}
\end{table}

We define the ensemble average of two-point meson correlators in the Euclidean space as
\beqs
C_{\mathcal{O}_M}(\vec{p},t)&\equiv&\sum_{\vec{x}} e^{-i\vec{p}\cdot \vec{x}} 
\langle 0 | \mathcal{O}_M(\vec{x},t) \mathcal{O}_M^\dagger(\vec{0},0) | 0\rangle\,,
\label{eq:corr}
\eeqs
where $\mathcal{O}_M$ denotes any of the meson interpolating operators in Table~\ref{Tab:mesons}. 
In the limit in which the Euclidean time $t$ is large,
and for vanishing three-momentum $\vec{p}$, the correlation function is dominated by 
the ground state. Its exponential  decay is controlled by the meson mass $m_M$, and can be approximated as
\beqs
C_{\mathcal{O}_M}(t)\xrightarrow{t\rightarrow \infty}\langle 0 |\mathcal{O}_M | M \rangle \langle 0 |\mathcal{O}_M | M \rangle^* 
\frac{1}{m_M L^3} \left[e^{-m_M t}+e^{-m_M(T-t)}\right]\,,
\label{eq:meson_corr}
\eeqs
where $T$ and $L$ are the temporal and the spatial extent of the lattice, respectively. 
In the meson states $|M\rangle$, we define $M=M^iT^i$, with $T^i$ the generators
of the group. 
Using this convention, the mesonic matrix elements are parameterised in terms
of decay constants $f_M$ and masses $m_M$ as\footnote{For comparison, with these conventions and normalisations, the corresponding experimental value of the pion decay constant in QCD  is $f_\pi \simeq 93\,{\rm MeV}$.}
\beqs
\langle 0 | \overline{Q_1} \gamma_5 \gamma_\mu Q_2 | PS \rangle &=& f_\pi p_\mu\,,\nn \\
\langle 0 | \overline{Q_1} \gamma_\mu Q_2 | V \rangle &=& f_\rho m_\rho \epsilon_\mu\,,\nn \\
\langle 0 | \overline{Q_1} \gamma_5 \gamma_\mu Q_2 | AV \rangle &=& f_{a_1} m_{a_1} \epsilon_\mu\,,
\label{eq:matrix_element}
\eeqs
where $\epsilon_\mu$ is the polarisation vector obeying $\epsilon_{\mu}p^{\mu}=0$ and $\epsilon_{\mu}\epsilon^{\mu}=+1$. 
Using Eq.~(\ref{eq:meson_corr}) and Eq.~(\ref{eq:matrix_element}), for vector and axial-vector mesons 
we can rewrite the correlation functions 
\beqs
C_{\mathcal{O}_V}(t) &\xrightarrow{t\rightarrow\infty}& 
\frac{m_\rho f_\rho^2}{L^3} \left[e^{-m_\rho t}+e^{-m_\rho(T-t)}\right], \nn \\
C_{\mathcal{O}_{AV}}(t) &\xrightarrow{t\rightarrow\infty}& 
\frac{m_{a_1} f_{a_1}^2}{L^3} \left[e^{-m_{a_1} t}+e^{-m_{a_1}(T-t)}\right].
\label{eq:vector_corr}
\eeqs

To calculate the decay constant of the pseudo-scalar meson, 
we additionally consider the following two-point correlation function 
\beqs
C_{\Pi}(\vec{p},t)&=&\sum_{\vec{x}} e^{-i\vec{p}\cdot \vec{x}} 
\langle 0 | [\overline{Q_1} \gamma_5 \gamma_\mu Q_2(\vec{x},t)]\,[\overline{Q_1} \gamma_5 Q_2(\vec{0},0)] | 0\rangle \nn \\
&\xrightarrow{t\rightarrow\infty}&
\frac{f_\pi \langle 0 | \mathcal{O}_{PS} | PS \rangle^*}{L^3}\left[
e^{-m_\pi t}-e^{-m_\pi (T-t)}\right] . 
\label{eq:axial_corr}
\eeqs
The pion mass $m_\pi$ and matrix element $\langle 0 | \mathcal{O}_{PS} | PS\rangle$ 
are obtained from the pion correlator, 
\beqs
C_{\mathcal{O}_{PS}}(t)
&\xrightarrow{t\rightarrow\infty}& 
\frac{|\langle 0 | \mathcal{O}_{PS} | PS \rangle|^2}{m_\pi L^3} \left[e^{-m_\pi t}+e^{-m_\pi(T-t)}\right].
\eeqs 

Meson decay constants computed on the lattice are matched to the continuum. 
In this work, we perform one-loop matching in  lattice perturbation theory. 
Because we are using Wilson fermions, the axial and vector currents are not conserved in the lattice theory,
and hence receive (finite) renormalization, that we write as
\beqs
f^{\rm ren}_\pi=Z_A f_\pi\,,~~~f^{\rm ren}_\rho = Z_V f_\rho\,,~~~{\rm and}~~~ f^{\rm ren}_{a_1}=Z_A f_{a_1}\,.
\eeqs
The pion decay constant $f_\pi$ is renormalised by $Z_A$, as the axial current is used 
to define $f_\pi$ in Eq.~(\ref{eq:matrix_element}). 
In the continuum limit, the renormalization constants are expected to be unity. 
The one-loop matching coefficients taken from~\cite{Martinelli:1982mw} are given by 
\beqs
Z_A&=&1+C(F)\left(\Delta_{\Sigma_1}+\Delta_{\gamma_5\gamma_\mu}\right)\frac{g^2}{16\pi^2}\,, \nn \\
Z_V&=&1+C(F)\left(\Delta_{\Sigma_1}+\Delta_{\gamma_\mu}\right)\frac{g^2}{16\pi^2}\,,
\label{eq:zfactor}
\eeqs
where $g$ is the coupling constant and the eigenvalue of the quadratic Casimir operator is $C(F)=5/4$ for  $Sp(4)$. 
The coefficients $\Delta_I$, relating the lattice computation with the continuum $\overline{\rm MS}$ regularisation scheme, 
result from one-loop integrals performed numerically, and are summarised in Table~\ref{Tab:matching}. 
{  
We notice that the coefficients reported here have been obtained by restricting the integrals within the first Brillouin zone.
We verified explicitly that the errors in numerical evaluation for these integrals are $2\%$ or less, and that there 
are no discernible finite-volume effects.  Therefore
we neglect the uncertainty on the $\Delta_i$ coefficients in the rest of this paper.
}
The coefficient $\Delta_{\Sigma_1}$ is taken from the wave-function renormalization of the external fermion lines, 
without taking into account the power-divergence contribution,
while the coefficients $\Delta_{\gamma_{\mu}}$ and $\Delta_{\gamma_5\gamma_{\mu}}$
are extracted  from the one-loop computations of the vertex functions. 

\begin{table}
\begin{center}
\begin{tabular}{|c|c|c|}
\hline\hline
~~~~~$\Delta_{\Sigma_1}$~~~~~ & ~~~~~$\Delta_{\gamma_\mu}$~~~~~ & ~~~~~$\Delta_{\gamma_5\gamma_\mu}$~~~~~\cr
\hline
$-12.82$ & $-7.75$ & $-3.00$\cr
\hline\hline
\end{tabular}
\end{center}
\caption{
Results of one-loop integrals for the matching coefficients in Eq.~(\ref{eq:zfactor}) at the choice of the Wilson parameter $r=1$,
and taken from~\cite{Martinelli:1982mw} .
}
\label{Tab:matching}
\end{table}

Wilson fermions receive quite large renormalisation 
and thus the perturbative expansion with the bare coupling is reliable only at very large values of the lattice
coupling $\beta=4N/g^2$. 
As in the continuum case, in lattice perturbation theory, the appropriate expansion parameter is the renormalised coupling $\bar{g}$ 
rather than the bare counterpart $g$. 
Following~\cite{Lepage:1992xa},
 we use the simple tadpole improved coupling defined by
\beqs
\bar{g}^2&\equiv& \frac{2Ng^2}{\langle {\Tr}\, \mathcal{P} \rangle}\,,
\eeqs
where $\mathcal{P}$ is the plaquette operator.

\subsection{Numerical results and EFT}
\label{Sec:quench}

\begin{table}
\begin{center}
\begin{tabular}{|c|c|c|c|c|c|c|c|}
\hline\hline
$\beta$ & $~a^2 t_0^p~$ & $~a^2 t_0^c~$ & $~a w_0^p~$ & $~a w_0^c~$ &
$\langle P\rangle$ & $~~Z_V~~$ & $~~Z_A~~$ \cr \hline
$7.62$ & $1.805(7)$ & $2.049(7)$ & $1.436(4)$ & $1.448(4)$ & $0.60190(19)$ & $0.71599(9)$ & $0.78157(7)$ \cr
$8.0$ & $4.899(18)$ & $5.115(19)$ & $2.300(6)$ & $2.308(6)$ & $0.63074(13)$ & $0.74185(5)$ & $0.80146(4)$ \cr
\hline\hline
\end{tabular}
\end{center}
\caption{
\label{Tab:zfactor}
Gradient-flow scales $t_0^p$, $t_0^c$, $w_0^p$ and $w_0^c$, 
plaquette values $\langle P\rangle$, and one-loop matching factors $Z_V$ and $Z_A$, computed for the two values of the lattice coupling $\beta$
used in subsequent exploratory quenched calculations.
The errors quoted arise from the stochastic determination of the tadpole-improved coupling.
}
\end{table}

As a first exploratory step towards understanding the qualitative features of mesons of the $Sp(4)$ theory,  
as well as to test the low-energy EFT and illustrate its use, 
we calculate the mesonic observables 
described in the previous section in the quenched approximation at $\beta=7.62$ and $8.0$. 
For each ensemble we generate $200$ gauge configurations on a lattice of size $48\times24^3$, using the HB algorithm. 
We first measure the plaquette values (which are used to compute the one-loop matching factors in Eq.~(\ref{eq:zfactor}) with the 
tadpole improved coupling $\bar{g}$), and the gradient flow scales $t_0$ and $w_0$. 
The numerical results are summarised in Table~\ref{Tab:zfactor}. 
The gradient flow scales are determined by setting $\mathcal{E}_0=\mathcal{W}_0=0.35$. 
As discussed in Section \ref{Sec:Phasespace}, we use two definitions of $G_{\mu\nu}$ 
denoted by the superscript $p$ for a plaquette and $c$ for a (four-plaquette) clover, 
with the difference of the two measuring the size of finite lattice spacing artefacts. 
We find that, compared to $t_{0}$, the difference between the clover and the plaquette regularisations for $w_{0}$
 is significantly smaller,  and thus 
we choose to use $w_0^c$ to convert lattice units to physical ones.

\begin{table}
\begin{center}
\begin{tabular}{|c|c|c|c|c|c|c|}
\hline \hline
~~$a m_0$~~ & ~~ $a^2 m_\pi^2$ ~~ & ~~ $a^2 m_\rho^2$ ~~ & ~~ $a^2 m_{a_1}^2$ ~~ & ~~ $a^2 f_\pi^2$ ~~ & ~~ $a^2 f_\rho^2$ ~~ & ~~ $a^2 f_{a_1}^2$ ~~ \\ \hline
$-0.65$ & 0.4325(5) & 0.5087(8) & 1.04(4) & 0.01451(9) & 0.0376(3) & 0.021(4) \\
$-0.7$ & 0.3042(4) & 0.3916(11) & 0.943(29) & 0.01246(8) & 0.0365(4) & 0.029(3) \\
$-0.73$ & 0.2318(4) & 0.3272(14) & 0.862(20) & 0.01101(8) & 0.0354(5) & 0.0313(20) \\
$-0.75$ & 0.1856(4) & 0.2875(15) & 0.831(16) & 0.00995(8) & 0.0346(4) & 0.0352(16) \\
$-0.77$ & 0.1409(4) & 0.2485(19) & 0.769(22) & 0.00879(8) & 0.0329(6) & 0.0350(22) \\
$-0.78$ & 0.1191(4) & 0.2312(21) & 0.796(17) & 0.00822(8) & 0.0327(6) & 0.0415(16) \\
$-0.79$ & 0.0977(4) & 0.2115(26) & 0.772(20) & 0.00760(8) & 0.0314(8) & 0.0418(19) \\
$-0.8$ & 0.0765(4) & 0.193(3) & 0.748(24) & 0.00698(8) & 0.0301(10) & 0.0417(23) \\
$-0.81$ & 0.0553(4) & 0.175(5) & 0.73(3) & 0.00635(9) & 0.0285(14) & 0.042(3) \\
$-0.815$ & 0.0446(4) & 0.166(7) & 0.70(4) & 0.00606(9) & 0.0280(19) & 0.040(4) \\
$-0.82$ & 0.0328(4) & 0.158(13) & 0.70(6) & 0.00572(15) & 0.028(4) & 0.041(6) \\ \hline \hline
\end{tabular}
\end{center}
\caption{%
\label{tab:quenched_b7.62}%
Masses and decay constants (squared) of pseudo-scalar ($\pi$), vector ($\r$), and axial-vector ($a_1$) mesons, as 
obtained from the  quenched calculations described in Sec.~\ref{Sec:quench}, 
for  $\beta=7.62$ on lattice of size $48\times 24^3$. }
\end{table}

\begin{table}
\begin{center}
\begin{tabular}{|c|c|c|c|c|c|c|}
\hline \hline
~~$a m_0$~~ & ~~ $a^2 m_\pi^2$ ~~ & ~~ $a^2 m_\rho^2$ ~~ & ~~ $a^2 m_{a1}^2$ ~~ & ~~ $a^2 f_\pi^2$ ~~ & ~~ $a^2 f_\rho^2$ ~~ & ~~ $a^2f_{a_1}^2$ ~~ \\ \hline
$-0.45$ & 0.5556(12) & 0.5837(13) & 0.868(13) & 0.00832(14) & 0.0153(3) & 0.0057(5) \\
$-0.5$ & 0.4244(11) & 0.4551(14) & 0.734(11) & 0.00771(14) & 0.0149(3) & 0.0073(6) \\
$-0.55$ & 0.3037(8) & 0.3383(12) & 0.593(11) & 0.00691(10) & 0.0145(3) & 0.0084(7) \\
$-0.6$ & 0.1937(8) & 0.2325(14) & 0.465(12) & 0.00567(8) & 0.0131(3) & 0.0096(8) \\
$-0.625$ & 0.1437(7) & 0.1862(16) & 0.405(13) & 0.00494(8) & 0.0125(4) & 0.0102(10) \\
$-0.64$ & 0.1156(7) & 0.1612(15) & 0.363(20) & 0.00449(7) & 0.0124(3) & 0.0099(18) \\
$-0.65$ & 0.0974(7) & 0.1448(16) & 0.349(15) & 0.00414(7) & 0.0120(3) & 0.0109(13) \\
$-0.66$ & 0.0812(5) & 0.1302(14) & 0.343(14) & 0.00397(5) & 0.0123(3) & 0.0128(12) \\
$-0.67$ & 0.0642(5) & 0.1179(18) & 0.318(18) & 0.00352(4) & 0.0123(4) & 0.0123(17) \\
$-0.68$ & 0.0479(4) & 0.1040(22) & 0.323(13) & 0.00319(4) & 0.0122(5) & 0.0150(11) \\
$-0.69$ & 0.0318(4) & 0.0907(28) & 0.304(19) & 0.00270(5) & 0.0115(5) & 0.0151(17) \\ \hline \hline
\end{tabular}
\end{center}
\caption{%
\label{tab:quenched_b8.0}%
Masses and decay constants (squared) of pseudo-scalar ($\pi$), vector ($\r$), and axial-vector ($a_1$) mesons, as 
obtained from the  quenched calculations described in Sec.~\ref{Sec:quench}, 
for  $\beta=8.0$ on lattice of size $48\times 24^3$.  }
\end{table}

The mesonic two-point correlation functions in Eq.~(\ref{eq:corr}) are measured using 
stochastic wall sources~\cite{Boyle:2008rh} at various quark masses. 
While at sufficiently large time we use the asymptotic expressions of 
two-point correlation functions in Eqs.~(\ref{eq:meson_corr}), (\ref{eq:vector_corr}) and~(\ref{eq:axial_corr})
to extract the meson masses and decay constants, we perform multi-exponential fits when 
the time extent is not large enough to reach the asymptotic region, according to
\beqs
C_{\mathcal{O}_M}(t)=\sum_{i=0}^\infty C_i \left(
e^{-E_i t}+e^{-E_i(N_t-t)}
\right)\,,
\eeqs
where $E_0=m_M<E_1<E_2<\cdots$. We find that the two-exponential fits are good enough to describe the numerical data 
in most cases. 
The numerical results (not renormalised)
 are summarised in Tables~\ref{tab:quenched_b7.62} and~\ref{tab:quenched_b8.0}. 
The statistical errors are estimated by using a standard bootstrapping technique (with about 250 bootstrap samples)
and a correlated fit with $\chi^2$ minimisation.

\begin{table}
\begin{center}
\begin{tabular}{|c|c|c|c|c|c|}
\hline\hline
 & ~~fit range~($m_0$)~~ & $~~~\chi^2/{\rm d.o.f}~~~$
& $~~~~~w_0^c~m_0^*~~~~~$ & $~~~~~w_0^c v~~~~~$ & $~~~~~w_0^c v_5~~~~~$ \cr \hline
$\beta=7.62$ & $[-0.82,\,-0.73]$ & $0.17$ & $-1.214(22)$ & $0.750(5)$ & $0.408(5)$\cr
$\beta=8.0$ & $[-0.69,\,-0.625]$ & $0.58$ & $-1.636(27)$ & $0.862(27)$ & $0.431(22)$ \cr
\hline\hline
\end{tabular}
\end{center}
\caption{
\label{Tab:GMOR_bare}
Results of the fit of the quenched meson data to  the GMOR relation in Eq.~(\ref{Eq:GMOR}), 
where the quark mass $m$ is replaced by the combination $w_0^c (m_0-m_0^*)$.
}
\end{table}

\begin{table}
\begin{center}
\begin{tabular}{|c|c|c|c|c|}
\hline\hline
 & ~~fit range~($m_0$)~~ & $~~~\chi^2/{\rm d.o.f}~~~$
& $~~~~~w_0^c\bar{v}~~~~~$ & $~~~~~w_0^c\bar{v}_5~~~~~$ \cr \hline
$\beta=7.62$ & $[-0.82,\,-0.75]$ & $0.19$ & $0.1839(17)$ & $0.1302(16)$\cr
$\beta=8.0$ & $[-0.69,\,-0.66]$ & $0.74$ & $0.1873(25)$ & $0.1276(34)$ \cr
\hline\hline
\end{tabular}
\end{center}
\caption{
\label{Tab:GMOR_ren}
Results of the fit of the quenched meson data to  the GMOR relation in Eq.~(\ref{Eq:GMOR}), where the quark mass $m$ is replaced by the combination
 $(w_0^c m_{\rm PS})^2$. 
The barred variables are defined as $\bar{v}=v/(2B)^{1/3}$ and $\bar{v}_5=v_5/(2B)$.
}
\end{table}

\begin{figure}
\begin{center}
\includegraphics{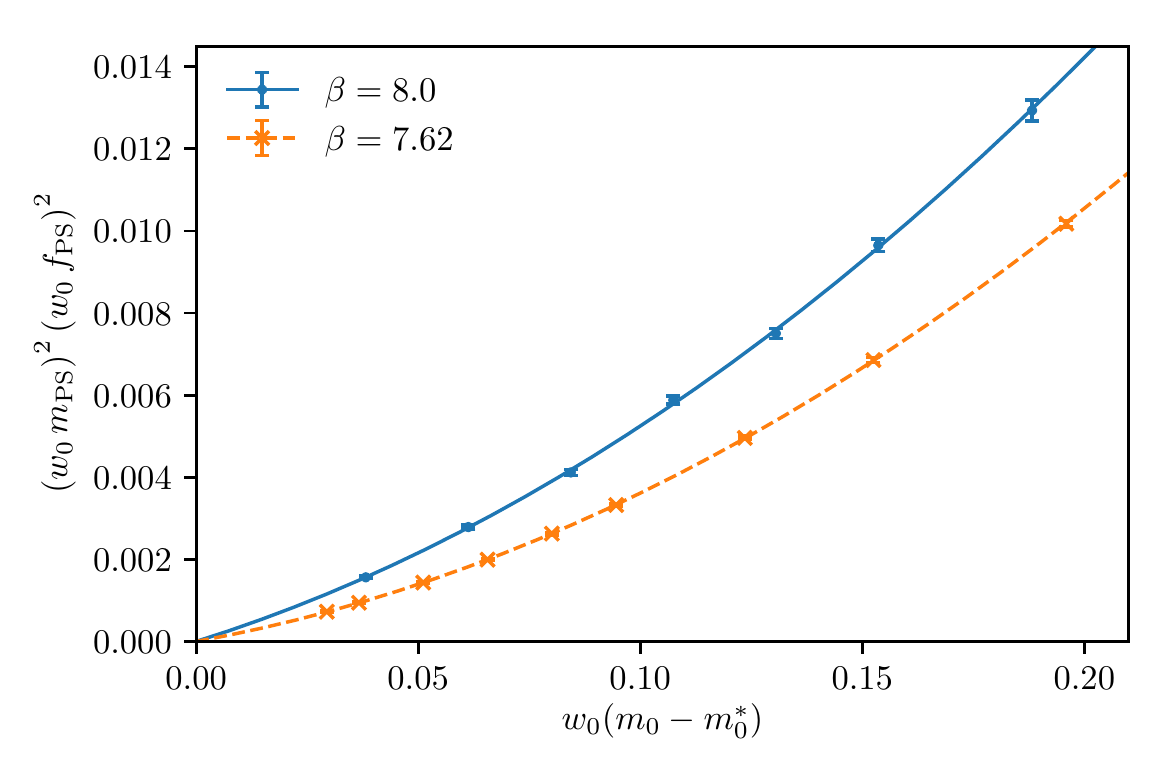}
\includegraphics{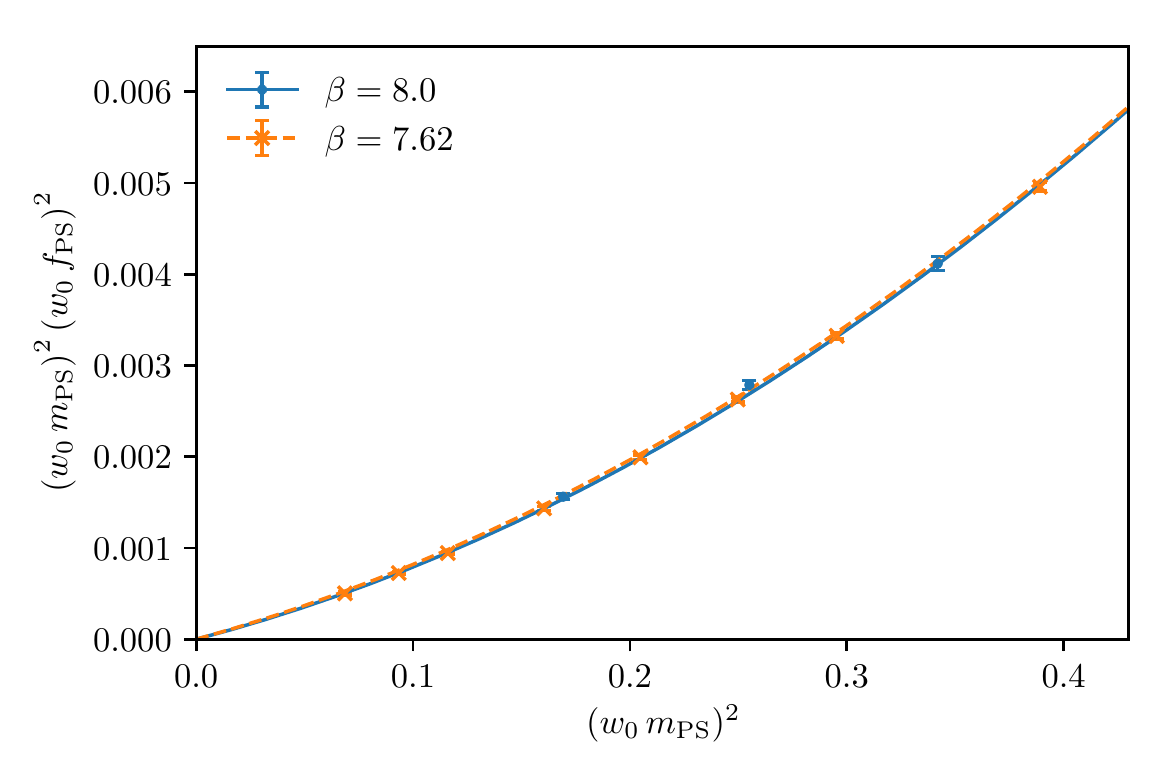}
\end{center}
\caption{
Tests of the GMOR relation in the quenched data. 
The solid and dashed lines are fit results for the cases of $\beta=8.0$ and $7.62$, respectively. 
In the top panel, we show the result of identifying the quark mass with the quantity $w_0 (m_0-m_0^*)$,
replaced in the lower panel by $(w_0 m_{\rm PS})^2$.
}
\label{Fig:GMOR}
\end{figure}

To analyse the numerical results using the continuum EFT developed in Sec.~\ref{Sec:EFT}, 
we reinstate the correct dimensionality by expressing all mass and decay constants in units of $w_0=w_{0}^{c}$.
The decay constants are further renormalised using perturbative one-loop matching. 

Let us first restrict our attention to the pseudo-scalar mesons and check the GMOR relation. 
In the upper panel of Fig.~\ref{Fig:GMOR} we plot $m_{PS}^2 f_{PS}^2$, 
against the quark mass, where the latter  is defined by the difference between the bare quark mass and 
the critical mass at which the mass of pseudo-scalar mesons vanishes. 
Even for the lightest masses $m_0$  available we do not find the expected leading-order linear behaviour. 
Therefore, we fit the data to the NLO results including the $v_5$ term in Eq.~(\ref{Eq:L}). 
The solid and dashed lines in Fig.~\ref{Fig:GMOR} are the fit results for $\beta=8.0$ and $7.62$, respectively, 
where the fitting ranges and the resulting fit parameters are found in Table~\ref{Tab:GMOR_bare}. 

As the quark mass is scheme-dependent, to compare the lattice results obtained with different coupling
we would have to properly renormalise the quark mass. 
Instead of doing so, we consider the GMOR relation with respect to the 
pseudo-scalar meson mass, $m_{PS}$, which is a scheme independent (physical) quantity, and show the result in 
the bottom panel of Fig.~\ref{Fig:GMOR}.
As shown in the fit results in Tab.~\ref{Tab:GMOR_ren}, 
we find that the parameters of the NLO GMOR relation between two lattices are statistically consistent, 
implying that  finite lattice artefacts affect this observable only in a  negligible way. 

\begin{figure}
\begin{center}
\includegraphics{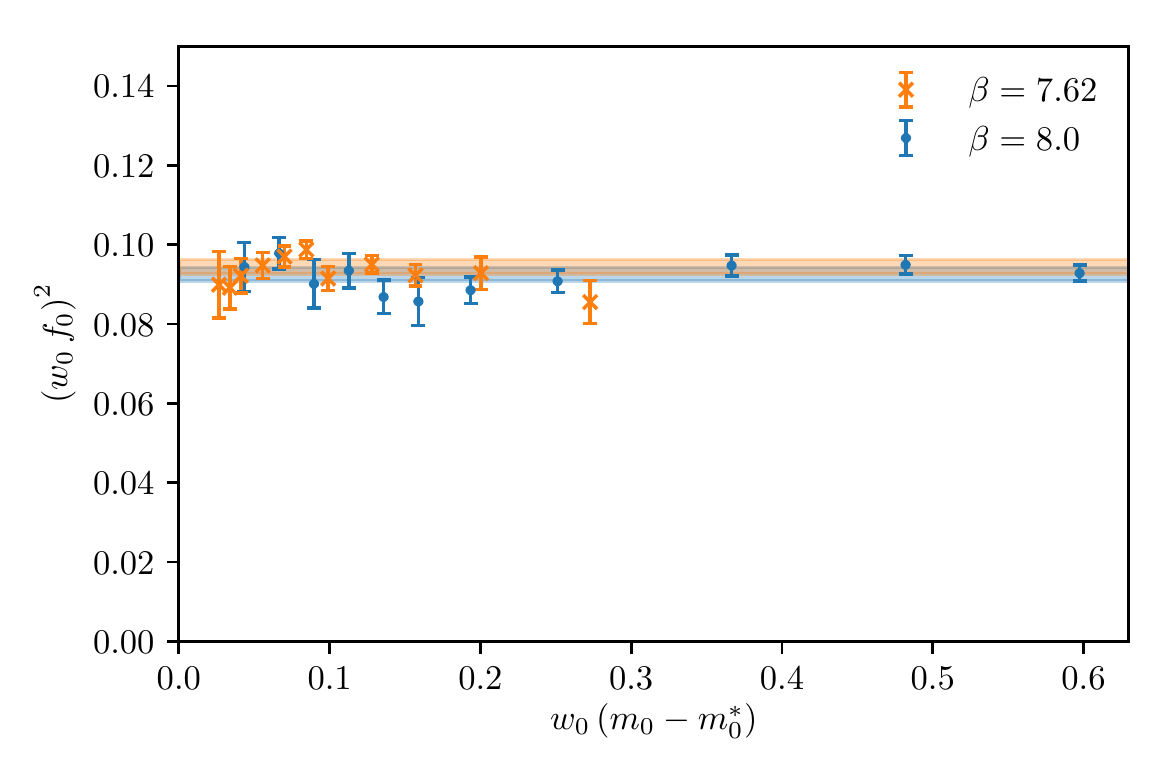}
\end{center}
\caption{
\label{Fig:f0}
The quantity $f_0^2=f_{\pi}^2(0)+f_{\rho}^2+f_{a_1}^2$, defined in Eq.~(\ref{Eq:Sigma}), measured in the quenched approximation
for varying values of  $w_0 (m_0-m_0^*)$.
}
\label{Fig:f0}
\end{figure}

One of the interesting quantities in the EFT is $f_0^2$, 
the sum of the squared decay constants of pseudo-scalar, vector, and axial-vector mesons, 
which is  independent of the  mass $m_0$ if one adopts Eq.~(\ref{Eq:L}). 
As shown in Fig.~\ref{Fig:f0}, the numerical results are in good agreement with such expectation. 
In the figure, blue circles and orange crosses represent the choices $\beta=7.62$ and for $\beta=8.0$, respectively. 
Notice that  we derived this relation from our NLO EFT at the tree level. 
In principle, one has to consider the contribution from the one-loop corrections, including chiral logarithms. 
We performed a constant fit to the results and {   obtained $f_0^2(\beta=8.0)=0.0926(17)$ and $f_0^2(\beta=7.62)=0.0944(18)$ }
denoted by blue and orange bands, respectively. 
As for the case of the GMOR relation, we find that lattice spacing artefacts in $f_0^2$ are negligible,
in the sense that no appreciable difference is measured.

\begin{figure}
\begin{center}
\includegraphics{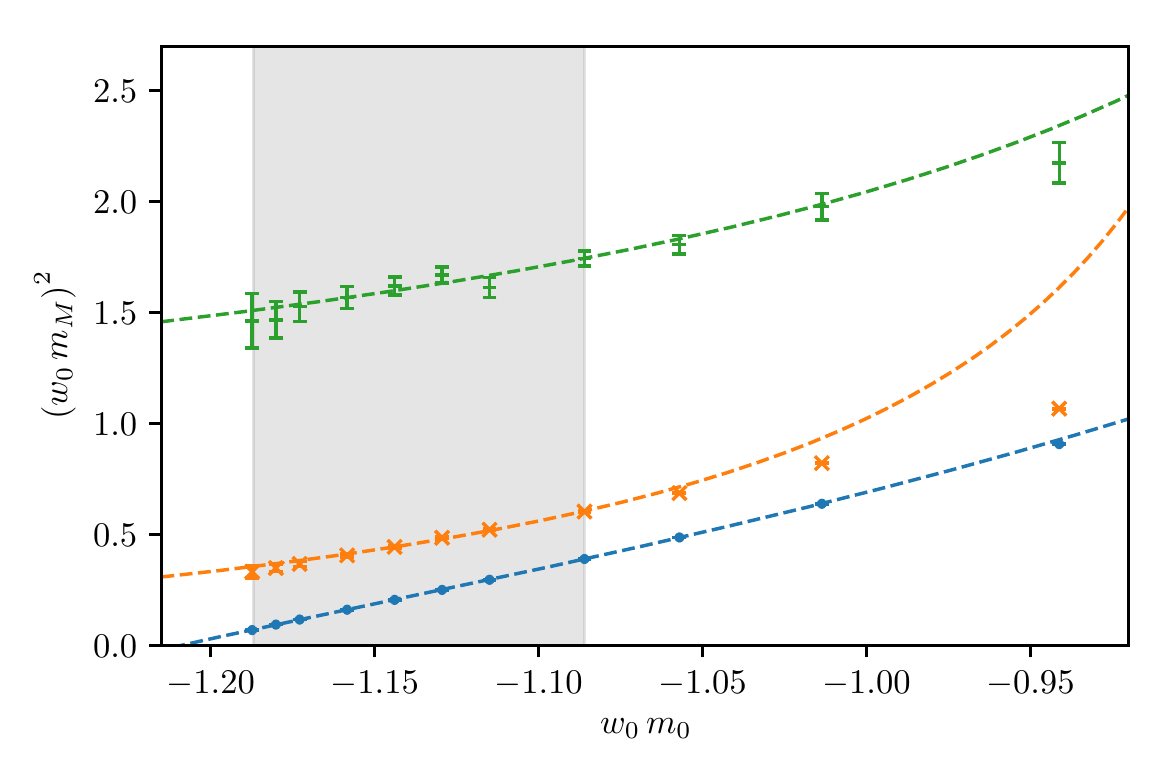}
\includegraphics{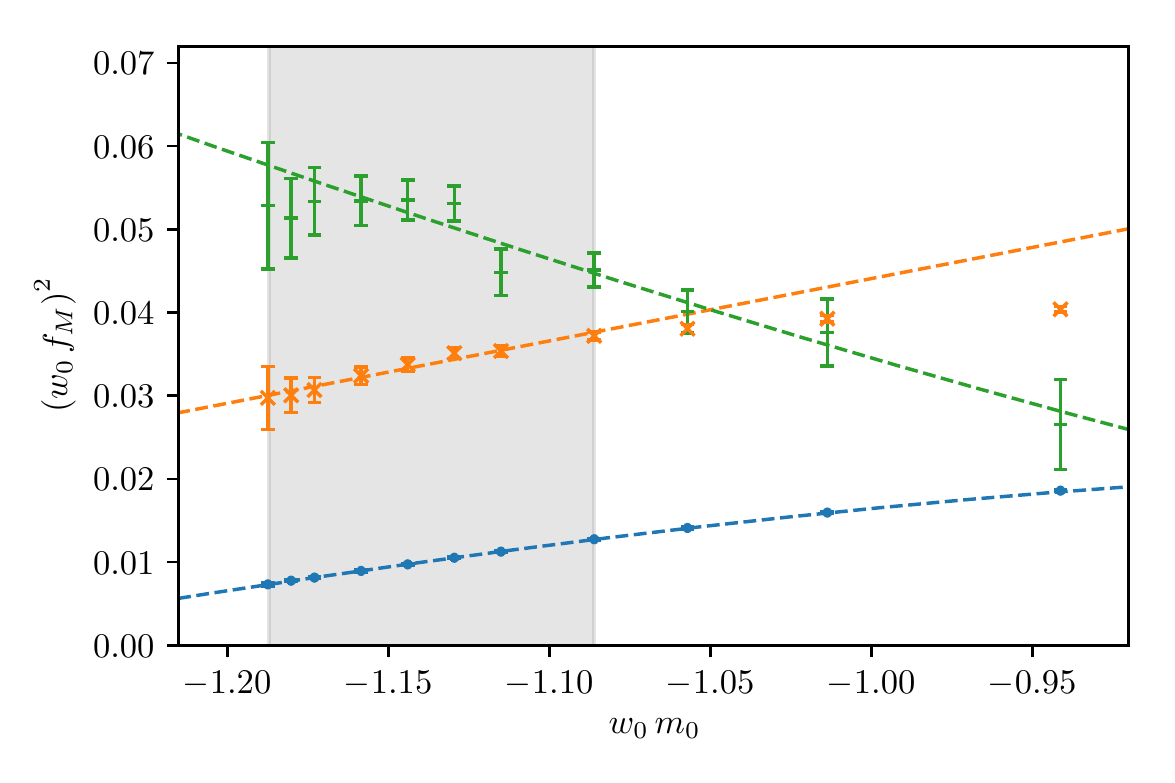}
\end{center}
\caption{
Plots of the meson masses (top panel) and decay constants (bottom panel) from the quenched calculation with $\beta=7.62$,
compared to the best-fit results based upon the EFT description in Sec.~\ref{Sec:vectors}. Blue circles, orange crosses, 
and green bars
represent for pseudo-scalar ($\pi$), vector ($\rho$), and axial-vector ($a_1$) mesons, respectively. 
For the EFT fits are based on the data in the shaded region.
}
\label{Fig:eft_fits_coarse}
\end{figure}

\begin{figure}
\begin{center}

\includegraphics{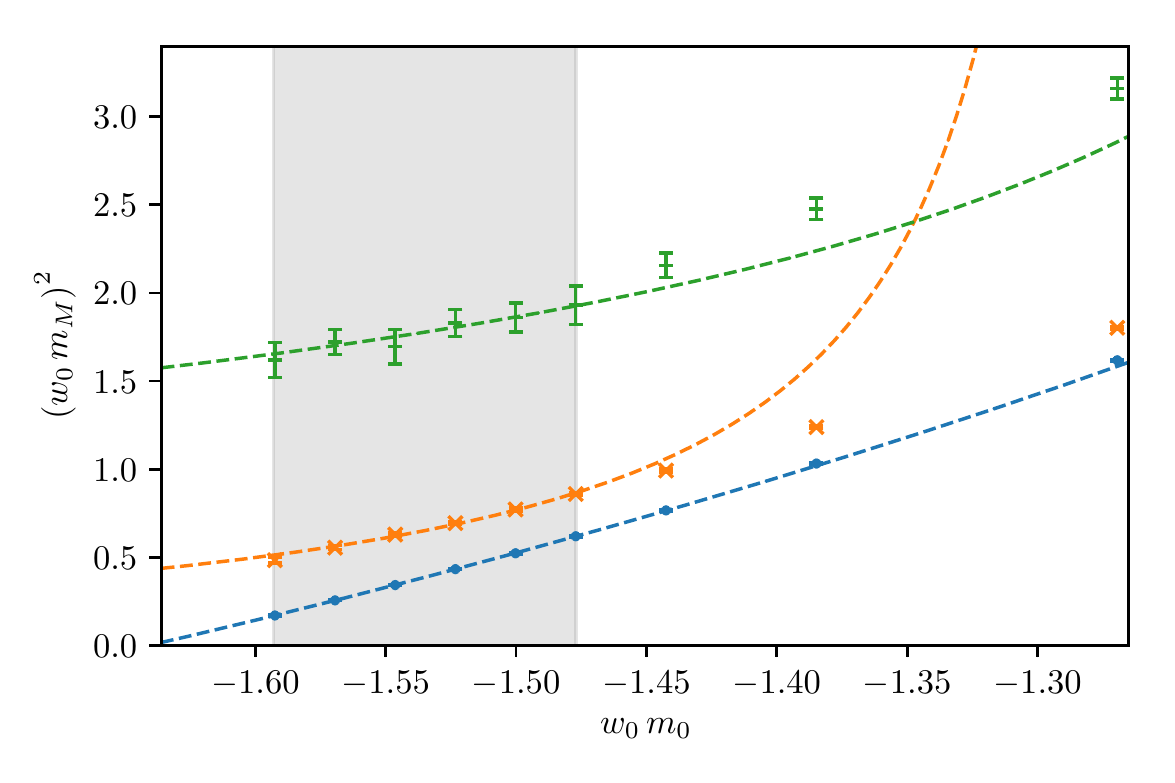}
\includegraphics{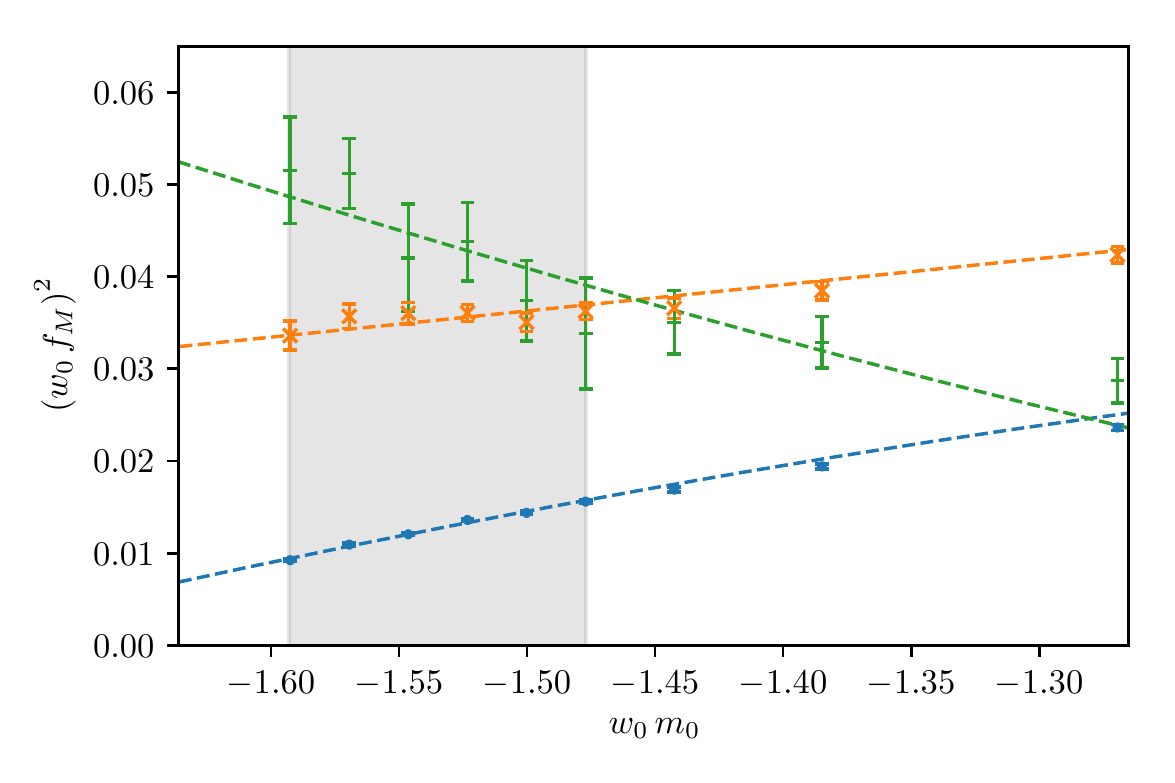}
\end{center}
\caption{
Plots of the meson masses (top panel) and decay constants (bottom panel) from the quenched calculation with $\beta=8.0$,
compared to the best-fit results based upon the EFT description in Sec.~\ref{Sec:vectors}. Blue circles, orange crosses, and green bars
represent for pseudo-scalar ($\pi$), vector ($\rho$), and axial-vector ($a_1$) mesons, respectively. 
For the EFT fits are based on the data in the shaded region.
}
\label{Fig:eft_fits_fine}
\end{figure}

We finally use the NLO EFT relations to construct a  global fit to the meson masses and decay constants. 
The results are illustrated in Figs.~\ref{Fig:eft_fits_coarse} and~\ref{Fig:eft_fits_fine}. 
We perform an uncorrelated fit to the data, restricted to the eight and six lightest masses  $m_0$ for $\beta=7.62$ and $8.0$, respectively.
The fitting range is shown as the shaded region in the figures. 
There are two main technical difficulties in this fit procedure. 
First of all, the parameter space is too large to determine the actual global minimum. 
In the NLO EFT we have thirteen fit parameters (including the critical bare mass $m_0^{\ast}$). 
The standard $\chi^2$ minimisation is not stable, and it typically 
yields two qualitatively very different results, one fairly linear and one exhibiting highly nonlinear behaviours with respect to the quark mass. 
We take the former as our best fit as the stability of the fits is better than the latter 
when we vary the fitting range. 
Secondly, statistical uncertainties vary widely for different types of mesons,
and as a result the pseudo-scalar mesons tightly constrain the fit results. 
Furthermore, the numerical data suffer from several systematics such as quenching and discretising effects.

\begin{figure}
\begin{center}
\includegraphics{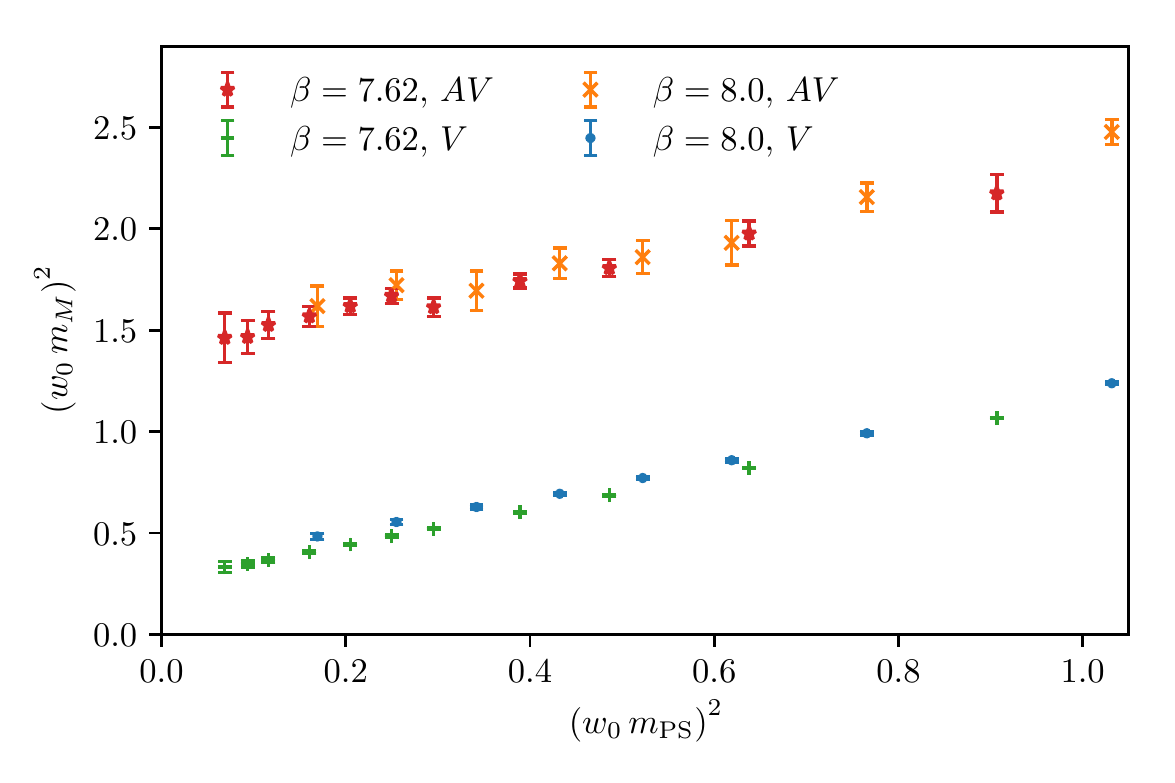}
\includegraphics{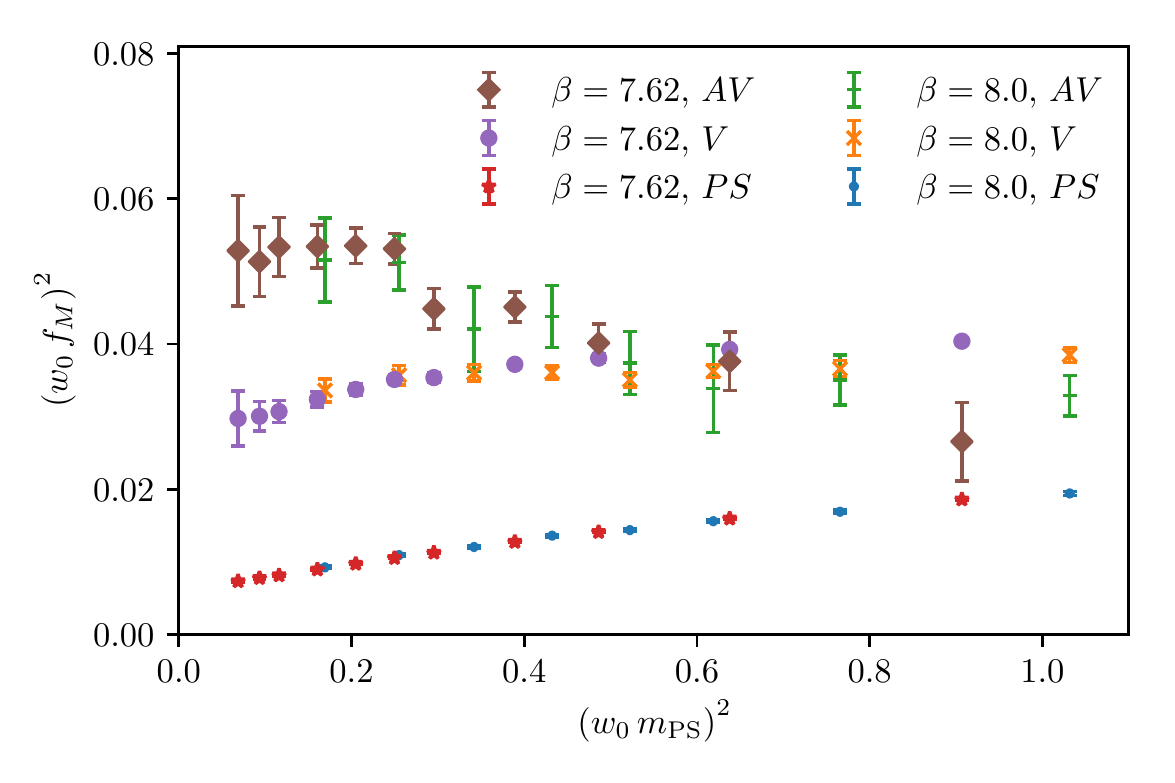}
\end{center}
\caption{
Comparison between the meson masses (top panel) and decay constants (bottom panel) in the quenched calculations with two different values of the lattice coupling $\beta$. 
The colour coding  for the pseudo-scalar, vector, and axial-vector mesons is explained by the legend.
The numerical results are shown as a function of the combination $(w_0 m_{\rm PS})^2$, highlighting the corresponding reduction of finite-spacing effects.
}
\label{Fig:EFT_mps}
\end{figure}

Undeterred by all these limitations and difficulties, 
since the purpose of this explorative study with quenched calculations is to show how our EFT works, 
 we attempted to perform the global fit to the central values using the standard $\chi^2$ minimisation.
The fit results are represented by the dashed lines in the figures. 
{   For completeness, we report the resulting values of the fit parameters  for $\beta=7.62$:
\beqs
&\kappa=-0.83,~y_3/w_0=-0.41,~y_4/w_0=3.14,~g_{\r}=1.93,~b=-0.38,~c=0.027,& \nn \\
&~f w_0=0.84,F w_0=0.57,~v_1 w_0=0.075,~v_2 w_0=-0.33,~v w_0=0.27,~v_5 w_0=0.40,& \nn \\
&~m_0^*w_0=-1.21,~\chi^2/{\textrm{d.o.f}}=0.11\,,&\nn
\eeqs
and for $\beta=8.0$:
\beqs
&\kappa=-0.90,~y_3/w_0=-0.26,~y_4/w_0=2.61,~g_{\r}=1.62,~b=-0.28,~c=0.012,& \nn \\
&~f w_0=1.06,F w_0=0.62,~v_1 w_0=0.028,~v_2 w_0=-0.31,~v w_0=0.28,~v_5w_0=0.48,& \nn \\
&~m_0^*w_0=-1.64,~\chi^2/{\textrm{d.o.f}}=0.99\,.&\nn
\eeqs}
We explicitly checked that all of these fits satisfy the unitary constraints. 
Notice that the values of $\chi^2/{\textrm{d.o.f}}$ are very reasonable,
as also shown by the figures. Yet, the fitting procedure is very rough, the comparison between quenched data and continuum EFT 
is not rigorous, and hence we do not include uncertainties on these
values of the couplings as a way to stress the fact that they should be used just for illustrative purposes.
With this illustrative values of the parameters, {   one finds that
$g_{\rho\pi\pi}^2/(48\,\pi)\sim 0.76$ (for $\beta=7.62$), and $g_{\rho\pi\pi}^2/(48\,\pi)\sim1.0$ (for $\beta=8.0$).}
Such  large values of $g_{\rho\pi\pi}$ are affected by the uncontrolled systematics originating from quenching effects, 
and hence should not be used beyond the illustrative purposes of this exercise.

In  future studies with dynamical calculations, a dedicated examination of the statistics and the systematics
of the EFT fits will be required to determine the corresponding low-energy constants in a meaningful way. 
Furthermore, we anticipate that it will be more involved to apply the continuum EFT result to the dynamical simulation, 
because the scale-setting procedure becomes more subtle. For instance, we observed in Sec.~\ref{sec:scale_setting} 
that the scale $w_0$  used in our analysis 
changes visibly as the quark mass is varied in the dynamical case.  
Finally, it would be interesting to compare the results for masses and decay constants to the calculation
 presented in~\cite{Bizot:2016zyu}, and a possible extension that includes the dependence on the quark mass $m_0$.

In order to investigate  finite lattice spacing artefacts, as we observed earlier 
it is more effective to plot the mesonic observables with respect to the pseudo-scalar meson mass, rather than  the bare quark mass.
The results are shown in Fig.~\ref{Fig:EFT_mps}. In this case, only the masses of vector and axial-vector mesons are plotted. 
As we already learned from the GMOR relation, discretisation effects for the pseudo-scalar mesons are negligible in their 
spectroscopy. Similar conclusion can be drawn for the axial-vector mesons, given the current uncertainties.
On the other hands, the masses and decay constants of vector mesons are affected significantly by lattice artefacts.

\section{Towards dynamical fermions}
\label{Sec:mass}
\begin{figure}
\begin{center}
\includegraphics{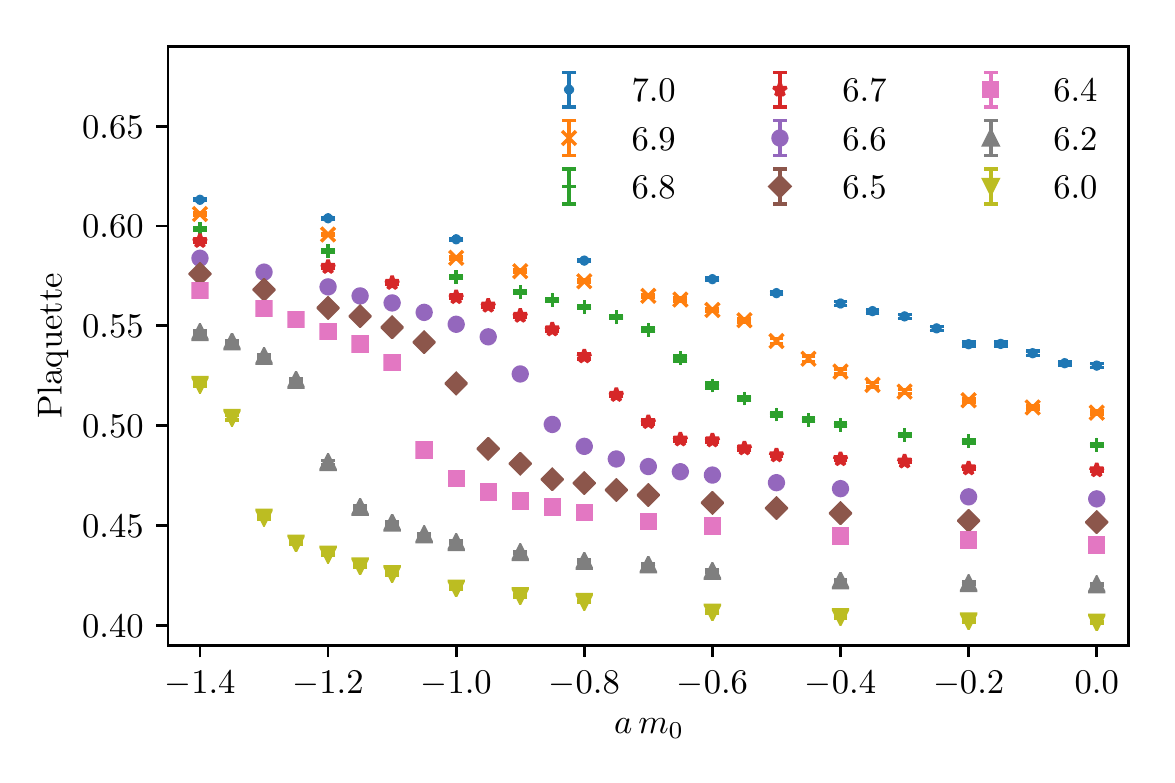}
\end{center}
\caption{
Value of the plaquette from a  lattice parameter scan with dynamical fermions for lattice size of $4^4$,
while varying the bare mass $a m_0$ and the coupling $\beta$. The colours refer to the  choices of $\beta$
indicated in the legend.
}
\label{Fig:plaq_mass_scan}
\end{figure}
\begin{figure}
\begin{center}
\includegraphics{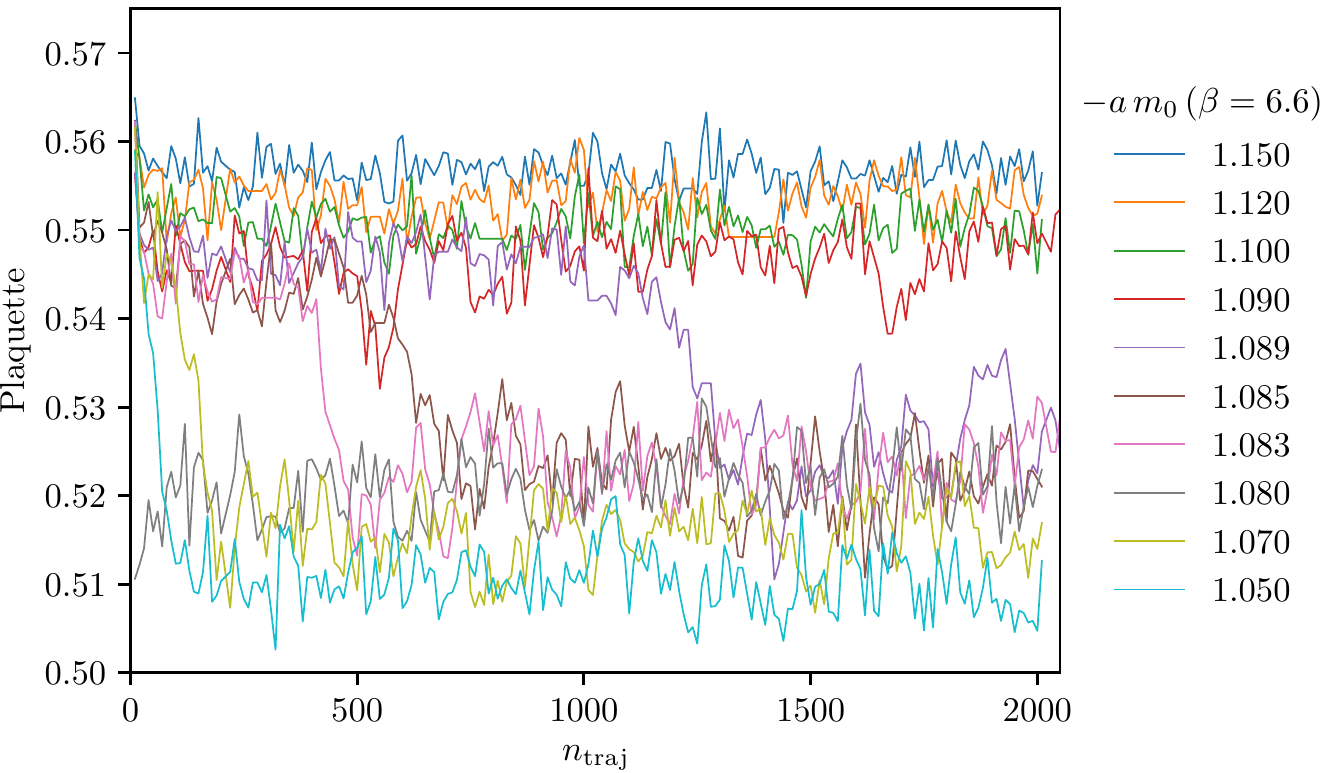}
\includegraphics{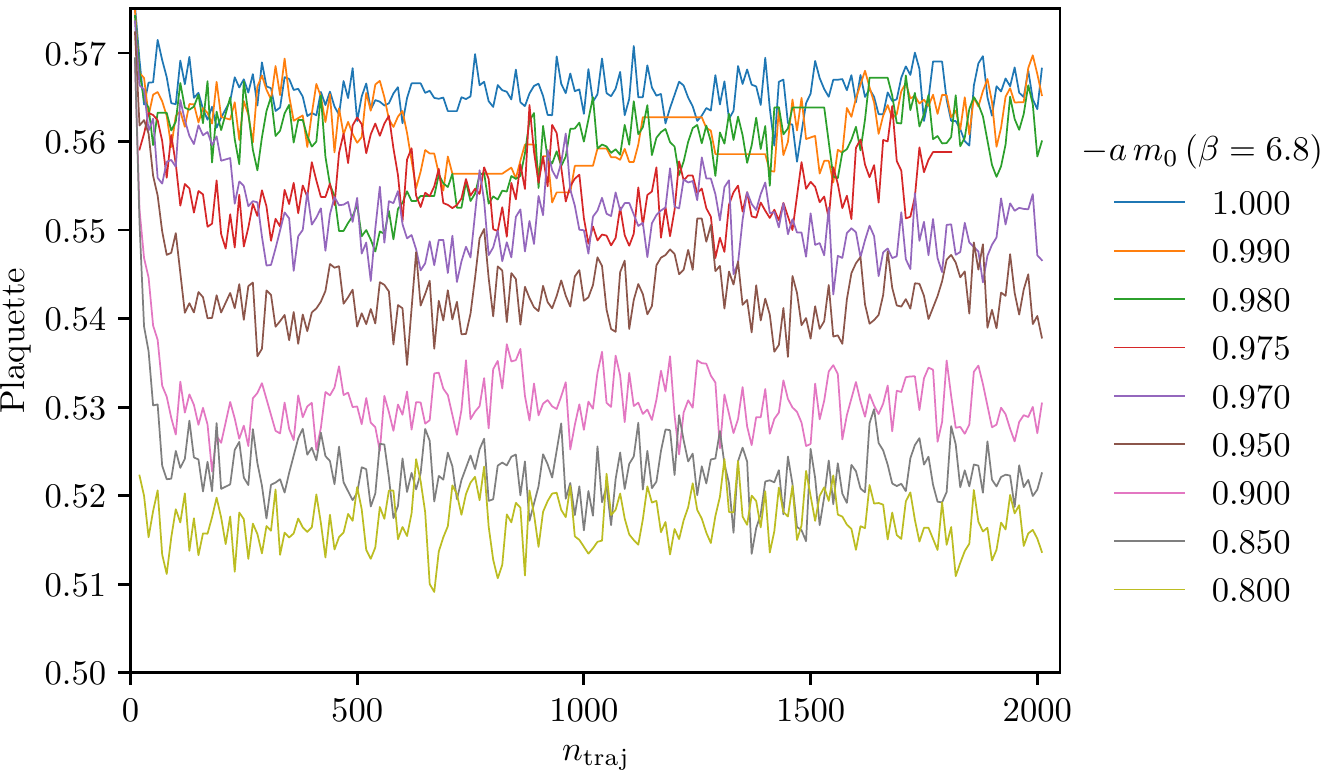}
\end{center}
\caption{
Trajectories of plaquette values for $\beta=6.6$ (top panel) and $\beta=6.8$ (bottom panel) on a lattice with size $8^4$, 
and for various values of the bare mass, as explained in the legend, from HMC calculations  with dynamical quarks.
}
\label{Fig:plaq_traj}
\end{figure}

\begin{figure}
\begin{center}
\includegraphics{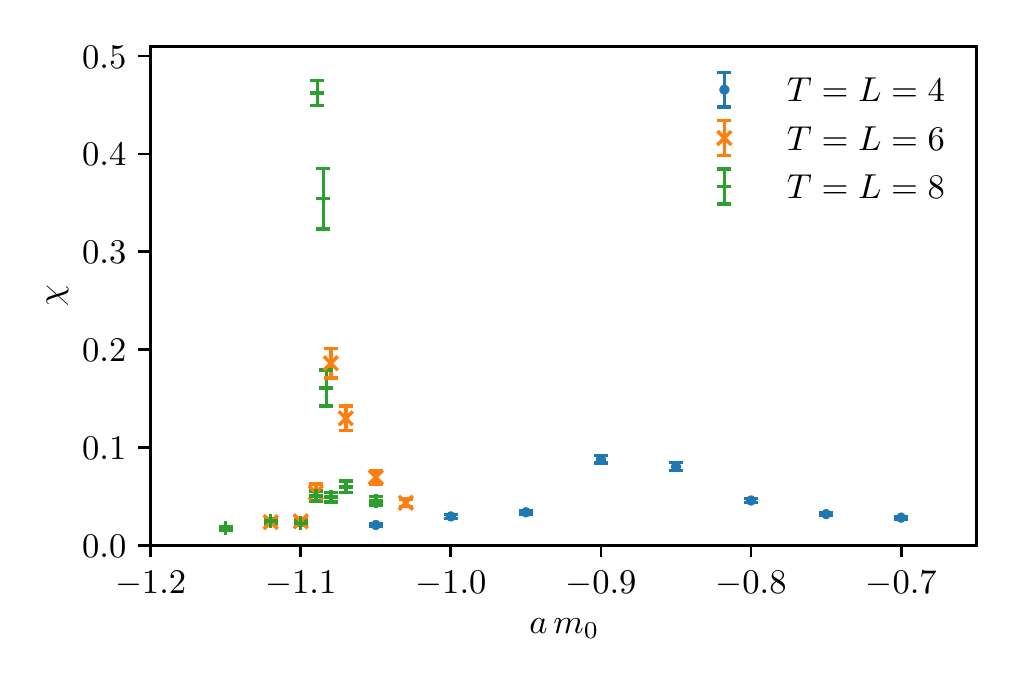}
\includegraphics{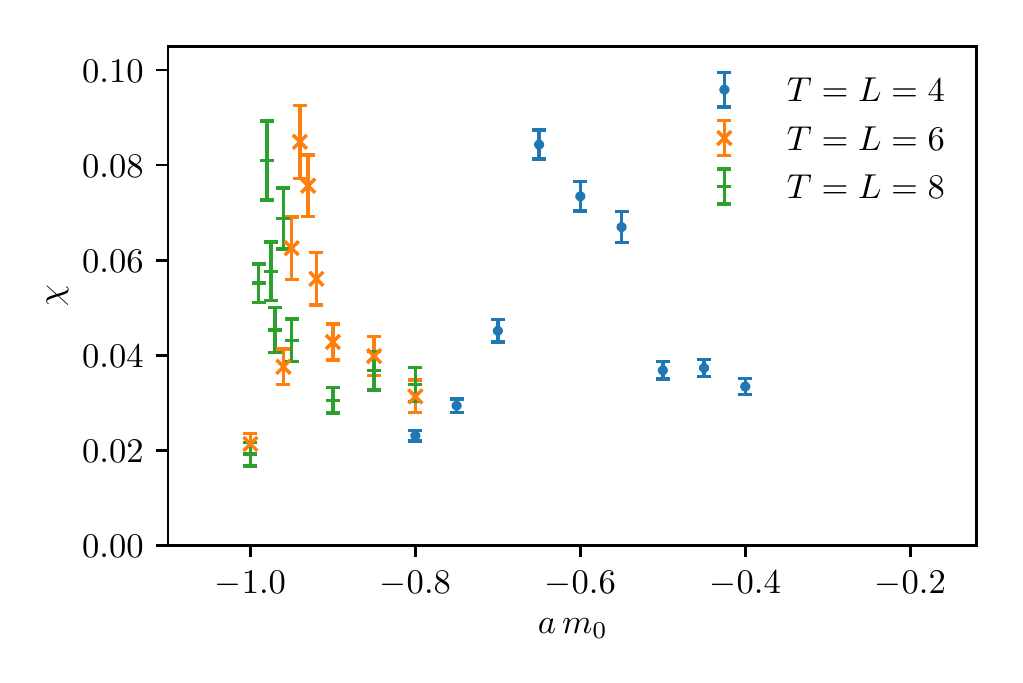}
\end{center}
\caption{
Plaquette susceptibilities $\chi$, measured in HMC calculations with dynamical quarks,  for $\beta=6.6$ (top panel) and $\beta=6.8$ (bottom panel),
as a function of the bare mass $a m_0$, for three values of the lattice size (see legend).
}
\label{Fig:plaq_suscept}
\end{figure}

\begin{figure}
\begin{center}
\includegraphics{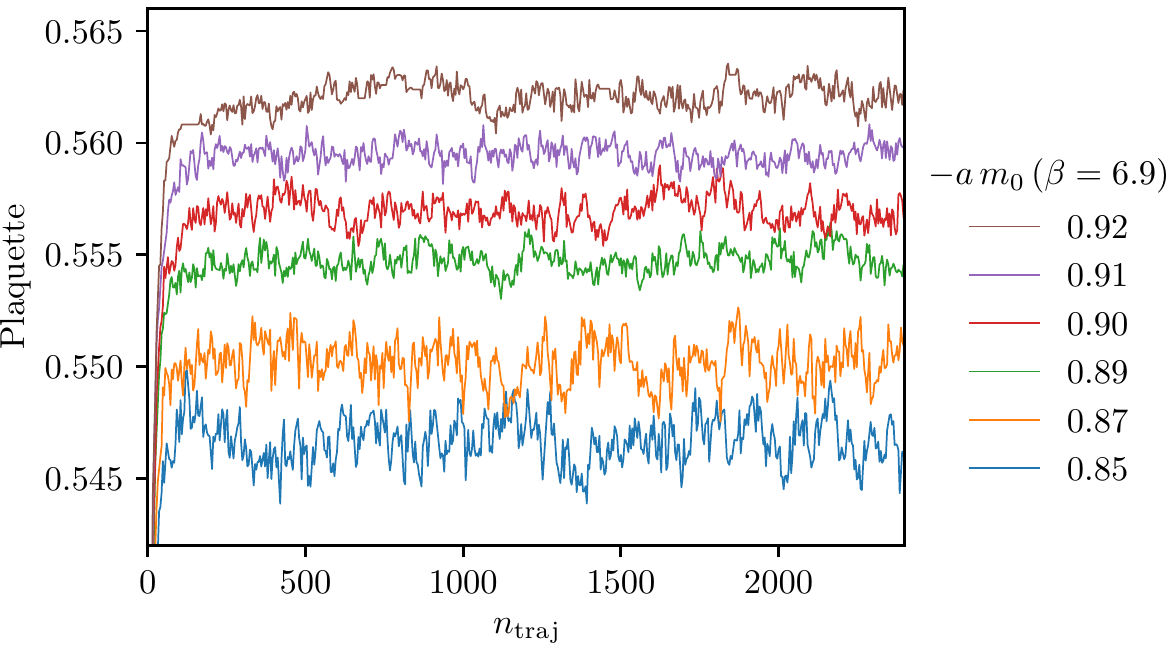}
\end{center}
\caption{
Trajectories of plaquette values for dynamical-fermion calculations at $\beta=6.9$.
Different colours represent  various fermion masses, as reported in the legend. 
}
\label{Fig:meff_dyn}
\end{figure}

\begin{figure}
\begin{center}
\includegraphics{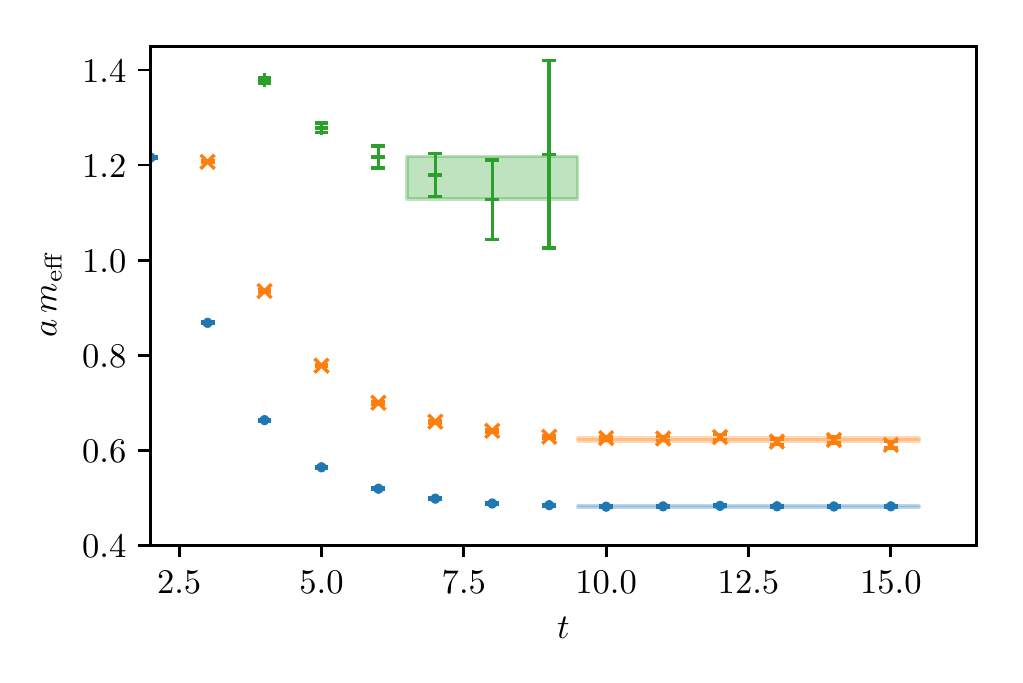}
\end{center}
\caption{
Effective mass plots for pseudo-scalar, vector, and axial-vector mesons at $\beta=6.9$ and $a m_0=-0.91$. 
The shaded regions denote the fitting ranges and the statistical uncertainties of fit results. 
}
\label{Fig:meff_dyn1}
\end{figure}

The study of strongly-coupled gauge theories on  discretised Euclidean space-time
 assumes  the existence of a proper continuum limit 
as the lattice spacing $a$ decreases, so that the  field-theory dynamics is recovered.
In order to avoid uncontrollable systematic effects in the continuum extrapolation, one must explore the lattice parameter space 
to identify any singularities or bulk transitions before carrying out detailed numerical studies 
of physical observables. 
In particular, a bulk phase where lattice discretisation effects
dominate the behaviour of the system is expected to be present at strong coupling, with the interesting physical region 
separated by this bulk region by a first-order phase transition, or a very sharp cross over.

The identification of  the associated (pseudo-)critical coupling is also strongly desired for practical purposes,
because with finite numerical resources one cannot reduce the lattice spacing to  arbitrarily small values, 
while at the same time using  large enough volumes. 
For the pure $Sp(4)$ lattice theory with the standard plaquette action 
the numerical study in~\cite{Holland:2003kg} shows the absence of bona-fide bulk phase transitions. 
To the best of our knowledge, no such a study for dynamical simulation with $N_f=2$ Wilson fermions
exists in the literature.

A possible choice of  order parameter associated with the lattice bulk transition 
is the expectation value of the plaquette. 
To have a rough mapping of the transition, 
we first scan the parameter space over the range of $\beta=[6.0,~7.0]$ using a $4^4$ lattice,
and show the results  
in Fig.~\ref{Fig:plaq_mass_scan}. 
For each lattice coupling, we calculate the average plaquette values and
vary the bare mass in steps of $0.1$, over the range $0.0\leq -a m_0 \leq 1.4$. 
Near the region in which the change of the  plaquette value is large as a function of $a m_0$, we add twice more data points, 
to increase the resolution. The abrupt change of the plaquette expectation value, visible at smaller values of $\beta$, 
strongly suggests the presence of a bulk phase transition.

To find more concrete evidence of the bulk phase and determine 
the phase boundary, we increase the volume to $6^4$ and $8^4$ 
for two lattice couplings, $\beta=6.6$ and $6.8$. 
In Fig.~\ref{Fig:plaq_traj} we plot the trajectories of the average plaquette
measured on a $8^4$ lattice with various values of bare quark mass close to  the transition. 
All configurations are generated from a cold start---the individual link is the unit matrix.
The top panel for $\beta=6.6$ shows evidence of metastability at the critical mass 
which is expected for a first-order bulk phase transition. 
By comparison, in the bottom panel of Fig.~\ref{Fig:plaq_traj}, obtained for $\beta=6.8$, 
 the plaquette values varies smoothly. 

These results are further supported by measuring the plaquette susceptibility
$
\chi= 
(\langle \mathcal{P}^2\rangle-\langle \mathcal{P}\rangle^2)\,V\,,
$
and investigating the  dependence of its maximum on the lattice four-volume $V$, as shown in Fig.~\ref{Fig:plaq_suscept}. 
The peaks of $\chi$ from the largest two volumes roughly scale with $V$ for $\beta=6.6$, 
while they are almost constant for $\beta=6.8$, indicating cross-over behaviour.

From the combination of all these numerical results, we find that a conservative estimate 
of the minimum value of $\beta$ that ensures the existence of
 the continuum limit is $\beta \geq 6.8$. 
Based on this finding we perform fully dynamical simulations at $\beta=6.9$ 
as a very preliminary study of the meson spectrum. 
We generate six ensembles, $a m_0=-0.85,\,-0.87$ on a $24\times12^3$ and $a m_0=-0.89,\,-0.9,\,-0.91,\,-0.92$ on a $32\times 16^3$ lattices, using the HMC algorithm.

In Fig.~\ref{Fig:meff_dyn}, we show the trajectories of the plaquette values. 
The asymptotic value of plaquette gradually increases as the bare fermion mass decreases. 
The typical thermalisation time appears to be $n_{\textrm{traj.}}\sim 300$, while the typical autocorrelation time 
ranges from $12$ to $32$, depending on the ensembles. 
The MD time steps for gauge and fermion actions are optimized such that 
the acceptance rate in the Metropolis test is in the range of $75-85\%$. 

Fig.~\ref{Fig:meff_dyn1} shows the effective masses for 
pseudo-scalar, vector, and axial-vector mesons at $a m_0=-0.91$, as an illustrative example. 
The pseudo-scalar and vector mesons clearly exhibit plateaux at large time,
starting around $t=10$ and persisting over six time slices. 
These two mesons are much lighter than the UV cutoff and we find that  $\frac{m_{\pi}}{m_{\rho}} \sim 0.8$.
The best estimation of the axial-vector meson mass is carried out by fitting 
the effective mass for $t=[7\,,\, 9]$,
but the resulting mass is already at the scale of the UV cutoff $\sim1/a$, as visible from Fig.~\ref{Fig:meff_dyn1}. 
More interesting numerical studies including the EFT fits are 
beyond the aims of this paper: 
a dedicated investigation of systematic effects, such as those relevant to the scale-setting and the continuum extrapolation, 
has to be carried out, before we perform a detailed analysis of confronting the lattice data with the EFT.

\section{Summary and Outlook}
\label{Sec:outlook}

With this paper (see also~\cite{Bennett:2017ttu,Bennett:2017tum,Bennett:2017kbp}), we have started a
 programme of  systematic lattice studies of the dynamics of $Sp(2N)$ gauge
theories with $N_f=2$ fundamental Dirac fermions and $N>1$.
As explained in the Introduction, we envisage developing this programme along several distinct lines,
of relevance in the contexts of composite-Higgs phenomenology, of composite fermions at strong coupling,
and of thermodynamics at finite $T$ and $\mu$. We are also interested in the pure Yang-Mills theories ($N_f=0$),
and in studying how the properties of these theories evolve as a function of $N$.
We outlined here the whole programme, and took some important steps along these lines. We focused for the time being
on $Sp(4)$, but constructed our numerical algorithms, data-analysis procedures and EFT treatment in such a way that 
they generalise straightforwardly to larger $N$.
We conclude the paper by summarising our findings, and how they allow us to proceed to the next steps in the near future.

We performed preliminary, technical analyses of the $Sp(4)$ lattice gauge theory, and the 
results are shown in Sections~\ref{Sec:Preliminary2} and~\ref{Sec:Phasespace}. 
We used two different lattice algorithms; we successfully checked that the Hybrid Monte Carlo and Heat Bath  both yield
results that are compatible with each other as well as with those reported previously in the literature, 
having introduced an adequate re-symplectization procedure.
We tested that the topological charge moves across sectors with dynamics suggesting good ergodicity properties
and that the distributions of the ensembles for various choices of the lattice parameters do 
not show appreciable indications of severe autocorrelation.
We also addressed the question of scale setting, by studying the gradient flow associated 
with the quantities $t_0$ and $w_0$
defined in the main body of the paper, and we found visible signals of quark-mass dependence.
From the field-theory perspective, this may not come as a surprise, given that the RG flow is two-dimensional and non-trivial.
Yet, it shows that in calculations with dynamical fermions one has to use extra caution in 
the process of extrapolating to the continuum limit. 
We expect that at least when the physical mass is
small compared to the confinement scale, and for lattice calculations performed 
close enough to the continuum limit, the RG flow be  driven mostly by the gauge coupling, with 
small dependence on the mass, as is the case of QCD~\cite{Borsanyi:2012zs}.
We would like to collect evidence of this with  values of lattice parameters
beyond those employed in this paper (but see also~\cite{Arthur:2016dir,Ayyar:2017qdf}).

The main body of this paper mostly  focused  on the dynamics of the glue.
In  Sec.~\ref{Sec:confinement} we studied in detail the pure (Yang-Mills) $Sp(4)$ theory, 
showed that it confines in a way that is compatible with the effective string description,
and performed the first detailed study of the spectrum of glueballs. We were able to perform the continuum-limit extrapolation
of the latter, hence providing a set of determinations for the physical masses  that is of quality 
comparable to the current state-of-the-art for other gauge theories.
We could hence compare the spectrum of $Sp(4)$ to that of $SU(N)$ gauge theories,
and in particular we found novel numerical support for two general expectations from the literature:  the ratio $R$
between the masses of the lightest spin-2 and spin-0 glueballs is independent of the gauge group~\cite{Athenodorou:2016ndx},
and the ratio of the mass of the lightest spin-0 glueball to the string tension obeys Casimir scaling~\cite{Hong:2017suj}.

The long-term objective of this programme is the investigation of whether
 composite-Higgs models of new physics based upon the $SU(4)/Sp(4)$ coset are realistic and predictive.
Having discussed the main features (and limitations) of the low-energy effective field theory 
description of pions, $\rho$, and $a_1$ mesons in Section~\ref{Sec:Preliminary1}, and postponing to the 
future the study of dynamical fermions, 
we performed a first, exploratory calculation of the masses and decay constants of the mesons
in the quenched approximation, and reported the results in Sec.~\ref{Sec:mesons}.
The main purpose of this study is to show that the whole technology works effectively.
We took particular care of precisely defining the operators of interest on the lattice, 
and of renormalizing the decay constants with
one-loop matching coefficients. We performed the calculations by varying the value of the bare mass
over a large range, while considering only two values of the lattice coupling $\beta$ and one choice of lattice volume.

We found several potentially interesting results for the mesons, although the preliminary nature of this quenched study 
implies that much caution has to be adopted.
While we found that the spectra and decay constants of pions, $\r$ and $a_1$ mesons 
can be fitted satisfactorily with the EFT description we provided, the spin-1 mesons are  heavy 
 with respect to the pion decay constant, and their coupling to the pions is large,
 hence bringing into serious question the reliability of the EFT description itself. 
 We do not know whether this feature persists also with dynamical fermions, yet
 we expect an improvement with larger values of $N$, and hence 
 find it encouraging that the fits 
 within quenched $Sp(4)$ work well.
We also found several other interesting features.
For example, the special combination $f_0$ (defined in Sec.~\ref{Sec:Preliminary1}) of the decay constants of the mesons
appears to be independent of the quark mass.  It will be interesting to test such features
beyond the quenched approximation, and possibly explaining them within field theory.

Finally, we uncovered evidence of a first-order (bulk) phase transition in the lattice theory 
with dynamical fermions, and presented
in Sec.~\ref{Sec:mass} the first
results of the coarse scanning of the parameter space, hence identifying 
regions that are safely connected to the field theory in the continuum limit.
We exemplified the calculations of the meson spectrum in the full dynamical theory for one choice of such parameters.
A much more extensive study of the spectra would be needed to match to the expectations from field theory, 
particularly because of the subtleties involved in taking the continuum limit for generic, 
non-trivial values of the fermion mass.
Having shown the feasibility of such a study with the instruments we put in place, we postpone to the future this extensive task.

The next steps of our programme will involve the following studies.
\begin{itemize}
\item We will compute the spectrum of glueballs in pure Yang-Mills for generic $Sp(2N)$. 
The HB algorithm has been already generalised to any $N$ and tested~\cite{Bennett:2017kbp}, and the process leading to 
the extraction of masses  has been shown here to be robust. 
This will allow us to put the level of understanding of the spectra of $Sp(2N)$ Yang-Mills 
theories on the same level as the $SU(N)$ ones.

\item The mass spectrum and decay constants of mesons will be studied with dynamical fermions, 
hence providing quantitative information of direct relevance to model-building and phenomenology in the context of 
composite-Higgs models of new physics.

\item We want to extend the present study to be of relevance to the context of composite fermions,
by generalising the underlying action to include fermionic matter in different representations. 
This is a novel direction for lattice studies, the very first such attempts  having appeared 
only recently~\cite{DeGrand:2015yna,Ayyar:2017qdf}.
We envision to perform  a preliminary study, possibly quenching part of the fermions, before attacking the
non-trivial (and model-dependent) problem of analysing the properties of fermionic composite states. 

\end{itemize}

Further in the future, we intend to extend the study of the mesons to other non-trivial dynamical properties 
of relevance to composite-Higgs models, such as the width
of the excited mesons, and the value of the condensates. 
We are also interested in extending to $Sp(2N)$ the study of the high temperature behaviour of the theory,
along the lines followed for $SU(2)\sim Sp(2)$ in~\cite{Lee:2017uvl}, and to introduce non-trivial  chemical potential.
Combinations of all these studies will provide a coherent framework within which
to gain new insight of relevance for field theory, model building, and thermodynamics in extreme conditions.

\vspace{1.0cm}
\begin{acknowledgments}
\end{acknowledgments}
MP would like to thank G.~Ferretti for useful discussions.
The work of EB  and DV has been funded in part by the Supercomputing Wales project, 
which is part-funded by the European Regional Development Fund (ERDF)
via Welsh Government.
The work of DKH and JWL  is supported in part by Korea Research Fellowship programme funded by the Ministry of Science,
 ICT and Future Planning through the National Research Foundation of Korea (2016H1D3A1909283) and under the framework of international 
 cooperation programme (NRF-2016K2A9A1A01952069).
The work of CJDL is supported by Taiwanese MoST grant 105-2628-M-009-003-MY4.
The work of BL is supported in part by the Royal Society and the Wolfson Foundation.
The work of BL, MP and DV has been supported in part by the STFC Consolidated Grants ST/L000369/1
and ST/P00055X/1.
\vspace{1.0cm}
\appendix
\section{Some useful elements of group theory}
\label{Sec:grouptheory}

We choose the generators of $SU(4)$ and of its $Sp(4)$ maximal subgroup 
as in Appendix B of~\cite{Lee:2017uvl}. We summarise  some useful properties of
the symplectic groups of  interest, which we
conventionally refer to  as $Sp(2N)$, and the real algebra of which is denoted $C_N$ in~\cite{Slansky:1981yr}.

The group $Sp(2N)$ is defined as the set of $2N\times 2N$ unitary matrices  $U$ with complex elements that satisfy the relation
\begin{equation}\label{eq:spcond1}
U \Omega U^T = \Omega,
\end{equation}
where $\Omega$ is the symplectic form, written---consistently with Eq.~(\ref{Eq:symplectic})---in $N\times N$ blocks as
\begin{equation}
\Omega = 
	\begin{bmatrix}
		0 & \mathbb{I}_N \\
		-\mathbb{I}_N & 0 \\
	\end{bmatrix}.
\end{equation}

As suggested by Eq.~(\ref{eq:spcond1}), $U$ may be written in block form as
\begin{equation}
\label{Eq:AB}
	U = 
	\begin{bmatrix}
		A & B \\
		-B^* & A^* \\
	\end{bmatrix}\,,
\end{equation}
where  $A$ and $B$  satisfy $A^\dag A + B^\dag B = \mathbb{I}$ and $A^TB = B^T A$.
From these relations, we can deduce many properties of $Sp(2N)$ matrices. 
Having unit determinant, the matrices of $Sp(2N)$ can be shown to  form a compact and simply connected subgroup of $SU(2N)$.
Moreover,
the structure in Eq.~(\ref{Eq:AB}) implies that 
the centre of the group is  
 isomorphic to $\mathbb{Z}_2$ for any $N$. 
Lastly, since $U^* = \Omega U \Omega^T$ and $\Omega \in Sp(2N)$, every representation of the group is equivalent to its complex conjugate. 
Thus $Sp(2N)$ has only pseudo-real representations, and charge conjugation is trivial. 

In model-building as well as numerical applications, a prominent role is played by the subgroups of $Sp(2N)$, 
especially those isomorphic to some $SU(N)$. In particular, one notices that $Sp(2N)\subset SU(2N)$, 
and that $Sp(2(N-1)) \subset Sp(2N)$. 
Starting with $Sp(2)\sim SU(2)$, this allows us to use the machinery already developed
 for the Monte Carlo simulation of $SU(2N)$ groups to the case of $Sp(2N)$. Particular 
 attention has to be given to the choice of subgroups, as we  further discuss in Appendix~\ref{Sec:projection}
in the  HB context.

 The subgroup structure of $Sp(2N)$ can be understood in terms of its algebra, to the study of which we now turn.
Locally, one can  represent a generic group element with the exponential map $U = \exp( \dot{\imath} H)$
 and impose the constraints of $Sp(2N)$. This is equivalent to taking only the generators of $SU(2N)$ that satisfy Eq.~(\ref{eq:spcond1}), 
i.e.~the hermitian traceless matrices with $H^* = \Omega H \Omega$, from which a block structure for $H$ follows,
\begin{equation}
	H = 
	\begin{bmatrix}
		A & B \\
		B^* & -A^* \\
	\end{bmatrix}.
\end{equation}

The properties $A=A^\dag$ and $B=B^T$ are a consequence of $H^\dag =H$. 
These conditions leave a total of $2N(N+1)$ degrees of freedom for $H$, which is also the dimension of the $Sp(2N)$ group.
The choice of generators that we use in this work is explicitly stated in~\cite{Lee:2017uvl}. 
The rank of the group is $N$, thus in $Sp(2N)$ we can find $N$ independent $SU(2)$ subgroups.
Once the elements of the algebra have been chosen, the $SU(2)$ subgroups of $Sp(2N)$ follow from their matrix structure
(see, once more,  Appendix~\ref{Sec:projection}).

\section{EFT and Technicolor}
\label{Sec:TC}

The 2-flavour $Sp(4)$ theory can be used as a technicolor (TC) model,
provided the embedding of the SM symmetries is such that the condensate breaks them,
and so can the EFT treatment we apply to the spin-1 states,
provided we identify the natural $SU(2)_L^t\times SU(2)_R^t$
symmetries acting on the left-handed and right-handed components of $Q^{i\,a}$
as the SM global symmetry (following the conventions in Appendix B in~\cite{Lee:2017uvl}).
We added the superscript $^t$  to the weakly-coupled gauge groups
to distinguish them from the embedding used in the body of this paper,  in the context of composite-Higgs models
The embedding is chosen so that $(SU(2)_L^t\times SU(2)_R^t )\cap Sp(4) \,= \,SU(2)$,
hence realising spontaneous symmetry breaking.
One can then match this model to the familiar electroweak chiral Lagrangian and its extensions,
based on the $(SU(2)\times SU(2))/SU(2)$ coset, which is practically advantageous as it allows to
re-use well known results.

Gauge invariance of the electroweak theory in this case requires setting $M=0$, so that the pions are massless,\footnote{To be rigorous, in the presence of a mass term
one should replace $M$ with a dynamical field
with infinitesimal kinetic term, hence reinstating explicitly gauge invariance. Because we focus only on the
transverse polarizations of the vectors, we do not worry about this otherwise important point.}
and to gauge $SU(2)_L^t\times U(1)_Y^t$, where the latter factor
is the subgroup of $SU(2)_R^t$ generated by the diagonal $t_R^3$.
The gauge couplings are $\tilde{g}$ and $\tilde{g}^{\prime}$, respectively.
By doing so, the exact symmetry is reduced from $SU(4)$ to $SU(2)^t_L\times U(1)^t_Y \times U(1)_B$,
with the last abelian factor denoting baryon number.
The condensate then breaks $SU(2)^t_L\times U(1)^t_Y  \rightarrow U(1)_{\rm e.m.}$,
and the $W$ and $Z$ bosons acquire a mass via the usual Higgs mechanism.

We discuss this TC model mostly for completeness,
and for technical reasons related to the calculation of the 2-point functions.
Among the phenomenological reasons why this is not a realistically viable TC model are the following.
\begin{itemize}
\item The spectrum contains two light (pseudo-)Goldstone bosons.
\item  Precision parameters such as $S$ might exceed experimental bounds.
\item The spectrum does not contain a light  scalar (Higgs) particle.
\item There is no high scale to suppress higher-dimensional operators and FCNC transitions.
\item There is no natural way to enhance the mass of the top quark.
\end{itemize}
We address here only the first two points, within the low-energy EFT, mostly for technical reasons
that are of interest also in the composite-Higgs framework.
We introduce the adjoint spurion $G$, formally transforming as
\beqs
G&\rightarrow & U_B G U_B^{\dagger}\,,
\eeqs
and fix it to
\beqs
G&=&\Lambda_G \,{\rm diag}\left\{\frac{}{}1\,,\,1\,,\,-1\,,\,-1\right\}\,,
\eeqs
We add to the Lagrangian density the additional (symmetry-breaking) term
\beqs
{\cal L}_G&=&\frac{\Lambda_G^2}{4}\Tr\left\{\frac{}{} G \left(S \Sigma S^T\right) G^{\ast} \left(S \Sigma S^T\right)^{\ast} \right\}\,,
\eeqs
in such a way as to induce the explicit breaking $SU(4)_B\rightarrow SU(2)_L^t\times SU(2)_R^t \times U(1)_B$.
By expanding explicitly we find that
\beqs
{\cal L}_{G}&=&\Lambda_G^4\,-\,\Lambda_{G}^4\left(\frac{2S}{F}+\frac{2C}{f}\right)^2\left((\bar\pi^{4})^{\,2}+(\bar\pi^{5})^{\,2}\frac{}{}\right)\,+\,\cdots\,,
\eeqs
where $C$ and $S$ have been defined in Eqs.~(\ref{Eq:defineS}) and~(\ref{Eq:defineC}).
We hence provided a mass for $\bar{\pi}^4$ and $\bar{\pi}^5$, while the massless  $\bar{\pi}^i$ fields, with $i=1,2,3$,
eventually become the longitudinal components of the $W$ and $Z$ bosons.
In the limit $\Lambda_G\rightarrow +\infty$, $\bar{\pi}^4$ and $\bar{\pi}^5$ decouple completely, while the dynamics of the spin-1 states
is unaffected, and the first of the phenomenological problems mentioned above is avoided.

We can now focus on the 2-point functions, by matching the theory defined by Eq.~(\ref{Eq:L}) onto the leading-order part of
the effective Lagrangian for the transverse components of the SM gauge bosons $V^i_{\mu}=(L^1_{\mu},L^2_{\mu},L^3_{\mu},R^3_{\mu})$,
which reads
\beqs
{\cal L}_{EFT}&=&\frac{1}{2}P^{\mu\nu}\Pi^{ij}(q^2)V_{\mu}^i(q)V_{\nu}^j(-q)\,,
\eeqs
where $P^{\mu\nu}=-q^{\mu}q^{\nu}/q^2+\eta^{\mu\nu}$, $q^{\mu}$ is the four-momentum, and all the dynamics is contained in the
non-trivial   
functions $\Pi^{ij}(q^2)$.

The functions $\Pi^{ij}$ are defined by matching the (gaussian) path integrals (at the tree-level). 
In practice, one takes the second derivatives in respect to the $15+3+1=19$ gauge bosons  $V_{\mu\,i}$
of the original theory $P^{\mu\nu}\Pi_T^{ij}\equiv P^{\mu\rho}\frac{\partial^2}{\partial V_{\rho i}\partial V_{\nu j}}{\cal L}=P^{\mu\nu}\left(q^2\delta^{ij}-({\cal M}^{2})^{\,ij}\right)$ 
(where ${\cal M}^2$ is the complete mass matrix of the vectors),
then one inverts it and retains only the $4\times 4$ sub-matrix
along the directions of $SU(2)_L^t\times U(1)_R^t$, and finally inverts it again to obtain  $\Pi^{ij}=\left(\left(\Pi_T^{-1}\right)_{L,R}\right)^{-1}$.

Focusing on the $1$ and $2$ components of $\Pi^{ij}$,  one
can write $\Pi^{11}(q^2)=\Pi^{22}(q^2)=q^2-\frac{\tilde{g}^2}{4}\Sigma(q^2)+{\cal O}(\tilde{g}^2/g_{\r}^2)$,
and hence obtain the Left-Left current-current correlator  in Eq.~(\ref{Eq:Sigma}):
\beqs
\Sigma(q^2)&\equiv&\lim_{\tilde{g}\rightarrow 0}\frac{4}{\tilde{g}^2}\left(\frac{}{}q^2-\Pi^{11}(q^2)\right)\,.
\eeqs

By expanding $\Pi^{ij}(q^2)$ in powers of $q^2$ one also obtains the precision electroweak parameters.
They are defined by first normalising the fields $V_{\mu i}$ so that 
$\Pi^{\prime\,ii}(0)\equiv\left.\frac{\di}{\di q^2}\Pi^{ii}(q^2)\right|_{q^2=0} =1$ for $i=1\,,\,\cdots\,,\,4$,
and then defining~\cite{Barbieri:2004qk}
\beqs
\hat{S}&\equiv&\frac{\tilde{g}}{\tilde{g}^{\prime}}\Pi^{\prime\,L^3R^3}(0)\,, \,\,\,\,\,\,\,\,\,\,\,\,\,\,\,
\hat{T}\,\equiv\,\frac{1}{M_W^2}\left(\frac{}{}\Pi^{\prime\,L^3L^3}(0)-\Pi^{\prime\,L^2L^2}(0)\right)\,.
\eeqs 

In the current case, one finds that $\hat{T}=0$, because of the custodial $SU(2)$, 
while the experimental bounds obtained from precision parameters 
are $\hat{S}<0.003$ (at $3\sigma$ c.l.)~\cite{Barbieri:2004qk}.

A subtlety should be noted here: the structure of the effective Lagrangian describing all the vectors in the  case of relevance for TC
should include additional terms in respect to Eq.~(\ref{Eq:L}), as in this case the condensate 
 breaks the $SU(2)_L^t$ symmetry. For example, this yields kinetic mixing between the $W$ and $B$ gauge bosons,
which contributes directly to the $\hat{S}$ parameter, and can be thought of as the contribution to $\hat{S}$ coming from 
heavier composite states that have been integrated out. We do not include such terms, because they go beyond the purposes of this paper.

If one applies the formalism described here to the Lagrangian density in Eq.~(\ref{Eq:L}), what results is not the $S$ 
parameter,\footnote{The normalization of $S$ by Peskin and Takeuchi~\cite{Peskin:1991sw} is such that
$\frac{S}{4\pi}\equiv \frac{4}{\tilde{g}^2}\hat{S}$, which yields a result independent of the SM gauge coupling $\tilde{g}$, 
contrary to the phenomenologically more convenient $\hat{S}$~\cite{Barbieri:2004qk}.}
but rather the contribution to $S$ coming from $\r$ and $a_1$ mesons only, which we called $S_0$.
We notice that the following sum rule (occasionally referred to as zeroth order in the literature) is verified exactly
within the framework defined by Eq.~(\ref{Eq:L}):
\beqs
\frac{S_0}{4\pi}&=&\frac{f_{\r}^2}{M_{\r}^2}-\frac{f_{a_1}^2}{M_{a_1}^2}\,.
  \label{Eq:S0}
\eeqs
We can express this result
explicitly in terms of the coefficients in the EFT Lagrangian density:
\beqs
\frac{S_0}{4\pi}&=&
   \frac{2}{{g_{\r}^2}} \left(1+\kappa+m\,y_3-\frac{(1-\kappa-m\,y_4) (-b {f^2}+{F^2}+2 {m}
   ({v_2}-{v_1}))^2}{((2b+4) {f^2}+2m v_2-b f^2+{F^2}+2 {m}(v_2- {v_1}))^2}\right)\,.
   \label{Eq:S}
\eeqs
We stress once more that this result holds only within the EFT in Eq.~(\ref{Eq:L}), but not 
in the fundamental theory,\footnote{
We refer the reader to Ref.~\cite{Appelquist:2010xv} for an example of such calculations on the lattice, 
specialised to  the $SU(3)$ gauge theories with $n_f=2$ and $6$ fermions 
in the fundamental representation, and to~\cite{Boyle:2009xi} for an earlier calculation of $S$ within QCD.}  for which one would have to 
replace the right-hand side of Eq.~(\ref{Eq:S0}) with a summation over all possible spin-1 states, not just 
the ground state. To include such contributions in the EFT would require introducing explicit symmetry-breaking terms
 in $\cal L$ of the form of kinetic mixing terms that are forbidden within the composite-Higgs scenario, and are hence omitted here.

We conclude with an exercise. By making use of the quenched calculations reported in subsection~\ref{Sec:quench},
and of the fits of the low-energy couplings in Eq.~(\ref{Eq:L}), {   we find that
$S_0/4\pi\sim0.049$ (for $\beta=7.62$), and $S_0/4\pi\sim0.041$ (for $\beta=8.0$)}.
The large values of $S_0$, in respect to QCD (for which the same treatment would yield $S_0/4\pi\sim0.025$,
obtained by just replacing the masses and decay constants from experiments~\cite{Patrignani:2016xqp}), 
might be due, at least in part, to the dimension of the gauge group $Sp(4)$
being larger than that of QCD. Yet these numerical results show a significant decrease
in approaching the continuum limit, and are affected by  quenching, hence they must be taken as
just for illustration purposes.

\section{Projecting on $Sp(4)$}
\label{Sec:projection}

The update of local gauge links must be supplemented 
by a  projection  onto the $Sp(4)$ group manifold,
to remove machine-precision effects that would bring the algorithm outside of the
group manifold. In this Appendix we explain how we implement this re-symplectisation procedure.

The generic $Sp(2N)$ elements can be represented by $2N\times 2N$ complex matrices composed of quaternions, 
\beq
Q(x,\mu)=Q_{0}(x,\mu)\otimes \mathbb{I}_2+Q_{1}(x,\mu)\otimes e_1+Q_{2}(x,\mu)\otimes e_2+Q_{3}(x,\mu)\otimes e_3,
\label{eq:quaternion_proj}
\eeq
where the quaternion units are
\beq
\mathbb{I}_2=\left( \begin{array}{cc} 1 & 0 \\ 0 & 1 \end{array} \right),~
e_1=\left( \begin{array}{cc} i & 0 \\ 0 & -i \end{array} \right),~
e_2=\left( \begin{array}{cc} 0 & 1 \\ -1 & 0 \end{array} \right),~
e_3=\left( \begin{array}{cc} 0 & i \\ i & 0 \end{array} \right).
\eeq
 $Q_i(x,\mu)$ are $N\times N$ real matrices. 
Given the updated link variables $U(x,\mu)$ in the HMC algorithm, we project it onto the quaternion basis 
and normalise it, hence obtaining a symplectic $U(x,\mu)$ ahead of the Metropolis test. 

\begin{figure}
\begin{center}
\includegraphics{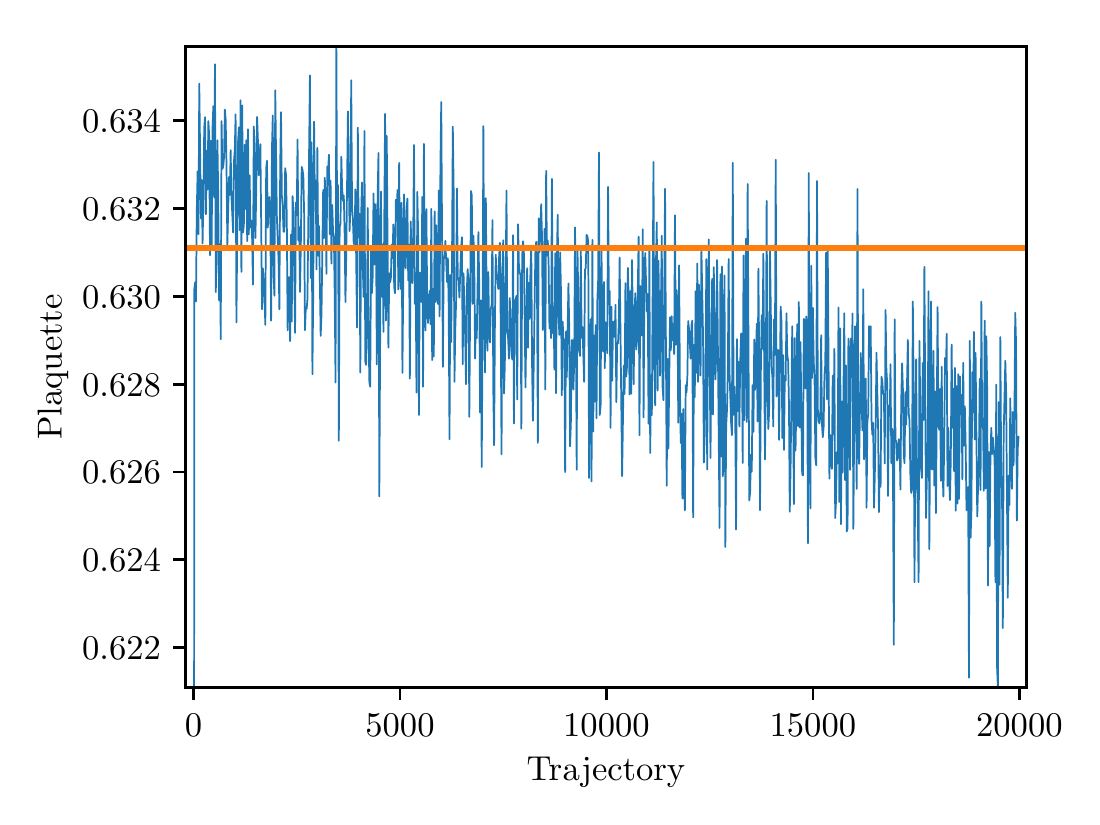}
\includegraphics{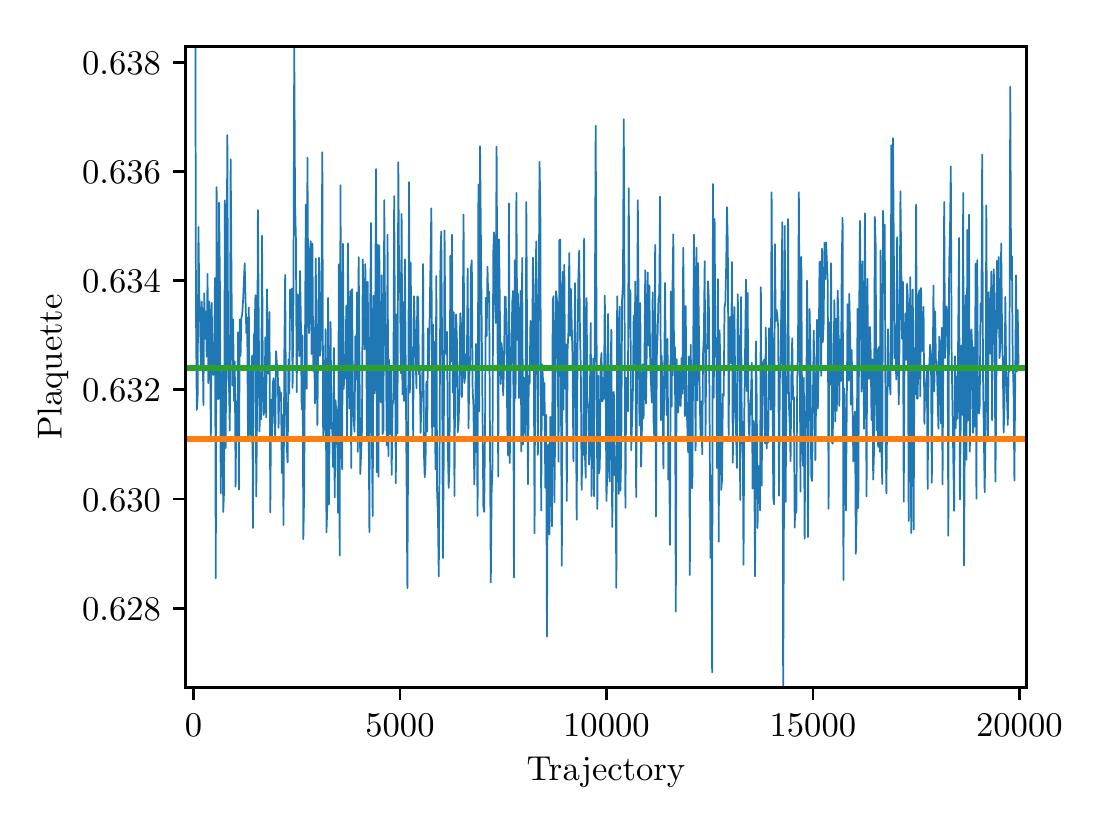}
\end{center}
\caption{
Trajectories of plaquette values without (top panel) and with (bottom panel) resymplectization, as described in Appendix~\ref{Sec:projection}. 
The orange band denotes the quenched result in~\cite{Holland:2003kg}, while the green band denotes the fit result. 
The lattice parameters are $\beta=8.0$, $m_0=1.0$, and $N_t\times N_s^3=8\times  8^3$. 
}
\label{Fig:plaq_resym}
\end{figure}

To gauge the potential size of the violation of symplectic conditions and the importance of the $Sp(4)$ projection, 
we plot in Fig.~\ref{Fig:plaq_resym} the trajectories of the plaquette values with (top) and without (bottom) re-symplectisation. 
The result shows that the re-symplectisation is successfully working and the plaquette values are stable.

In the HB calculations, we use instead a variant of the (modified) Gram-Schmidt algorithm. 
A re-symplectisation algorithm that inherits the numerical stability of the Gram-Schmidt process and that can be applied to any $N$ is 
obtained by noting that, owing to the general form of $Sp(2N)$ matrices, once the first $N$ columns of the matrix are known, the remaining ones can be computed from
\begin{equation}
	col_{j + N} = - \Omega~col_{j}^*.
\end{equation}

After normalising the first column, one can obtain the $(N+1)$-th. 
The second column is then obtained by orthonormalisation with respect to the first \emph{and} the $(N+1)$-th. 
Repeating the process for every column, 
one obtains an $Sp(2N)$ matrix.

\begin{thebibliography}{99}

\bibitem{Holland:2003kg} 
  K.~Holland, M.~Pepe and U.~J.~Wiese,
  Nucl.\ Phys.\ B {\bf 694}, 35 (2004)
  doi:10.1016/j.nuclphysb.2004.06.026
  [hep-lat/0312022].



\bibitem{Aad:2012tfa} 
  G.~Aad {\it et al.} [ATLAS Collaboration],
  Phys.\ Lett.\ B {\bf 716}, 1 (2012)
  doi:10.1016/j.physletb.2012.08.020
  [arXiv:1207.7214 [hep-ex]].
 
  \bibitem{Chatrchyan:2012xdj} 
  S.~Chatrchyan {\it et al.} [CMS Collaboration],
  Phys.\ Lett.\ B {\bf 716}, 30 (2012)
  doi:10.1016/j.physletb.2012.08.021
  [arXiv:1207.7235 [hep-ex]].

\bibitem{Kaplan:1983fs} 
  D.~B.~Kaplan and H.~Georgi,
  Phys.\ Lett.\  {\bf 136B}, 183 (1984).
  doi:10.1016/0370-2693(84)91177-8

\bibitem{Georgi:1984af} 
  H.~Georgi and D.~B.~Kaplan,
  Phys.\ Lett.\  {\bf 145B}, 216 (1984).
  doi:10.1016/0370-2693(84)90341-1


\bibitem{Dugan:1984hq} 
  M.~J.~Dugan, H.~Georgi and D.~B.~Kaplan,
  Nucl.\ Phys.\ B {\bf 254}, 299 (1985).
  doi:10.1016/0550-3213(85)90221-4

\bibitem{Agashe:2004rs} 
  K.~Agashe, R.~Contino and A.~Pomarol,
  Nucl.\ Phys.\ B {\bf 719}, 165 (2005)
  doi:10.1016/j.nuclphysb.2005.04.035
  [hep-ph/0412089].


\bibitem{Katz:2005au} 
  E.~Katz, A.~E.~Nelson and D.~G.~E.~Walker,
  JHEP {\bf 0508}, 074 (2005)
  doi:10.1088/1126-6708/2005/08/074
  [hep-ph/0504252].

\bibitem{Contino:2006qr} 
  R.~Contino, L.~Da Rold and A.~Pomarol,
  Phys.\ Rev.\ D {\bf 75}, 055014 (2007)
  doi:10.1103/PhysRevD.75.055014
  [hep-ph/0612048].

\bibitem{Barbieri:2007bh} 
  R.~Barbieri, B.~Bellazzini, V.~S.~Rychkov and A.~Varagnolo,
  Phys.\ Rev.\ D {\bf 76}, 115008 (2007)
  doi:10.1103/PhysRevD.76.115008
  [arXiv:0706.0432 [hep-ph]].


  
  \bibitem{Lodone:2008yy} 
  P.~Lodone,
  JHEP {\bf 0812}, 029 (2008)
  doi:10.1088/1126-6708/2008/12/029
  [arXiv:0806.1472 [hep-ph]].
  
  \bibitem{Ferretti:2013kya} 
  G.~Ferretti and D.~Karateev,
  JHEP {\bf 1403}, 077 (2014)
  doi:10.1007/JHEP03(2014)077
  [arXiv:1312.5330 [hep-ph]].
  
  \bibitem{Cacciapaglia:2014uja} 
  G.~Cacciapaglia and F.~Sannino,
  JHEP {\bf 1404}, 111 (2014)
  doi:10.1007/JHEP04(2014)111
  [arXiv:1402.0233 [hep-ph]].

 \bibitem{Arbey:2015exa} 
  A.~Arbey, G.~Cacciapaglia, H.~Cai, A.~Deandrea, S.~Le Corre and F.~Sannino,
  Phys.\ Rev.\ D {\bf 95}, no. 1, 015028 (2017)
  doi:10.1103/PhysRevD.95.015028
  [arXiv:1502.04718 [hep-ph]].
  
\bibitem{Vecchi:2015fma} 
  L.~Vecchi,
  JHEP {\bf 1702}, 094 (2017)
  doi:10.1007/JHEP02(2017)094
  [arXiv:1506.00623 [hep-ph]].

\bibitem{Panico:2015jxa} 
  G.~Panico and A.~Wulzer,
  Lect.\ Notes Phys.\  {\bf 913}, pp.1 (2016)
  doi:10.1007/978-3-319-22617-0
  [arXiv:1506.01961 [hep-ph]].

\bibitem{Ferretti:2016upr} 
  G.~Ferretti,
  JHEP {\bf 1606}, 107 (2016)
  doi:10.1007/JHEP06(2016)107
  [arXiv:1604.06467 [hep-ph]].


\bibitem{Agugliaro:2016clv} 
  A.~Agugliaro, O.~Antipin, D.~Becciolini, S.~De Curtis and M.~Redi,
  Phys.\ Rev.\ D {\bf 95}, no. 3, 035019 (2017)
  doi:10.1103/PhysRevD.95.035019
  [arXiv:1609.07122 [hep-ph]].
  
  {  
  \bibitem{Alanne:2017rrs} 
  T.~Alanne, D.~Buarque Franzosi and M.~T.~Frandsen,
  Phys.\ Rev.\ D {\bf 96}, no. 9, 095012 (2017)
  doi:10.1103/PhysRevD.96.095012
  [arXiv:1709.10473 [hep-ph]].
  }
  
  \bibitem{Feruglio:2016zvt} 
  F.~Feruglio, B.~Gavela, K.~Kanshin, P.~A.~N.~Machado, S.~Rigolin and S.~Saa,
  JHEP {\bf 1606}, 038 (2016)
  doi:10.1007/JHEP06(2016)038
  [arXiv:1603.05668 [hep-ph]].
  
  \bibitem{Fichet:2016xvs} 
  S.~Fichet, G.~von Gersdorff, E.~Pontón and R.~Rosenfeld,
  JHEP {\bf 1609}, 158 (2016)
  doi:10.1007/JHEP09(2016)158
  [arXiv:1607.03125 [hep-ph]].
  
  \bibitem{Galloway:2016fuo} 
  J.~Galloway, A.~L.~Kagan and A.~Martin,
  Phys.\ Rev.\ D {\bf 95}, no. 3, 035038 (2017)
  doi:10.1103/PhysRevD.95.035038
  [arXiv:1609.05883 [hep-ph]].
  
  {  
  \bibitem{Alanne:2017ymh} 
  T.~Alanne, D.~Buarque Franzosi, M.~T.~Frandsen, M.~L.~A.~Kristensen, A.~Meroni and M.~Rosenlyst,
  arXiv:1711.10410 [hep-ph].
  }
  
  \bibitem{Csaki:2017cep} 
  C.~Csaki, T.~Ma and J.~Shu,
  Phys.\ Rev.\ Lett.\  {\bf 119}, no. 13, 131803 (2017)
  doi:10.1103/PhysRevLett.119.131803
  [arXiv:1702.00405 [hep-ph]].
  
  \bibitem{Chala:2017sjk} 
  M.~Chala, G.~Durieux, C.~Grojean, L.~de Lima and O.~Matsedonskyi,
  JHEP {\bf 1706}, 088 (2017)
  doi:10.1007/JHEP06(2017)088
  [arXiv:1703.10624 [hep-ph]].
  
  



\bibitem{Gripaios:2009pe} 
  B.~Gripaios, A.~Pomarol, F.~Riva and J.~Serra,
  JHEP {\bf 0904}, 070 (2009)
  doi:10.1088/1126-6708/2009/04/070
  [arXiv:0902.1483 [hep-ph]].

\bibitem{Barnard:2013zea} 
  J.~Barnard, T.~Gherghetta and T.~S.~Ray,
  JHEP {\bf 1402}, 002 (2014)
  doi:10.1007/JHEP02(2014)002
  [arXiv:1311.6562 [hep-ph]].



\bibitem{Lewis:2011zb} 
  R.~Lewis, C.~Pica and F.~Sannino,
  Phys.\ Rev.\ D {\bf 85}, 014504 (2012)
  doi:10.1103/PhysRevD.85.014504
  [arXiv:1109.3513 [hep-ph]].


  
  \bibitem{Hietanen:2014xca} 
  A.~Hietanen, R.~Lewis, C.~Pica and F.~Sannino,
  JHEP {\bf 1407}, 116 (2014)
  doi:10.1007/JHEP07(2014)116
  [arXiv:1404.2794 [hep-lat]].
 
 
  
         \bibitem{Arthur:2016dir} 
  R.~Arthur, V.~Drach, M.~Hansen, A.~Hietanen, C.~Pica and F.~Sannino,
  Phys.\ Rev.\ D {\bf 94}, no. 9, 094507 (2016)
  doi:10.1103/PhysRevD.94.094507
  [arXiv:1602.06559 [hep-lat]].
  
  \bibitem{Arthur:2016ozw} 
  R.~Arthur, V.~Drach, A.~Hietanen, C.~Pica and F.~Sannino,
  arXiv:1607.06654 [hep-lat].
  
    \bibitem{Pica:2016zst} 
  C.~Pica, V.~Drach, M.~Hansen and F.~Sannino,
  EPJ Web Conf.\  {\bf 137}, 10005 (2017)
  doi:10.1051/epjconf/201713710005
  [arXiv:1612.09336 [hep-lat]].



  
 \bibitem{Detmold:2014kba} 
  W.~Detmold, M.~McCullough and A.~Pochinsky,
  Phys.\ Rev.\ D {\bf 90}, no. 11, 114506 (2014)
  doi:10.1103/PhysRevD.90.114506
  [arXiv:1406.4116 [hep-lat]].

\bibitem{Lee:2017uvl} 
  J.~W.~Lee, B.~Lucini and M.~Piai,
  JHEP {\bf 1704}, 036 (2017)
  doi:10.1007/JHEP04(2017)036
  [arXiv:1701.03228 [hep-lat]].
  
  \bibitem{Cacciapaglia:2015eqa} 
  G.~Cacciapaglia, H.~Cai, A.~Deandrea, T.~Flacke, S.~J.~Lee and A.~Parolini,
  JHEP {\bf 1511}, 201 (2015)
  doi:10.1007/JHEP11(2015)201
  [arXiv:1507.02283 [hep-ph]].
 

\bibitem{Hong:2017smd} 
  D.~K.~Hong,
  arXiv:1703.05081 [hep-ph].
  
    {  
  \bibitem{Golterman:2017vdj} 
  M.~Golterman and Y.~Shamir,
  arXiv:1707.06033 [hep-ph].
  }
  
 \bibitem{Bizot:2016zyu} 
  N.~Bizot, M.~Frigerio, M.~Knecht and J.~L.~Kneur,
  Phys.\ Rev.\ D {\bf 95}, no. 7, 075006 (2017)
  doi:10.1103/PhysRevD.95.075006
  [arXiv:1610.09293 [hep-ph]].
 
 

\bibitem{Lucini:2012gg} 
  B.~Lucini and M.~Panero,
  Phys.\ Rept.\  {\bf 526}, 93 (2013)
  doi:10.1016/j.physrep.2013.01.001
  [arXiv:1210.4997 [hep-th]].

\bibitem{Lucini:2004my}
B.~Lucini, M.~Teper and U.~Wenger,
  JHEP {\bf 0406}, 012 (2004)
  doi:10.1088/1126-6708/2004/06/012
  [hep-lat/0404008].

\bibitem{Athenodorou:2015nba} 
  A.~Athenodorou, R.~Lau and M.~Teper,
  Phys.\ Lett.\ B {\bf 749}, 448 (2015)
  doi:10.1016/j.physletb.2015.08.023
  [arXiv:1504.08126 [hep-lat]].


\bibitem{Lau:2017aom} 
  R.~Lau and M.~Teper,
  JHEP {\bf 1710}, 022 (2017)
  doi:10.1007/JHEP10(2017)022
  [arXiv:1701.06941 [hep-lat]].





  

\bibitem{Athenodorou:2016ndx} 
  A.~Athenodorou, E.~Bennett, G.~Bergner, D.~Elander, C.-J.~D.~Lin, B.~Lucini and M.~Piai,
  JHEP {\bf 1606}, 114 (2016)
  doi:10.1007/JHEP06(2016)114
  [arXiv:1605.04258 [hep-th]].
 
\bibitem{Hong:2017suj} 
  D.~K.~Hong, J.~W.~Lee, B.~Lucini, M.~Piai and D.~Vadacchino,
  Phys.\ Lett.\ B {\bf 775}, 89 (2017)
  doi:10.1016/j.physletb.2017.10.050
  [arXiv:1705.00286 [hep-th]].
 
\bibitem{Bando:1984ej} 
  M.~Bando, T.~Kugo, S.~Uehara, K.~Yamawaki and T.~Yanagida,
  Phys.\ Rev.\ Lett.\  {\bf 54}, 1215 (1985).
  doi:10.1103/PhysRevLett.54.1215
\bibitem{Casalbuoni:1985kq} 
  R.~Casalbuoni, S.~De Curtis, D.~Dominici and R.~Gatto,
  Phys.\ Lett.\  {\bf 155B}, 95 (1985).
  doi:10.1016/0370-2693(85)91038-X
\bibitem{Bando:1987br} 
  M.~Bando, T.~Kugo and K.~Yamawaki,
  Phys.\ Rept.\  {\bf 164}, 217 (1988).
  doi:10.1016/0370-1573(88)90019-1
  \bibitem{Casalbuoni:1988xm} 
  R.~Casalbuoni, S.~De Curtis, D.~Dominici, F.~Feruglio and R.~Gatto,
  Int.\ J.\ Mod.\ Phys.\ A {\bf 4}, 1065 (1989).
  doi:10.1142/S0217751X89000492
\bibitem{Harada:2003jx} 
  M.~Harada and K.~Yamawaki,
  Phys.\ Rept.\  {\bf 381}, 1 (2003)
  doi:10.1016/S0370-1573(03)00139-X
  [hep-ph/0302103].

\bibitem{Georgi:1989xy} 
  H.~Georgi,
  Nucl.\ Phys.\ B {\bf 331}, 311 (1990).
  doi:10.1016/0550-3213(90)90210-5
  
  
\bibitem{Appelquist:1999dq} 
  T.~Appelquist, P.~S.~Rodrigues da Silva and F.~Sannino,
  Phys.\ Rev.\ D {\bf 60}, 116007 (1999)
  doi:10.1103/PhysRevD.60.116007
  [hep-ph/9906555].
  
\bibitem{Piai:2004yb} 
  M.~Piai, A.~Pierce and J.~G.~Wacker,
  hep-ph/0405242.

  
  \bibitem{Franzosi:2016aoo} 
  D.~Buarque Franzosi, G.~Cacciapaglia, H.~Cai, A.~Deandrea and M.~Frandsen,
  JHEP {\bf 1611}, 076 (2016)
  doi:10.1007/JHEP11(2016)076
  [arXiv:1605.01363 [hep-ph]].

{  
\bibitem{DeGrand:2016pgq} 
  T.~DeGrand, M.~Golterman, E.~T.~Neil and Y.~Shamir,
  Phys.\ Rev.\ D {\bf 94}, no. 2, 025020 (2016)
  doi:10.1103/PhysRevD.94.025020
  [arXiv:1605.07738 [hep-ph]].
}

\bibitem{Luscher:1991cf} 
  M.~Luscher,
  Nucl.\ Phys.\ B {\bf 364}, 237 (1991).
  doi:10.1016/0550-3213(91)90584-K
\bibitem{Feng:2010es} 
  X.~Feng, K.~Jansen and D.~B.~Renner,
  Phys.\ Rev.\ D {\bf 83}, 094505 (2011)
  doi:10.1103/PhysRevD.83.094505
  [arXiv:1011.5288 [hep-lat]].
 



\bibitem{DelDebbio:2008zf} 
  L.~Del Debbio, A.~Patella and C.~Pica,
  Phys.\ Rev.\ D {\bf 81}, 094503 (2010)
  doi:10.1103/PhysRevD.81.094503
  [arXiv:0805.2058 [hep-lat]].

\bibitem{Cabibbo:1982zn} 
  N.~Cabibbo and E.~Marinari,
  Phys.\ Lett.\  {\bf 119B}, 387 (1982).
  doi:10.1016/0370-2693(82)90696-7



{  
\bibitem{Takaishi:1999bi} 
  T.~Takaishi,
  Comput.\ Phys.\ Commun.\  {\bf 133}, 6 (2000)
  doi:10.1016/S0010-4655(00)00161-2
  [hep-lat/9909134].
}

{  
\bibitem{Creutz:1988wv} 
  M.~Creutz,
  Phys.\ Rev.\ D {\bf 38}, 1228 (1988).
  doi:10.1103/PhysRevD.38.1228
}

\bibitem{Luscher:2009eq} 
  M.~Luscher,
  Commun.\ Math.\ Phys.\  {\bf 293}, 899 (2010)
  doi:10.1007/s00220-009-0953-7
  [arXiv:0907.5491 [hep-lat]].


\bibitem{Luscher:2010iy} 
  M.~Lüscher,
  JHEP {\bf 1008}, 071 (2010)
  Erratum: [JHEP {\bf 1403}, 092 (2014)]
  doi:10.1007/JHEP08(2010)071, 10.1007/JHEP03(2014)092
  [arXiv:1006.4518 [hep-lat]].
 

\bibitem{Luscher:2011bx} 
  M.~Luscher and P.~Weisz,
  JHEP {\bf 1102}, 051 (2011)
  doi:10.1007/JHEP02(2011)051
  [arXiv:1101.0963 [hep-th]].
 
  
  \bibitem{Fujikawa:2016qis} 
  K.~Fujikawa,
  JHEP {\bf 1603}, 021 (2016)
  doi:10.1007/JHEP03(2016)021
  [arXiv:1601.01578 [hep-lat]].
  
  \bibitem{Borsanyi:2012zs} 
  S.~Borsanyi {\it et al.},
  JHEP {\bf 1209}, 010 (2012)
  doi:10.1007/JHEP09(2012)010
  [arXiv:1203.4469 [hep-lat]].
  
  
  
\bibitem{Fodor:2014cpa} 
  Z.~Fodor, K.~Holland, J.~Kuti, S.~Mondal, D.~Nogradi and C.~H.~Wong,
  JHEP {\bf 1409}, 018 (2014)
  doi:10.1007/JHEP09(2014)018
  [arXiv:1406.0827 [hep-lat]].
  
\bibitem{Ramos:2015baa} 
  A.~Ramos and S.~Sint,
  Eur.\ Phys.\ J.\ C {\bf 76}, no. 1, 15 (2016)
  doi:10.1140/epjc/s10052-015-3831-9
  [arXiv:1508.05552 [hep-lat]].

\bibitem{Lin:2015zpa} 
  C.-J.~D.~Lin, K.~Ogawa and A.~Ramos,
  JHEP {\bf 1512}, 103 (2015)
  doi:10.1007/JHEP12(2015)103
  [arXiv:1510.05755 [hep-lat]].
  
\bibitem{Luscher:2011kk} 
  M.~Luscher and S.~Schaefer,
  JHEP {\bf 1107}, 036 (2011)
  doi:10.1007/JHEP07(2011)036
  [arXiv:1105.4749 [hep-lat]].


  
\bibitem{Galletly:2006hq}
 D.~Galletly, M.~Gurtler, R.~Horsley, H.~Perlt, P.~E.~L.~Rakow, G.~Schierholz, A.~Schiller and T.~Streuer,
  Phys.\ Rev.\ D {\bf 75}, 073015 (2007)
  doi:10.1103/PhysRevD.75.073015
  [hep-lat/0607024].
  
  \bibitem{Luscher:1980fr}
   M.~Luscher, K.~Symanzik and P.~Weisz,
  Nucl.\ Phys.\ B {\bf 173}, 365 (1980).
  doi:10.1016/0550-3213(80)90009-7
  
  \bibitem{Polchinski:1991ax}
  J.~Polchinski and A.~Strominger,
  Phys.\ Rev.\ Lett.\  {\bf 67}, 1681 (1991).
  doi:10.1103/PhysRevLett.67.1681
  
    \bibitem{Luscher:1980ac}
     M.~Luscher,
  Nucl.\ Phys.\ B {\bf 180}, 317 (1981).
  doi:10.1016/0550-3213(81)90423-5
    
   \bibitem{Lucini:2001nv}
B.~Lucini and M.~Teper,
  Phys.\ Rev.\ D {\bf 64}, 105019 (2001)
  doi:10.1103/PhysRevD.64.105019
  [hep-lat/0107007].
  
   \bibitem{Necco:2001xg}
    S.~Necco and R.~Sommer,
  Nucl.\ Phys.\ B {\bf 622}, 328 (2002)
  doi:10.1016/S0550-3213(01)00582-X
  [hep-lat/0108008].
     
   \bibitem{Athenodorou:2010cs}
   A.~Athenodorou, B.~Bringoltz and M.~Teper,
  JHEP {\bf 1102}, 030 (2011)
  doi:10.1007/JHEP02(2011)030
  [arXiv:1007.4720 [hep-lat]].
  
     \bibitem{Luscher:2004ib}
      M.~Luscher and P.~Weisz,
  JHEP {\bf 0407}, 014 (2004)
  doi:10.1088/1126-6708/2004/07/014
  [hep-th/0406205].
     
       \bibitem{Aharony:2009gg} 
 O.~Aharony and E.~Karzbrun,
  JHEP {\bf 0906}, 012 (2009)
  doi:10.1088/1126-6708/2009/06/012
  [arXiv:0903.1927 [hep-th]].

      \bibitem{Drummond:2004yp}
       J.~M.~Drummond,
  hep-th/0411017.
      
    \bibitem{HariDass:2006sd}
    N.~D.~Hari Dass and P.~Matlock,
  hep-th/0606265.
        
   \bibitem{Drummond:2006su}
    J.~M.~Drummond,
  hep-th/0608109.
          
\bibitem{Dass:2006ud}
 N.~D.~H.~Dass and P.~Matlock,
  hep-th/0611215.

\bibitem{Aharony:2013ipa}
 O.~Aharony and Z.~Komargodski,
  JHEP {\bf 1305}, 118 (2013)
  doi:10.1007/JHEP05(2013)118
  [arXiv:1302.6257 [hep-th]].

\bibitem{Dubovsky:2015zey}
S.~Dubovsky and V.~Gorbenko,
  JHEP {\bf 1602}, 022 (2016)
  doi:10.1007/JHEP02(2016)022
  [arXiv:1511.01908 [hep-th]].


\bibitem{Berg:1982kp} 
  B.~Berg and A.~Billoire,
  Nucl.\ Phys.\ B {\bf 221}, 109 (1983).
  doi:10.1016/0550-3213(83)90620-X
  
  \bibitem{Morningstar:1997ff} 
  C.~J.~Morningstar and M.~J.~Peardon,
  Phys.\ Rev.\ D {\bf 56}, 4043 (1997)
  doi:10.1103/PhysRevD.56.4043
  [hep-lat/9704011].

\bibitem{Morningstar:1999rf} 
  C.~J.~Morningstar and M.~J.~Peardon,
  Phys.\ Rev.\ D {\bf 60}, 034509 (1999)
  doi:10.1103/PhysRevD.60.034509
  [hep-lat/9901004].



\bibitem{Hoek:1986nd}
 J.~Hoek, M.~Teper and J.~Waterhouse,
  Nucl.\ Phys.\ B {\bf 288}, 589 (1987).
  doi:10.1016/0550-3213(87)90230-6

\bibitem{Lucini:2010nv}
  B.~Lucini, A.~Rago and E.~Rinaldi,
  JHEP {\bf 1008}, 119 (2010)
  doi:10.1007/JHEP08(2010)119
  [arXiv:1007.3879 [hep-lat]].





\bibitem{Lucini:2001ej}
 B.~Lucini and M.~Teper,
  JHEP {\bf 0106}, 050 (2001)
  doi:10.1088/1126-6708/2001/06/050
  [hep-lat/0103027].



\bibitem{Bernard:1992mk} 
  C.~W.~Bernard and M.~F.~L.~Golterman,
  Phys.\ Rev.\ D {\bf 46}, 853 (1992)
  doi:10.1103/PhysRevD.46.853
  [hep-lat/9204007].

\bibitem{Sharpe:1992ft} 
  S.~R.~Sharpe,
  Phys.\ Rev.\ D {\bf 46}, 3146 (1992)
  doi:10.1103/PhysRevD.46.3146
  [hep-lat/9205020].

\bibitem{Sharpe:2000bc} 
  S.~R.~Sharpe and N.~Shoresh,
  Phys.\ Rev.\ D {\bf 62}, 094503 (2000)
  doi:10.1103/PhysRevD.62.094503
  [hep-lat/0006017].

\bibitem{Sharpe:2001fh} 
  S.~R.~Sharpe and N.~Shoresh,
  Phys.\ Rev.\ D {\bf 64}, 114510 (2001)
  doi:10.1103/PhysRevD.64.114510
  [hep-lat/0108003].

\bibitem{Bernard:1993sv} 
  C.~W.~Bernard and M.~F.~L.~Golterman,
  Phys.\ Rev.\ D {\bf 49}, 486 (1994)
  doi:10.1103/PhysRevD.49.486
  [hep-lat/9306005].
  
  \bibitem{Martinelli:1982mw} 
  G.~Martinelli and Y.~C.~Zhang,
  Phys.\ Lett.\  {\bf 123B}, 433 (1983).
  doi:10.1016/0370-2693(83)90987-5

\bibitem{Lepage:1992xa} 
  G.~P.~Lepage and P.~B.~Mackenzie,
  Phys.\ Rev.\ D {\bf 48}, 2250 (1993)
  doi:10.1103/PhysRevD.48.2250
  [hep-lat/9209022].
  
  
\bibitem{Boyle:2008rh} 
  P.~A.~Boyle, A.~Juttner, C.~Kelly and R.~D.~Kenway,
  JHEP {\bf 0808}, 086 (2008)
  doi:10.1088/1126-6708/2008/08/086
  [arXiv:0804.1501 [hep-lat]].


  
\bibitem{Bennett:2017ttu} 
  E.~Bennett, D.~K.~Hong, J.~W.~Lee, C.-J.~D.~Lin, B.~Lucini, M.~Piai and D.~Vadacchino,
  arXiv:1710.06941 [hep-lat].

  \bibitem{Bennett:2017tum} 
  E.~Bennett, D.~K.~Hong, J.~W.~Lee, C.-J.~D.~Lin, B.~Lucini, M.~Piai and D.~Vadacchino,
  arXiv:1710.06715 [hep-lat].
  
\bibitem{Bennett:2017kbp} 
  E.~Bennett, D.~K.~Hong, J.~W.~Lee, C.-J.~D.~Lin, B.~Lucini, M.~Piai and D.~Vadacchino,
  arXiv:1710.07043 [hep-lat].

  \bibitem{Ayyar:2017qdf} 
  V.~Ayyar, T.~DeGrand, M.~Golterman, D.~C.~Hackett, W.~I.~Jay, E.~T.~Neil, Y.~Shamir and B.~Svetitsky,
  arXiv:1710.00806 [hep-lat].
  
\bibitem{DeGrand:2015yna} 
  T.~DeGrand and Y.~Shamir,
  Phys.\ Rev.\ D {\bf 92}, no. 7, 075039 (2015)
  doi:10.1103/PhysRevD.92.075039
  [arXiv:1508.02581 [hep-ph]].


\bibitem{Slansky:1981yr} 
  R.~Slansky,
  Phys.\ Rept.\  {\bf 79}, 1 (1981).
  doi:10.1016/0370-1573(81)90092-2
  
\bibitem{Barbieri:2004qk} 
  R.~Barbieri, A.~Pomarol, R.~Rattazzi and A.~Strumia,
  Nucl.\ Phys.\ B {\bf 703}, 127 (2004)
  doi:10.1016/j.nuclphysb.2004.10.014
  [hep-ph/0405040].
  

\bibitem{Peskin:1991sw} 
  M.~E.~Peskin and T.~Takeuchi,
  Phys.\ Rev.\ D {\bf 46}, 381 (1992).
  doi:10.1103/PhysRevD.46.381

\bibitem{Appelquist:2010xv} 
  T.~Appelquist {\it et al.} [LSD Collaboration],
  Phys.\ Rev.\ Lett.\  {\bf 106}, 231601 (2011)
  doi:10.1103/PhysRevLett.106.231601
  [arXiv:1009.5967 [hep-ph]].
     
 \bibitem{Boyle:2009xi} 
  P.~A.~Boyle {\it et al.} [RBC and UKQCD Collaborations],
  Phys.\ Rev.\ D {\bf 81}, 014504 (2010)
  doi:10.1103/PhysRevD.81.014504
  [arXiv:0909.4931 [hep-lat]].
 
\bibitem{Patrignani:2016xqp} 
  C.~Patrignani {\it et al.} [Particle Data Group],
  Chin.\ Phys.\ C {\bf 40}, no. 10, 100001 (2016).
  doi:10.1088/1674-1137/40/10/100001

\end{thebibliography}
\end{document}